\documentclass[letterpaper,aps,prd,groupedaddress,superscriptaddress,preprintnumbers,floatfix,nofootinbib]{revtex4}  
\usepackage{graphicx,comment}  
\usepackage{bm}        
\usepackage{amssymb}   

\usepackage{subfigure}
\usepackage{multirow}
\usepackage{color}

\usepackage{amsfonts}    
\usepackage{amsmath}
\usepackage{dcolumn}

\begin{document}

\preprint{
\vbox{
\hbox{MIT-CTP/5069}
}}

\title{Gluon gravitational form factors of the nucleon and the pion from lattice QCD}

\author{P.~E.~Shanahan}\affiliation{Center for Theoretical Physics, Massachusetts Institute of Technology, Cambridge, MA 02139, U.S.A.}
\affiliation{Perimeter Institute for Theoretical Physics, Waterloo, Ontario N2L 2Y5, Canada}
\author{W.~Detmold}      \affiliation{Center for Theoretical Physics, Massachusetts Institute of Technology, Cambridge, MA 02139, U.S.A.}
	
\begin{abstract}
A future Electron-Ion Collider will enable the gluon contributions to the gravitational form factors of the proton to be constrained experimentally for the first time. Here, the first calculation of these form factors from lattice Quantum Chromodynamics is presented. The calculations use a larger-than-physical value of the light quark mass corresponding to $m_\pi \sim 450$ MeV. All three form factors, which encode the momentum-dependence of the lowest moment of the spin independent gluon generalised parton distributions and are related to different components of the energy-momentum tensor, are resolved. In particular, the gluon $D$-term form factor, related to the pressure distribution inside the nucleon, is determined for the first time. The gluon contributions to the two gravitational form factors of the pion are also determined, and are compared to existing lattice determinations of the quark contributions to the gravitational form factors and to phenomenology. 
\end{abstract}

\maketitle

\section{Introduction}

A defining challenge for hadronic physics research is to achieve a quantitative understanding of the structure of the proton and other hadrons in terms of their fundamental quark and gluon constituents. 
Generalised parton distributions (GPDs)~\cite{Mueller:1998fv,Ji:1996nm,Radyushkin:1997ki} provide a framework for a three-dimensional encoding of this structure. These distributions (see Refs.~\cite{Ji:1998pc,Diehl:2003ny,Belitsky:2005qn} for reviews) combine and generalise the features of elastic form factors, which describe the charge and magnetisation distributions of the hadron as seen by a photon of a given virtuality, and parton densities, which describe the longitudinal partonic composition of a fast moving hadron. Moreover, the nucleon GPDs encode, for example, the nucleon mass and spin, the quark and gluon contributions to the nucleon angular momentum, and various inter- and multi-parton correlations. They are also directly related to the `mechanical properties' of the nucleon system, such as the pressure distributions and shear forces~\cite{Polyakov:2002wz,Polyakov:2002yz,Polyakov:2018zvc}. Given their importance in hadron structure, there are significant experimental and theory efforts targeted at determining GPDs, especially for the proton.

In particular, proton and neutron quark GPDs have been constrained in limited kinematic regions by deeply-virtual Compton scattering (DVCS) and deeply-virtual meson production (DVMP) experiments at Thomas Jefferson National Accelerator Facility (JLab) \cite{Niccolai:2016sdf}, HERA \cite{Capua:2012sd}, and COMPASS \cite{Sandacz:2018zsy} (Refs.~\cite{Favart:2015umi,dHose:2016mda,Kumericki:2016ehc} summarise the world datasets). Ongoing studies at COMPASS and within the 12 GeV program at JLab will significantly improve extractions of these quantities. The quark GPDs have also been studied theoretically in a number of frameworks, including determinations of their connection to different experiments and evolution with factorisation and renormalisation scales~\cite{Mueller:1998fv,Ji:1996nm,Radyushkin:1997ki,Ji:1998pc,Diehl:2003ny,Belitsky:2005qn}, estimates of their forms in hadronic models (see e.g. Refs.~\cite{Goeke:2001tz,Kumericki:2016ehc} for reviews), and calculations of their lowest few Mellin moments for the pion \cite{Brommel:2007xd} and nucleon \cite{Gockeler:2003jfa,Hagler:2003jd,Diehl:2005jf,Gockeler:2005cj,Gockeler:2006zu,Hagler:2007xi,Bratt:2010jn,Alexandrou:2011nr} in lattice QCD (LQCD) (see e.g. Ref.~\cite{Hagler:2009ni} for a review). 
Gluon GPDs, on the other hand, are as-yet unknown from experiment or theory~\cite{Boer:2011fh}. Performing first measurements of these quantities is a key goal of the planned Electron-Ion Collider (EIC)~\cite{Accardi:2012qut,Kalantarians:2014eda}, and theory constraints on the gluon GPDs will provide important information as the physics case for an EIC is refined.

This manuscript presents the first LQCD determination of the complete set of gluon gravitational form factors (GFFs) of the nucleon, which are defined as the lowest moments of the spin-independent gluon GPDs. The calculations are undertaken with a larger-than-physical value of the light quark mass that corresponds to a pion mass $m_\pi \sim 450$~MeV. All three gluon GFFs of the nucleon are determined at discrete values of the squared momentum transfer $t$ up to $|t| \sim$ 2~GeV$^2$, as are the two gluon GFFs of the pion. The results are presented in a modified minimal subtraction ($\overline{\rm MS}$) scheme at a renormalisation scale $\mu=2$ GeV,  by performing a non-perturbative renormalisation using a RI-MOM scheme~\cite{Martinelli:1994ty} and a perturbative matching to $\overline{\rm MS}$. Mixing of the gluon GFFs with the corresponding quark distributions is neglected; lattice perturbation theory calculations~\cite{Alexandrou:2016ekb} indicate that such effects are likely to be small compared with the statistical and other systematic uncertainties of this study.
For both the pion and the nucleon, the gluon momentum fraction is found to be approximately 0.5--0.6, somewhat larger than the phenomenological value in both cases, while the fractional contributions of gluons to the nucleon lightcone momentum and angular momentum are found to be  consistent within uncertainties. The $|t|$-dependences of two of the three nucleon gluon GFFs are consistent with dipole forms, while the third GFF shows no $|t|$-dependence and is consistent with zero. Comparing the results with those of previous calculations of the quark GFFs using similar lattice discretisations and at similar values of the quark masses reveals that the gluon radii defined as the slopes of the nucleon gluon GFFs at $t=0$ are larger than the corresponding quark radii for each form factor. 
The pion gluon GFFs also have dipole-like dependences on $|t|$, but are consistent with the corresponding quark GFFs within uncertainties, revealing no clear ordering of pion quark and gluon radii. Compared with the nucleon gluon GFFs, the pion gluon GFFs define consistently smaller radii, consistent with the ordering of the nucleon and pion charge radii (defined from the electric form factors) determined experimentally.

In the following section, the quark and gluon GPDs and GFFs of the nucleon and pion are defined. Section \ref{sec:latt} details the LQCD calculations that are performed to extract the gluon GFFs, while the results of those calculations are presented in Section~\ref{sec:results}. The extracted gluon GFFs are compared with the corresponding quark GFFs, which have been previously calculated using similar lattice discretisations at quark masses corresponding to a similar value of the pion mass. Earlier LQCD results for the gluon momentum fractions of the nucleon and pion, defined as the forward limits of the appropriate GFFs, are also collated for comparison. Finally, Sec.~\ref{sec:conc} highlights the conclusions that can be drawn from this study.

\section{Gluon GPDs and GFFs}

\subsection{Nucleon}

GPDs encode the three-dimensional distribution of quarks and gluons in the nucleon~\cite{Mueller:1998fv,Ji:1996nm,Radyushkin:1997ki,Ji:1998pc,Belitsky:2005qn,Diehl:2003ny}. 
In the deep inelastic regime, the leading contributions arise from the lowest-twist operators. For the nucleon, the leading spin-independent quark and gluon distributions are twist-two~\cite{Hoodbhoy:1998vm,Ji:2013dva,Zhang:2018diq}, and, following the conventions of Ref.~\cite{Diehl:2003ny}, can be expressed in terms of matrix elements of non-local light-ray operators as:
\begin{align}\nonumber 
 \int_{-\infty}^\infty  {d\lambda \over 2\pi} &e^{i\lambda x}
\langle p',s'|\bar\psi_q(-{\frac{\lambda}{2} n})\gamma^\mu {\cal U}_{\left[-\frac{\lambda}{2}n,\frac{\lambda}{2}n\right]}
\psi_q(\frac{\lambda}{2} n)|p,s \rangle\\\label{eq:GPD1}
&= H_q(x,\xi,t) \bar U(p',s')\gamma^\mu U(p,s)  + E_q(x,\xi,t) \bar U(p',s'){i\sigma^{\mu\nu}
	\Delta_{\nu}
	\over 2M}U(p,s) + ...  \ , \\
 \int_{-\infty}^\infty { d\lambda\over 2\pi} &e^{i\lambda x}
\langle p',s'|G_a^{\{\mu\alpha}(-{\lambda\over2}n)
 \left[{\cal U}^{(A)}_{\left[-\frac{\lambda}{2}n,\frac{\lambda}{2}n\right]}\right]_{ab}
G_{b\alpha}^{~~\nu\}}({\lambda\over2}n)
 |p,s\rangle   
\nonumber \\\label{eq:GPD2}  & = 
\frac{1}{2}\left(H_g(x,\xi,t) \bar U(p',s') P^{\{\mu} \gamma^{\nu \}} U(p,s)
+ E_g(x,\xi,t) \bar U(p',s'){ P^{\{\mu} i\sigma^{\nu\}\alpha }
	\Delta_\alpha \over2M} U(p,s)\right) + ...\ ,
\end{align} 
where $\psi_q$ is a quark field of flavour $q$, $G^a_{\mu\nu}=(\partial_\mu A^a_\nu - \partial_\nu A^a_\mu +g f^{abc}A_\mu^b A_\nu^c)$ is the gluon field strength tensor built from the gluon field $A^a_\mu$, and the ellipses denote structures with twist greater than two. Here, $n^\mu$ is a light-like vector with $n^2=0$, the momenta and spins of the initial and final nucleons are $(p,s)$ and $(p',s')$ respectively, and it is convenient to define $P= \frac12(p^\prime + p)$, $\Delta = p^\prime-p$, $t=\Delta^2$, Bjorken $x=\frac{1}{2}\Delta^2/p\cdot\Delta$, and skewness $\xi=-\frac{1}{2}n\cdot \Delta/n\cdot P$. The path-ordered gauge links in the fundamental and adjoint representations are
\begin{eqnarray}
{\cal U}_{[z_2,z_1]} = {\cal P} {\rm exp}\left[ig\int_{z_1}^{z_2} d\lambda^\prime n^\mu A_\mu^a(\lambda^\prime n)\,t_a\right]
,\qquad
\left[{\cal U}^{(A)}_{[z_2,z_1]}\right]_{bc} = {\cal P} {\rm exp}\left[g  f_{abc} \int_{z_1}^{z_2} d\lambda^\prime n^\mu A_\mu^a(\lambda^\prime n)\right],
\end{eqnarray}
where $t_a$ are SU(3) generators in the fundamental representation and $f_{abc}$ are the structure constants defining the adjoint representation.
The inclusion of the gauge links in Eqs.~\eqref{eq:GPD1} and \eqref{eq:GPD2} ensures the gauge-invariance of these expressions (in the case of the gluon operator, alternate gauge link choices are also possible~\cite{Boer:2016dlh}).
Braces denote symmetrisation and trace-subtraction in the free indices, i.e., $a_{\{\mu} b_{\nu\}}=\frac{1}{2} (a_\mu b_\nu + a_\nu b_\mu)-\frac{1}{4}g_{\mu\nu}a_\alpha b^\alpha$, and the covariant normalisation of states $\langle p^\prime,s^\prime|\,p,s\rangle=2p^0\,(2\pi)^3\delta_{s^\prime s}\delta^{(3)}(\bm{p^\prime}-\bm{p})$ is used along with the spinor normalisation $\bar U(p,s)\, U(p,s) =2M_N  $. In the forward limit, the distributions $H_a(x,0,0)$, for $a=\{u,\,d,\ldots, g\}$, define the familiar unpolarised quark and gluon PDFs, i.e., $H_q(x,0,0)=q(x)$ and $H_g(x,0,0)=xg(x)$.

The operator product expansion (OPE) relates the Bjorken-$x$ (Mellin) moments of the GPDs $H_a(x,\xi,t)$ and $E_a(x,\xi,t)$ to matrix elements of local twist-two operators. The focus of this work is on the lowest moments of the spin-independent gluon GPDs, which are related to the nucleon matrix element of the gluon contribution to the (traceless, symmetric) energy-momentum tensor (EMT)\footnote{The gluon contribution to the EMT can be determined from a canonically normalised action though the Belinfante procedure \cite{Jackiw}.}, and are encoded in three scalar GFFs that are functions of $t$~\cite{Polyakov:2018zvc}:
\begin{align}
\langle p^\prime,s^\prime| G^a_{\{\mu\alpha}{G}_{~~\nu\}}^{a \alpha} |p,s\rangle
= \bar U(p',s') {\cal F}_{\mu\nu}[A_g,B_g,D_g] U(p,s),
\label{Eq:EMT-FFs-spin-12}
\end{align}
where (again, following the conventions of Ref.~\cite{Diehl:2003ny})
\begin{eqnarray}
{\cal F}_{\mu\nu}[A_g,B_g,D_g] =
A_g(t)\,\gamma_{\{\mu} P_{\nu\}}
+ B_g(t)\,\frac{i\,P_{\{\mu}\sigma_{\nu\}\rho}\Delta^\rho}{2M_N}
+ D_g(t)\,\frac{\Delta_{\{\mu}\Delta_{\nu\}}}{4M_N}\,.
\label{eq:Fdef}
\end{eqnarray}
An exactly analogous decomposition exists for matrix elements of the quark contribution of flavour $q$ to the traceless part of the EMT:
\begin{align}
\langle p^\prime,s^\prime| \overline\psi_q \gamma_{\{\mu} i\overset{\text{\tiny$\leftrightarrow$}}{D}_{\nu\}}\psi_q|p,s\rangle
= \bar U(p',s') {\cal F}_{\mu\nu}[A_q,B_q,D_q] U(p,s).
\label{Eq:EMT-FFs-spin-12-quark}
\end{align}
For each $q=\{u,\,d,\dots\}$, the GFFs are related to the lowest Mellin moments of the relevant 
unpolarised GPDs defined in Eq.~\eqref{eq:GPD1}:
\begin{align}
\int_{-1}^1{\rm d}x\;x\, H_q(x,\xi,t) = A_q(t) + \xi^2 D_q(t) \,, \quad \quad
\int_{-1}^1{\rm d}x\;x\, E_q(x,\xi,t) = B_q(t) - \xi^2 D_q(t) \,,
\label{Eq:GPD-Mellin} 
\end{align}
and similarly the gluon GFFs are related to the GPDs defined in Eq.~\eqref{eq:GPD2}:
\begin{align}
\int_{0}^1{\rm d}x\;  H_g(x,\xi,t) = A_g(t) + \xi^2 D_g(t) \,, \quad \quad
\int_{0}^1{\rm d}x\;  E_g(x,\xi,t) = B_g(t) - \xi^2 D_g(t) \,.
\end{align}

Since the quark and gluon pieces of the EMT are not separately conserved, the individual form factors $A_a(t)$, $B_a(t)$ and $D_a(t)$ are scale- and scheme-dependent, although the total form factors $A(t)$, $B(t)$, $D(t)$, where $X(t)\equiv \sum_a X_a(t)$ with $a=\{u,\,d,\dots,g\}$, are renormalisation-scale invariant. 
The GFFs $A_a(t)$ encode the distribution of the nucleon's momentum among its constituents (and momentum conservation implies $A(0)=1$), while the angular momentum distributions are described by $J_a(t) = \frac{1}{2}(A_a(t)+B_a(t))$ (and total spin constrains $J(0)=\frac{1}{2}$). The $D_a(t)$ terms encode the shear forces acting on the quarks and gluons in the nucleon while their sum $D(t)$ determines the pressure distribution~\cite{Polyakov:2002wz,Polyakov:2002yz,Polyakov:2018zvc}.

\subsection{Pion}

The spin-independent pion GPDs are defined by pion matrix elements of the lowest-twist light-ray quark and gluon operators: 
\begin{equation}
\label{EQ:GPD-spin-0}
\int_{-\infty}^\infty \frac{d\lambda}{2\pi}e^{i\lambda x} 
\langle p'|\bar\psi_q(-{\frac{\lambda}{2} n})\gamma^\mu {\cal U}_{\left[-\frac{\lambda}{2}n,\frac{\lambda}{2}n\right]} 
\psi_q(\frac{\lambda}{2} n)|p \rangle
= 2 P^\mu H_q^{(\pi)}(x,\xi,t) + \ldots
\end{equation}
for $q=\{u,d,\ldots\}$, and 
\begin{equation}
\label{EQ:gGPD-spin-0}
\int_{-\infty}^\infty \frac{d\lambda}{2\pi}e^{i\lambda x} 
\langle p'| G_a^{\{\mu\alpha}(-{\lambda\over2}n)\hspace{-1mm}
\left[{\cal U}^{(A)}_{\left[-\frac{\lambda}{2}n,\frac{\lambda}{2}n\right]}\right]_{ab}\hspace{-2.5mm}
G_{b\alpha}^{~~\nu\}}({\lambda\over2}n)
|p \rangle
= P^{\{\mu} P^{\nu\}}H_g^{(\pi)}(x,\xi,t) + \ldots,
\end{equation}
where the notation is as in Eqs.~\eqref{eq:GPD1} and \eqref{eq:GPD2}. A covariant normalisation of pion states has been used: $\langle p^\prime|\,p\rangle=2p^0\,(2\pi)^3\delta^{(3)}(\bm{p^\prime}-\bm{p})$. The lowest moments of these GPDs are related to the pion matrix elements of the quark and gluon pieces of the traceless EMT, which are described by two scalar GFFs for each flavour $a$, labelled $A^{(\pi)}_a(t)$ and $D^{(\pi)}_a(t)$. Precisely,
\begin{equation}\label{Eq:EMT-FFs-spin-0}
\langle p^{\,\prime\,}|G^a_{\{\mu\alpha}{G}_{~~\nu\}}^{a\alpha} |p\rangle = 
2 P_{\{\mu}P_{\nu\}}\, A^{(\pi)}_g(t) + 
\frac12\Delta_{\{\mu}\Delta_{\nu\}}\,D^{(\pi)}_g(t) \equiv {\cal K}_{\mu\nu}[A_g^{(\pi)},D_g^{(\pi)}]
 \,,
\end{equation}
and similarly for the quark operators,
\begin{equation}\label{Eq:EMT-FFs-spin-0-quark}
\langle p^{\,\prime\,}|\overline\psi_q \gamma_{\{\mu}i\overset{\text{\tiny$\leftrightarrow$}}{D}_{\nu\}} \psi_q| p\rangle 
=  {\cal K}_{\mu\nu}[A_q^{(\pi)},D_q^{(\pi)}]
 \,.
\end{equation}
 Just as for the nucleon, the GFFs which describe pion matrix elements of the EMT correspond to the quark and gluon gravitational form factors of the pion, and can be expressed as Mellin moments of the pion GPDs:
\begin{eqnarray}
\int_{-1}^1 dx \,x \,H^{(\pi)}_q(x,\xi,t)= A^{(\pi)}_q(t)  + \xi^2 D^{(\pi)}_q(t) \,,
\qquad
\int_{0}^1 dx \, H^{(\pi)}_g(x,\xi,t)= A^{(\pi)}_g(t)  + \xi^2 D^{(\pi)}_g(t) \,.
\end{eqnarray}
The forward limit $A_a^{(\pi)}(0)$ encodes the light-cone momentum fraction of the pion carried by parton $a$. 
The GFFs $D_a^{(\pi)}(t)$ are related to the pressure and shear distributions in the pion~\cite{Polyakov:2002wz,Polyakov:2002yz,Polyakov:2018zvc}.

\section{Lattice QCD calculation}
\label{sec:latt}

In this work, a single ensemble of isotropic gauge-field configurations is used to determine the matrix elements corresponding to the gravitational form factors of the nucleon and pion, Eqs.~(\ref{Eq:EMT-FFs-spin-12}) and (\ref{Eq:EMT-FFs-spin-0}), respectively. 
Simulations are performed with $N_f=2+1$ flavours of  quarks, with quark masses chosen such that $m_\pi\sim450(5)$~MeV. A clover-improved quark action~\cite{Sheikholeslami:1985ij} and L\"uscher-Weisz gauge action~\cite{LuscherWeisz} are used, with the clover coefficient set equal to its tree-level tadpole-improved value. The configurations have dimensions $L^3\times T=32^3\times 96$, with lattice spacing $a=0.1167(16)$~fm~\cite{stefan}. Details of this ensemble are given in Table~\ref{tab:configs} and in Ref.~\cite{PhysRevD.92.114512}. Subsections \ref{sec:operators}, \ref{sec:NPR} and \ref{sec:MEs} define the Euclidean-space gluon operators studied here, detail the renormalisation prescription, and outline the extraction of the gluon GFFs from Euclidean correlation functions, respectively.
\begin{table*}
	\begin{tabular}{ccccccccccccccc}\toprule
		 $L/a$ & $T/a$ & $\beta$ & $am_l$ & $am_s$ & $a$~(fm) & $L$~(fm) & $T$~(fm) & $m_\pi$~(MeV) & $m_K$~(MeV)  & $m_\pi L$ & $m_\pi T$ & $N_\textrm{cfg}$ & $N_\textrm{meas}$ \\ \hline
		 32 & 96 & 6.1 & -0.2800 & -0.2450 & 0.1167(16) &  3.7 &11.2  & 450(5) & 596(6)  & 8.5 & 25.6 & 2821 & 203 \\\toprule
	\end{tabular}
	\caption{\label{tab:configs}
		LQCD simulation details. The gauge configurations have dimensions $L^3\times T$, lattice spacing $a$, and bare quark masses $a m_q$ (in lattice units). An average of $N_\textrm{meas}$ light-quark sources were used to perform measurements on each of $N_\textrm{cfg}$ configurations, generated in two streams with samples separated by 10 hybrid Monte-Carlo trajectories in each stream.}
\end{table*}

\subsection{Operators}
\label{sec:operators}

To determine the spin-independent gluon GFFs, matrix elements of the gluon operators\footnote{Since the L\"uscher-Weisz gauge action is used in this work, the Belinfante procedure~\cite{Jackiw} produces a gluon EMT that has an additional contribution that is higher-order in $a$. This term is neglected in the present work.}
\begin{eqnarray}
{\cal O}_{\mu\nu}= G^a_{\alpha\{\mu}{G}_{\nu\}}^{a~ \alpha},
\label{eq:glueop}
\end{eqnarray}
are constructed, where the brackets denote symmetrisation and tracelessness in the $\mu$ and $\nu$ indices by $a_{\{\mu} b_{\nu\}}=\frac{1}{2} (a_\mu b_\nu + a_\nu b_\mu)-\frac{1}{4}g_{\mu\nu}a_\alpha b^\alpha$. In Euclidean space, the unrenormalised gluon operators are defined using the clover definition of the discretised  Euclidean-space field-strength tensor
\begin{equation}
G_{\mu\nu}^{(E)}(x) = \frac{1}{8}\left(P_{\mu\nu}(x)-P^\dagger_{\mu\nu}(x)\right),
\end{equation}
derived from the combination of plaquettes
\begin{align}\nonumber
P_{\mu\nu}(x) = & \,U_\mu(x) U_\nu(x+\mu)U_\mu^\dagger(x+\nu) U_\nu^\dagger(x)\\\nonumber
& + U_\nu(x)U^\dagger_\mu(x-\mu+\nu)U^\dagger_\nu(x-\mu)U_\mu(x-\mu)\\\nonumber
& + U^\dagger_\mu(x-\mu)U^\dagger_\nu(x-\mu-\nu)U_\mu(x-\mu-\nu)U_\nu(x-\nu)\\
& + U^\dagger_\nu(x-\nu)U_\mu(x-\nu)U_\nu(x-\nu+\mu)U^\dagger_\mu(x),
\end{align}
which are in turn built from gauge link fields that have been subject to Wilson flow to flow-time $\mathfrak{t} =1.0$ \cite{Luscher:2010iy} in order to increase the signal-to-noise ratio of the calculation. In a previous study of these operators in a $\phi$ meson~\cite{Detmold:2016gpy,Detmold:2017oqb}, the effects of different flow times and different choices of smearing prescription on the bare matrix elements have been found to be mild. Since a non-perturbative renormalisation procedure is used here (discussed in the next section), the differences between bare matrix elements calculated with different smearing prescriptions will be compensated for by differences in the renormalisation.

Because of the reduced symmetry of the lattice geometry, the discretised operators transform in particular representations of the hypercubic group $H(4)$. Specifically, the operators in Eq.~\eqref{eq:glueop} subduce into traceless, symmetric representations of $H(4)$, two of which do not mix with same or lower-dimension operators (labelled $\tau_1^{(3)}$ and $\tau_3^{(6)}$ in the notation of  Refs.~\cite{Mandula:1983ut,Gockeler:1996mu}). In Minkowski space\footnote{The Euclidean operators are related to these by $G^{(E)}_{ij}  = G_{ij}$ for  $i,j\in \{1,2,3\}$, and $G^{(E)}_{4j}  = (-i)G_{0j}$.}, a basis of operators in the three-dimensional $\tau_1^{(3)}$ representation is
\begin{eqnarray}\label{eq:oprep}
{\mathcal{O}}^{\tau_1^{(3)}}_{{1}} = \frac{1}{2}\left({\mathcal{O}}_{11}+{\mathcal{O}}_{22}-{\mathcal{O}}_{33}+{\mathcal{O}}_{00}\right), \hspace{4mm} {\mathcal{O}}^{\tau_1^{(3)}}_{{2}} =  \frac{1}{\sqrt{2}} \left({\mathcal{O}}_{33}+{\mathcal{O}}_{00}\right),
\hspace{4mm} {\mathcal{O}}^{\tau_1^{(3)}}_{{3}} =  \frac{1}{\sqrt{2}} \left({\mathcal{O}}_{11}-{\mathcal{O}}_{22}\right), 
\end{eqnarray}
while a basis for the six-dimensional $\tau_3^{(6)}$ representation is: 
	\begin{align}\label{eq:oprep2}
	{\mathcal{O}}^{\tau_3^{(6)}}_{i=\{1,\ldots, 6\}}= & \left\{\frac{(-i)^{\delta_{\nu 0}}}{\sqrt{2}} \left({\mathcal{O}}_{\mu\nu}+{\mathcal{O}}_{\nu\mu}\right), \hspace{2mm} 0\le \mu < \nu \le 3\right\}.
	\end{align}
All basis operators in each of these two representations are studied here. Within each representation, the renormalisation of the different operators are related by symmetries, while  the renormalisations of operators in the two different representations are only constrained to be the same in the continuum limit. Studying both representations thus permits a test of the discretisation artefacts in this calculation.

\subsection{Renormalisation}
\label{sec:NPR}

The unrenormalised operators in Eq.~\eqref{eq:glueop} mix with the flavour-singlet quark operators ${\mathcal{Q}}_{\mu\nu}=\sum_{q\in\{u,d,s\}}\overline{\psi}_q\gamma_{\{\mu} i\overset{\text{\tiny$\leftrightarrow$}}{D}_{\nu\}}\psi_q$ such that the renormalised gluon operator ${\mathcal{O}}^{\text{ren.}}_{\mu\nu}$ (in any particular scheme) is described by  ${\mathcal{O}}^{\text{ren.}}_{\mu\nu} = Z^{gg} \mathcal{{O}}_{\mu\nu}+ Z^{gq}\mathcal{{Q}}_{\mu\nu}$. It was shown in Ref.~\cite{Alexandrou:2016ekb} that the mixing of the quark operator into the gluon operator is a few-percent effect, using a one-loop perturbative renormalisation procedure and a similar action to the one used here. Consequently, this mixing is assumed to be negligible relative to the statistical uncertainties of this calculation and is neglected here.

The bare lattice operators described in the previous section are renormalised via a non-perturbative RI-MOM prescription~\cite{Martinelli:1993dq,Martinelli:1994ty}, similar to that recently investigated for gluon operators in Refs.~\cite{Yang:2018bft,Yang:2018nqn}. A perturbative matching is used to relate the renormalised operators to the $\overline{\text{MS}}$ scheme. A bare lattice operator $\mathcal{O}^\text{latt}$ is thus renormalised as\footnote{In the general case this is a matrix equation that accounts for mixing among a set of bare operators $\mathcal{O}_i^\text{latt}$.}:
\begin{equation}\label{eq:renormfull}
\mathcal{O}^{\overline{\text{MS}}}(\mu^2) =Z^{\overline{\text{MS}}}_\mathcal{O}(\mu^2)\mathcal{O}^\text{latt} = \mathcal{R}^{\overline{\text{MS}}}(\mu^2,\mu_R^2)Z^\text{RI-MOM}_{\mathcal{O}}(\mu_R^2)\mathcal{O}^\text{latt}. 
\end{equation}
The conversion factor $\mathcal{R}^{\overline{\text{MS}}}(\mu^2,\mu_R^2)$ from the RI-MOM scheme to $\overline{\text{MS}}$ is calculated in continuum perturbation theory~\cite{Yang:2016xsb}, while the RI-MOM renormalisation constant $Z^\text{RI-MOM}_{\mathcal{O}}(\mu_R^2)$ is determined non-perturbatively by imposing the condition
\begin{equation}\label{eq:renorm}
Z_g(p^2)Z^\text{RI-MOM}_\mathcal{O}(p^2)  \Lambda_\mathcal{O}^\text{bare}(p)  \left(\Lambda_\mathcal{O}^\text{tree}(p)\right)^{-1} \bigg|_{p^2=\mu_R^2}=1,
\end{equation}
which relates the bare and tree-level amputated Green's functions $\Lambda_\mathcal{O}^\text{bare/tree}(p)$ for the operator $\mathcal{O}$ in a Landau-gauge--fixed gluon state of momentum $p^2=\mu_R^2$. 
Here, $Z^{1/2}_g(p^2)$ denotes the gluon field renormalisation.

For the particular operator of interest, $\mathcal{O}_{\mu\nu}$, the tree-level amputated Green's function can be expressed as~\cite{Collins:1994ee}:
\begin{align}\nonumber
\Lambda_\mathcal{O}^\text{tree}(p)=\langle \mathcal{O}_{\mu\nu}{\rm Tr}[A_\sigma(p)  A_\tau(-p)]\rangle^\text{tree}_\text{amp.} = \frac{N_c^2-1}{2}( &2p_\mu p_\nu g_{\sigma\tau}-p_\tau p_\nu g_{\sigma\mu}-p_\tau p_\mu g_{\sigma\nu}-p_\sigma p_\nu g_{\tau\mu} \\\label{eq:Otree}
&-p_\sigma p_\mu g_{\tau \nu} + p_\sigma p_\tau g_{\mu\nu} - p^2(g_{\sigma\tau}g_{\mu\nu}-g_{\sigma\mu}g_{\tau\nu}-g_{\sigma\nu}g_{\tau\mu})).
\end{align}
As discussed in Ref.~\cite{Collins:1994ee}, and also noted in Ref.~\cite{Yang:2018bft}, only the first structure in this expression is protected from mixing with the gauge-variant parts of the energy-momentum tensor. Consequently, choosing renormalisation conditions that only involve this term allows a purely multiplicative renormalisation procedure even for gauge-fixed states. The operators in Eqs.~\eqref{eq:oprep} and \eqref{eq:oprep2} can be arranged into the forms\footnote{For the operators in representation $\mathfrak{R}=\tau_1^{(3)}$, defined in  Eq.~\eqref{eq:oprep}, $\mathcal{O}^{\tau_1^{(3)}}_2$, $\mathcal{O}^{\tau_1^{(3)}}_3$, and the combination $(1/\sqrt{2})\mathcal{O}^{\tau_1^{(3)}}_1+(1/2)\left(\mathcal{O}^{\tau_1^{(3)}}_2+\mathcal{O}^{\tau_1^{(3)}}_3\right)$ take this form. }
\begin{align}
\hat{\mathcal{O}}^\mathfrak{R}_{\alpha\beta} = \frac{1}{\sqrt{2}}\begin{cases}
(\mathcal{O}_{\alpha\alpha}+g_{\beta\beta}\mathcal{O}_{\beta\beta}) \hspace{3mm} &\mathfrak{R}=\tau_1^{(3)},\\[2mm]
(-i)^{\delta_{\nu 4}}(\mathcal{O}_{\alpha\beta}+\mathcal{O}_{\beta\alpha}) \hspace{3mm} & \mathfrak{R}=\tau_3^{(6)},
\end{cases}
\end{align} 
with no summation over repeated indices implied. 
For these operators, the constraint $\sigma=\tau\ne\alpha\ne\beta$ is sufficient to isolate the desired term in Eq.~\eqref{eq:Otree}, leading to
\begin{equation}\label{eq:tree}
\Lambda_{\hat{\mathcal{O}}}^\text{tree}(p)=\langle \hat{\mathcal{O}}^\mathfrak{R}_{\alpha\beta}\text{Tr}[A_\tau(p)  A_\tau(-p)]\rangle^\text{tree}_\text{amp.}\big|_{\tau\ne\alpha\ne\beta} = \frac{N_c^2-1}{\sqrt{2}}g_{\tau\tau}\begin{cases}( p_\alpha^2+g_{\beta\beta}p_\beta^2) \hspace{3mm} &\mathfrak{R}=\tau_1^{(3)},\\[2mm]
2(-i)^{\delta_{\nu 4}}\,p_\alpha p_\beta \hspace{3mm} & \mathfrak{R}=\tau_3^{(6)}.
\end{cases}
\end{equation}

\begin{table*}
	\begin{tabular}{cccccc}\toprule
		$L/a$ & $T/a$ & $\beta$ & $am_l$ & $am_s$ & $N_\textrm{cfg}$ \\ \hline
		12 & 24 & 6.1 & -0.2800 & -0.2450 & 24600  \\\toprule
	\end{tabular}
	\caption{\label{tab:configsRenorm}
		Details of the gauge ensemble used to determine the non-perturbative operator renormalisation. Other than the lattice volume $L^3\times T$, which is smaller, the parameters are the same as those of the ensemble (detailed in Table~\ref{tab:configs}) used for the primary calculation. A total of $N_\textrm{cfg}$ configurations are used, generated in 50 streams of configurations, with samples separated by 10 hybrid Monte-Carlo trajectories in each stream.}
\end{table*}

In general, the construction of the forward, amputated bare Green's function $\Lambda_\mathcal{O}^\text{bare}(p)$ will depend on the operator $\mathcal{O}$. For the operators $\hat{\mathcal{O}}_{\alpha\beta}$ considered here, with the same conditions on the external states as applied in Eq.~\eqref{eq:tree}, the additional condition $p_\tau=0$ (where $\tau$ is the Lorentz index of the external gluon fields) permits a simple form:
\begin{align}\label{eq:bare}\nonumber
\Lambda_{\hat{\mathcal{O}}}^\text{bare}(p) = &\frac{\langle\hat{\mathcal{O}}_{\alpha\beta}\text{Tr}[A_\tau(p)A_\tau(-p)]\rangle  (N_c^2-1)^2}{4\langle\text{Tr}[A_\tau(p)A_\tau(-p)]\rangle^2}\bigg|_{p_\tau=0,\tau\ne \alpha \ne \beta}\\
=& \frac{{p}^2 \langle\hat{\mathcal{O}}_{\alpha\beta}\text{Tr}[A_\tau(p)A_\tau(-p)]\rangle (N_c^2-1)}{2Z_g(p^2)\langle\text{Tr}[A_\tau(p)A_\tau(-p)]\rangle}\bigg|_{p_\tau=0,\tau\ne \alpha \ne \beta},
\end{align}
where no summation over repeated indices is implied, and where the second line follows by substitution of the trace of the gluon propagator 
\begin{equation}\label{eq:gluonprop}
D_{\mu\nu}(p) = \langle {\rm Tr}[A_\mu(p)A_\nu(-p)]\rangle = Z_g(p^2)\frac{N_c^2-1}{ 2{p}^2}\left(g_{\mu\nu}-\frac{{p}_\mu {p}_\nu}{{p}^2}\right)
\end{equation}
for one of the gluon terms in the denominator.
The RI-MOM renormalisation constant $Z^\text{RI-MOM}_{\hat{\mathcal{O}}}(p^2)$ can thus be determined by combining Eqs.~\eqref{eq:tree} and \eqref{eq:bare} as prescribed by Eq.~\eqref{eq:renorm}, and taking $N_c=3$:
\begin{align}\label{eq:Zinv}
\left(Z^\text{RI-MOM}_{\hat{\mathcal{O}}}(\mu_R^2)\right)^{-1}=&\frac{4 {p}^2 \langle\hat{\mathcal{O}}_{\alpha\beta}\text{Tr}[A_\tau(p)A_\tau(-p)]\rangle}{\Lambda_{\hat{\mathcal{O}}}^\text{tree}({p})\langle\text{Tr}[A_\tau(p)A_\tau(-p)]\rangle}\bigg|_{{p}^2=\mu_R^2,\tau\ne \alpha \ne \beta,p_\tau=0}.
\end{align}

The gluon three-point and two-point functions that appear in this expression are computed on the ensemble detailed in Table~\ref{tab:configsRenorm}, which has the same bare parameters, but smaller lattice volume and an order of magnitude more configurations, as the ensemble used for the main calculation.
The gluon fields are computed from Landau-gauge--fixed links (gauge-fixed using an iterative procedure with tolerance $10^{-5}$) $U_\mu(x)$:
\begin{equation}
A^\text{latt}_\mu(x+a\hat{e}_\mu /2) = \frac{1}{2 i g_0}\left[\left( U_\mu(x) - U^\dagger_\mu(x)\right) - \frac{1}{N_c} \text{Tr}\left(U_\mu(x) - U^\dagger_\mu(x)\right)\right],
\end{equation}
which holds up to $\mathcal{O}(a^2)$ corrections. Momentum-space lattice gluon fields are defined by the discrete Fourier transform:
\begin{equation}\label{eq:A}
A^\text{latt}_\mu(p) = \sum_x e^{-i {p}\cdot(x+a\hat{e}_\mu/2)}A^\text{latt}_\mu(x+a\hat{e}_\mu/2), \hspace{3mm} \text{with} \hspace{3mm} p_\mu = \frac{2\pi n_\mu}{a L_\mu}, \hspace{3mm} n_\mu=\{0,\ldots,L_\mu-1\},
\end{equation}
where $L_\mu$ denotes the number of lattices sites in dimension $\mu$ and where the discretised momenta accessible on the finite lattice volume are
\begin{equation}\label{eq:bdymom}
\tilde{p}_\mu = \frac{2}{a}\text{sin}\left(\frac{{p}_\mu a}{2}\right).
\end{equation}

Gluon two-point functions $D_{\tau\tau}(\tilde{p})=\langle \text{Tr}[A_\tau({p})A_\tau(-{p})]\rangle$ are constructed for all four-momenta $\tilde{p}_\mu$ corresponding to $p_\mu$ with $\sum_\mu n_\mu^2 \le 36$ (Eq.~\eqref{eq:A}). Correlation functions are calculated both with and without Wilson flow (to flow time $\mathfrak{t} =1.0$) applied to the gluon fields; determinations of $Z^\text{RI-MOM}_{\hat{\mathcal{O}}}(\mu_R^2)$ using flowed or unflowed fields in the propagators will agree up to discretisation artefacts, and comparing the two determinations provides a measure of such effects.
Gluon three-point functions $\langle\hat{\mathcal{O}}_{\alpha\beta}\text{Tr}[A_\tau(p)A_\tau(-p)]\rangle$ are constructed on each configuration by correlating the gluon two-point functions with the operators $\hat{\mathcal{O}}_{\alpha\beta}$, computed as described in Section~\ref{sec:operators} and projected to zero four-momentum, and subtracting the vacuum contribution.

At each unique squared four-momentum $\tilde{p}^2$, the right-hand side of Eq.~\eqref{eq:Zinv} is computed for each corresponding $\tilde{p}$, for all operators in a given representation $\mathfrak{R}\in\{\tau_1^{(3)},\tau_3^{(6)}\}$, and for all allowed choices of the Lorentz index $\tau$ of the external gluon states. Fits to these results are performed in a correlated manner to determine the RI-MOM renormalisation factor $Z_{\mathfrak{R}}^\text{RI-MOM}(\tilde{p}^2)$ for that scale and representation. 
The correlations are propagated using the bootstrap resampling procedure described in Sec.~\ref{sec:MEs}. Choices of the number of bootstraps $N_\text{boot}$ from 200 to 1000 are tested and found to give consistent results and uncertainties. As discussed in Ref.~\cite{Gockeler:2010yr}, combining data from all operators in a given irreducible representation of the hypercubic group, as is done here, in general reduces the amount of $O(4)$ violation and produces a smoother dependence of the common renormalisation factor on the scale $\tilde{p}^2$ than choosing a single operator.

In addition to the RI-MOM factors $Z_{\mathfrak{R}}^\text{RI-MOM}(\tilde{p}^2)$ for the two representations, the complete multiplicative renormalisation constant $Z_\mathfrak{R}^{\overline{\text{MS}}}(\mu=2~\text{GeV})=\mathcal{R}^{\overline{\text{MS}}}(\mu=2~\text{GeV},\tilde{p}^2)\,Z_\mathfrak{R}^\text{RI-MOM}(\tilde{p}^2)$ includes a perturbative matching factor which converts from the RI-MOM renormalisation at scale  $\tilde{p}^2$ to the $\overline{\text{MS}}$ scheme at $\mu=2$~GeV. In this work, the 1-loop expression for this matching, derived in Ref.~\cite{Yang:2016xsb}, is used:
\begin{equation}
\mathcal{R}^{\overline{\text{MS}}}(\mu^2,\mu_R^2) = 1-\frac{g^2 N_f}{16 \pi^2}\left(\frac{2}{3}\text{log}(\mu^2/\mu_R^2)+\frac{10}{9}\right)-\frac{g^2 N_c}{16\pi^2}\left(\frac{4}{3}-2\xi+\frac{\xi^2}{4}\right).
\end{equation}
For these calculations in the Landau gauge, $\xi=0$, $N_c=3=N_f$, and $g^2$ is defined by $\alpha(\mu_{\overline{\rm MS}})$ evaluated to three loops~\cite{Herren:2017osy,Schmidt:2012az,Chetyrkin:2000yt}. 

\begin{figure}[!t]
	\centering
	\subfigure[]{
		\includegraphics[width=0.47\textwidth]{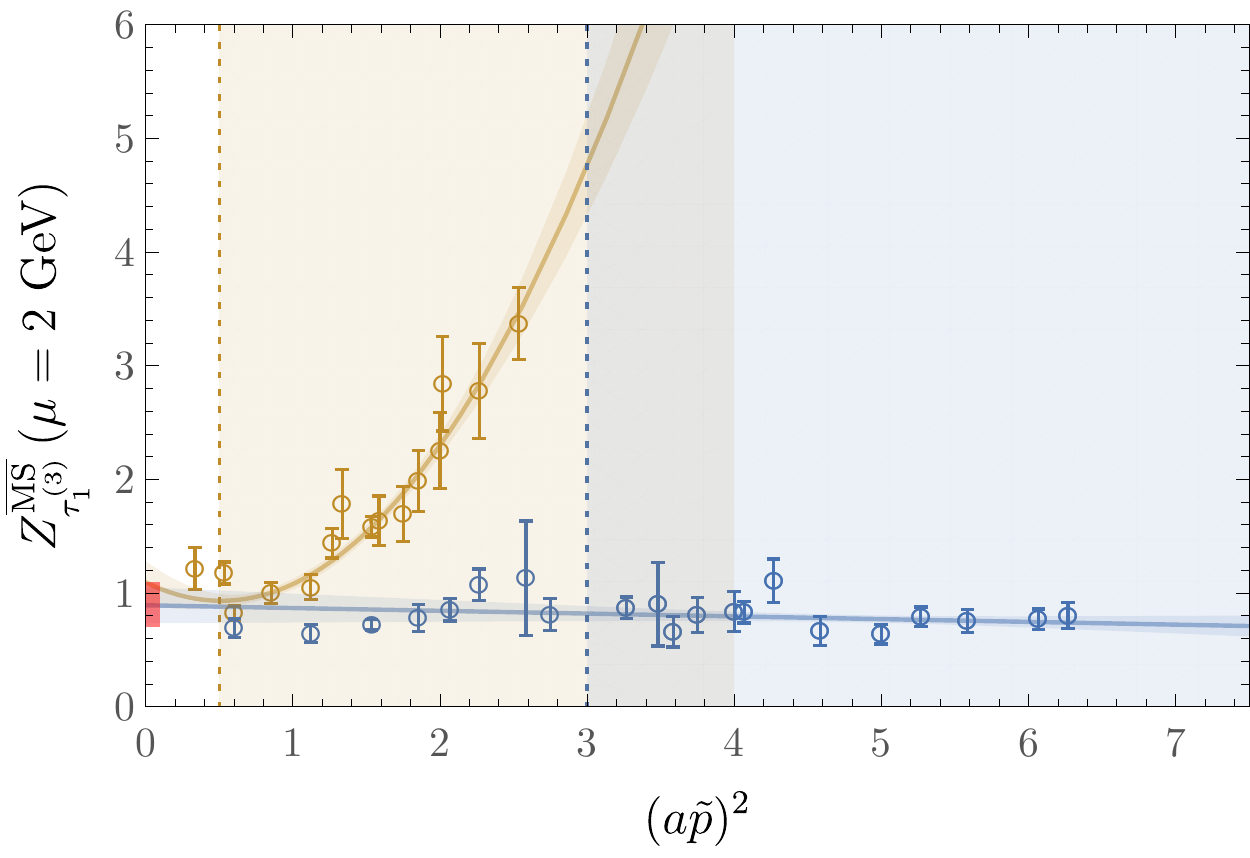}}\hspace{5mm}
	\subfigure[]{
		\includegraphics[width=0.47\textwidth]{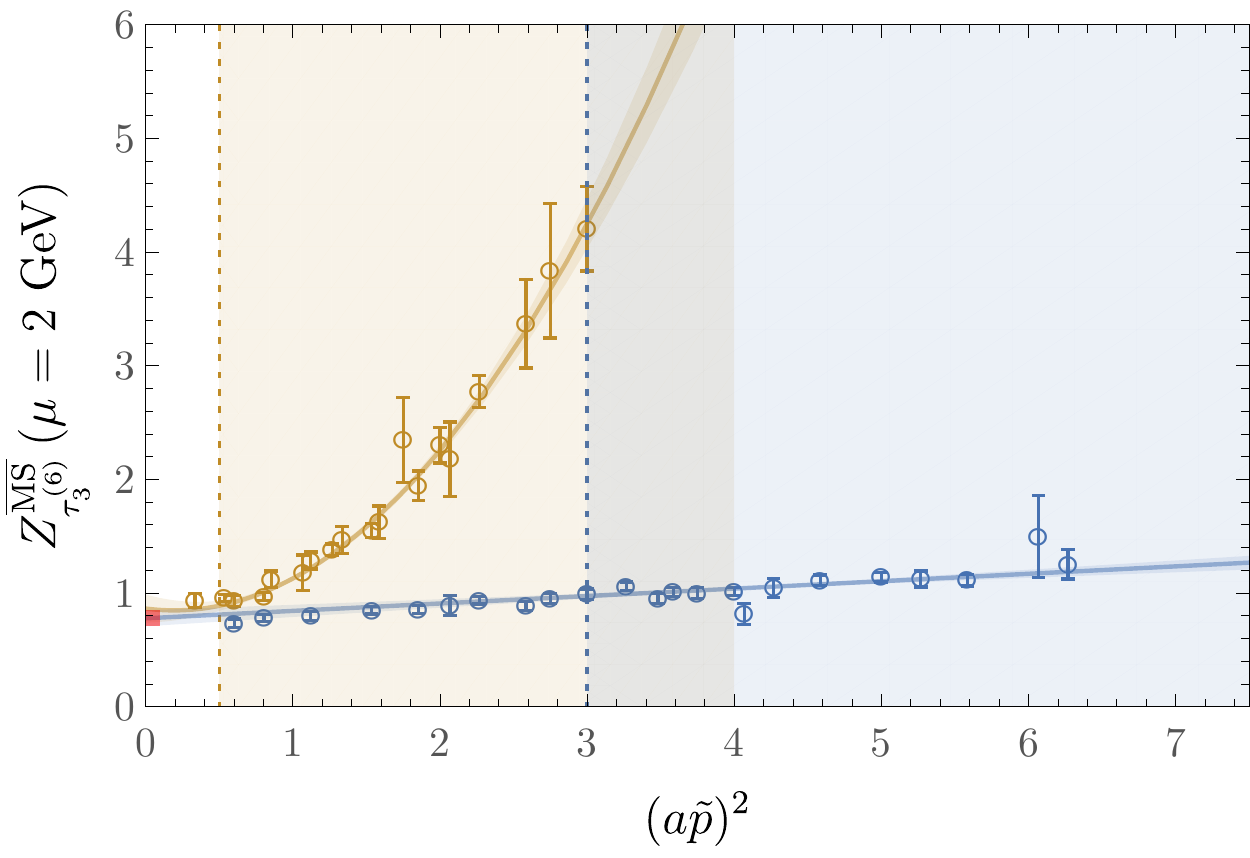}}
	\caption{The $\overline{\text{MS}}$ renormalisation constant for gluon operators in the two irreducible representations of the hypercubic group considered here, Eqs.~\eqref{eq:oprep} and \eqref{eq:oprep2}, calculated on the ensemble detailed in Table~\ref{tab:configsRenorm}, with cuts on four-momenta such that $\sum_\mu \tilde{p}_\mu^4 / \left( \sum_\mu \tilde{p}_\mu^2\right)^2 < 0.5$. The orange diamonds and blue circles denote results obtained using gauge fields with and without Wilson flow in the construction of the two- and three-point gluon correlation functions. The corresponding orange and blue shaded regions denote the fit ranges of the displayed fit bands to each dataset, which are quadratic (orange) and linear (blue) in $(a\tilde{p})^2$, respectively. The red shaded area on each figure denotes the final value and uncertainty for each renormalisation constant, which includes a systematic uncertainty arising from different choices of hypercubic cut and fit range in $(a\tilde{p})^2$, as described in the text.}
	\label{fig:Zinvlatt}
\end{figure}

The extracted renormalisation constants $Z_\mathfrak{R}^{\overline{\text{MS}}}(\mu=2~\text{GeV})$, determined from the RI-MOM factors $Z_{\mathfrak{R}}^\text{RI-MOM}(\tilde{p}^2)$ at a range of scales $(a\tilde p)^2$, are displayed in Fig.~\ref{fig:Zinvlatt}. In the absence of discretisation artefacts and in the perturbative regime, each renormalisation constant would be independent of the intermediate scale $(a\tilde p)^2$. It is apparent that the results obtained using Wilson-flowed fields in the gluon two-point functions have smaller discretisation artefacts than those with unflowed fields: while the former are consistent with a linear form in $(a\tilde{p})^2$ at large scales, the latter display significant quadratic effects. Nevertheless, in the limit $a\rightarrow 0$, the renormalisation constants constructed using flowed and unflowed gauge fields in the gluon propagators agree for each representation $\mathfrak{R}$. The results for the $\tau_1^{(3)}$ representation display larger discretisation artefacts and statistical fluctuations than those for the $\tau_3^{(6)}$ representation.

Values for the constants $Z_\mathfrak{R}^{\overline{\text{MS}}}(\mu=2~\text{GeV})$ that are used to renormalise the bare lattice results for the GFFs are taken from the $a=0$ extrapolations of linear fits in $(a\tilde{p})^2$ to the renormalisation constants constructed using flowed gauge fields with different intermediate scales $(a\tilde{p})^2$. 
The significance of discretisation artefacts is assessed by taking various cuts on the four-momenta included in the fits, such that $\sum_\mu \tilde{p}_\mu^4 / \left( \sum_\mu \tilde{p}_\mu^2\right)^2 < X$ for  $\{0.3<X<0.5\}$. There is insufficient data to constrain the extrapolation for smaller $X$, while the fit quality decreases for larger $X$, indicating significant contamination from discretisation effects when more momentum components are included. For each cut, linear fits in $(a\tilde{p})^2$ over all fit ranges with a lower bound of $(a\tilde{p})^2\ge 1$ are performed.
The standard deviation of the variation of the central values over all fits with acceptable $\chi^2/\text{d.o.f}$ for all cuts is included in quadrature with the statistical uncertainty of the best-fit extrapolation to $a=0$. The resulting values of the renormalisation constants for the respective representations are:
 \begin{align}\label{eq:Z1}
 Z^{\overline{\text{MS}}}_{\tau_1^{(3)}}(\mu=2~\text{GeV})&=0.9(2); \\\label{eq:Z2} Z^{\overline{\text{MS}}}_{\tau_2^{(6)}}(\mu=2~\text{GeV})&=0.78(7).
\end{align}

\subsection{Matrix Elements}
\label{sec:MEs}

Bare matrix elements of the operators in Sec.~\ref{sec:operators} are extracted from ratios of two- and three-point correlation functions built from quark propagators originating from an APE-smeared~\cite{Falcioni:1984ei} source and having either an APE-smeared or point sink\footnote{ Sources or sinks are smeared with 35 steps of gauge-invariant Gaussian smearing with smearing parameter $\rho=4.7$.}. The two sets of resulting correlation functions are labelled as smeared-smeared (SS) and smeared-point (SP), respectively.
For the nucleon, two-point correlation functions are defined as 
\begin{eqnarray}
C_{s }^{\rm{2pt}}(\vec{p},t_f; \vec{x}_0, t_0) &=& \sum_{\vec{x}} e^{-i \vec{p}\cdot\vec{x}} \left(\Gamma_s\right)_{\alpha\beta}\langle 0 | \chi_{\beta}(\vec{x},t_f)\overline\chi_{\alpha}(\vec{x}_0,t_0) | 0 \rangle  \nonumber \\
&\stackrel{t_f\gg t_0}{\longrightarrow}&
 \frac{\sqrt{Z^\ast(p) \tilde{Z}(p)}}{2E^{(N)}_{\vec{p}}}   {\rm Tr}[ \Gamma_s (p\!\!\!/ +M_N)  )]  
 e^{-E^{(N)}_{\vec{p}} (t_f-t_0)} +\ldots\,.
\label{eq:c2ptprot}
\end{eqnarray}
Here ($\vec{x}_0,t_0$) denotes the source position, $\vec{p}$ is the chosen momentum projection, and $\chi_\alpha(\vec{x},t) = \epsilon^{ijk} (\psi^i_u(\vec{x},t) C\gamma_5 \psi^j_d(\vec{x},t))   \psi^k_{u;\alpha}(\vec{x},t)$ is an interpolating operator for the nucleon with a given spinor index $\alpha$.  The matrix $\Gamma_s=(1+\gamma_4)(1+(-1)^s \gamma_1\gamma_2)$ selects the positive energy component of the nucleon and projects its spin ($s=\{0,1\}$ corresponding to spin \{up, down\}), $E^{(N)}_{\vec{p}} =\sqrt{M_N^2 + |\vec{p}\,|^2}$ is the nucleon energy for a given momentum $\vec{p}$, and $Z(p)$ $(\tilde{Z}(p))$ controls the overlap factor of the source (sink) interpolating operator onto the nucleon state (the source and sink overlaps are distinct for the SP correlation functions). The ellipsis denotes contributions from higher excitations, which are exponentially suppressed for $t_f\gg t_0$. 

Similarly, the pion two-point correlation function is defined by
\begin{eqnarray}
C_{(\pi)}^{\rm{2pt}}(\vec{p},t_f; \vec{x}_0, t_0) &=& \sum_{\vec{x}} e^{-i \vec{p}\cdot\vec{x}} \langle 0 | \chi_{(\pi)}(\vec{x},t_f)\chi_{(\pi)}^\dagger(\vec{x}_0,t_0) | 0 \rangle \nonumber \\
&\stackrel{t_f\gg t_0}{\longrightarrow}&
\frac{\sqrt{Z_\pi^\ast(p) \tilde Z_\pi(p)}}{2E^{(\pi)}_{\vec{p}}} 
e^{-E^{(\pi)}_{\vec{p}} (t_f-t_0)} +\ldots\,,
\label{eq:c2ptpi}
\end{eqnarray}
where  $E^{(\pi)}_{\vec{p}} =\sqrt{m_\pi^2 + |\vec{p}\,|^2}$ and  $Z_\pi(p)$ $(\tilde Z_\pi(p))$ controls the overlap factor of the source (sink) interpolating operator onto the pion states. The interpolating operator is $\chi_{(\pi)} (\vec{x},t) = \overline \psi_u(\vec{x},t) \gamma_5 \psi_d(\vec{x},t)$, constructed both with and without APE smearing as described for the nucleon. 

\begin{figure}
	\centering
	\subfigure[Pion]{
	\includegraphics[width=0.45\linewidth]{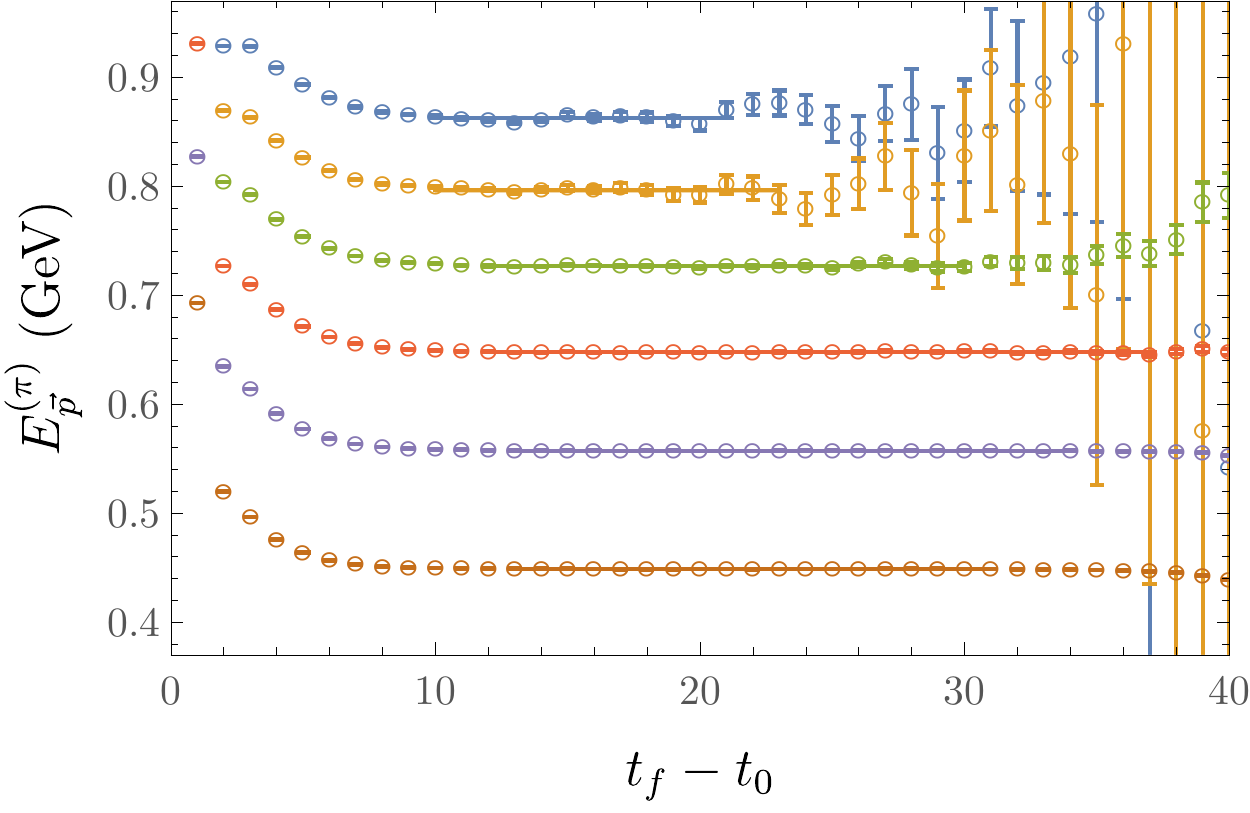}}\hspace{5mm}
\subfigure[Nucleon]{
	\includegraphics[width=0.45\linewidth]{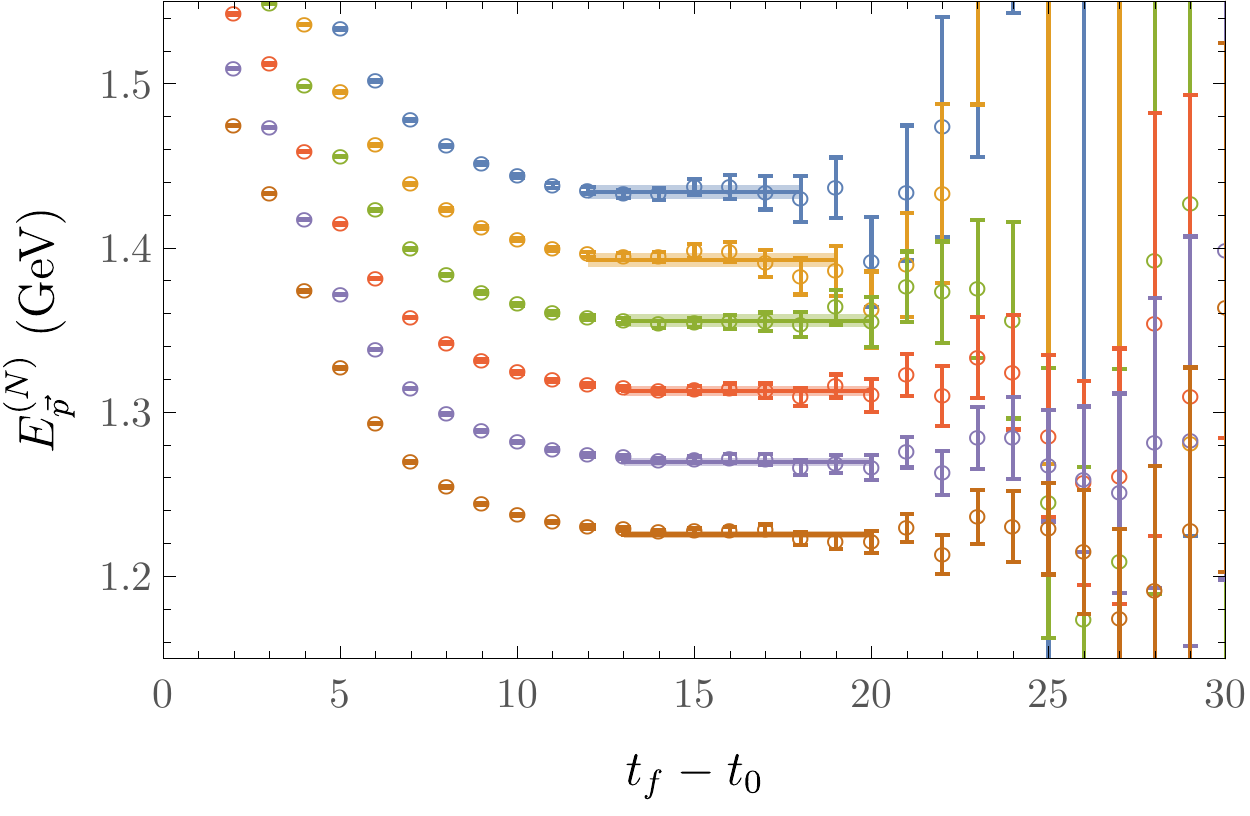}}
	\caption{Effective mass plots (Eq.~\eqref{eq:effmass}) formed from SP correlation functions for the pion and nucleon. The shaded bands show constant fits to the data for each $|\vec{p}\,|^2 \leq 5 (2\pi/L)^2$, as described in the text. The effective masses generated with SS correlation functions are similar and result in energy extractions that are consistent with those shown for the SP case within uncertainties.}
	\label{fig:2pts}
\end{figure}

\begin{figure}
	\centering
	\subfigure[Pion]{
	\includegraphics[width=0.45\linewidth]{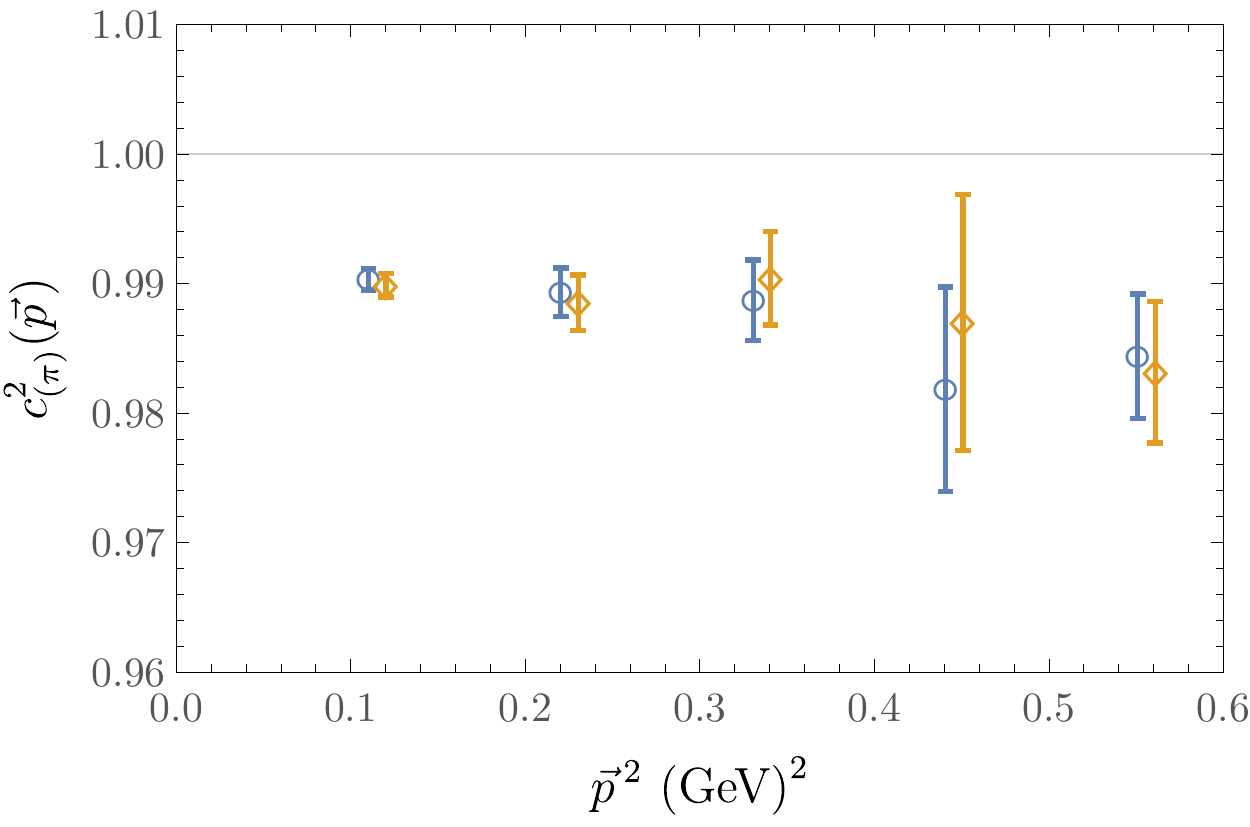}}\hspace{5mm}
\subfigure[Nucleon]{
	\includegraphics[width=0.45\linewidth]{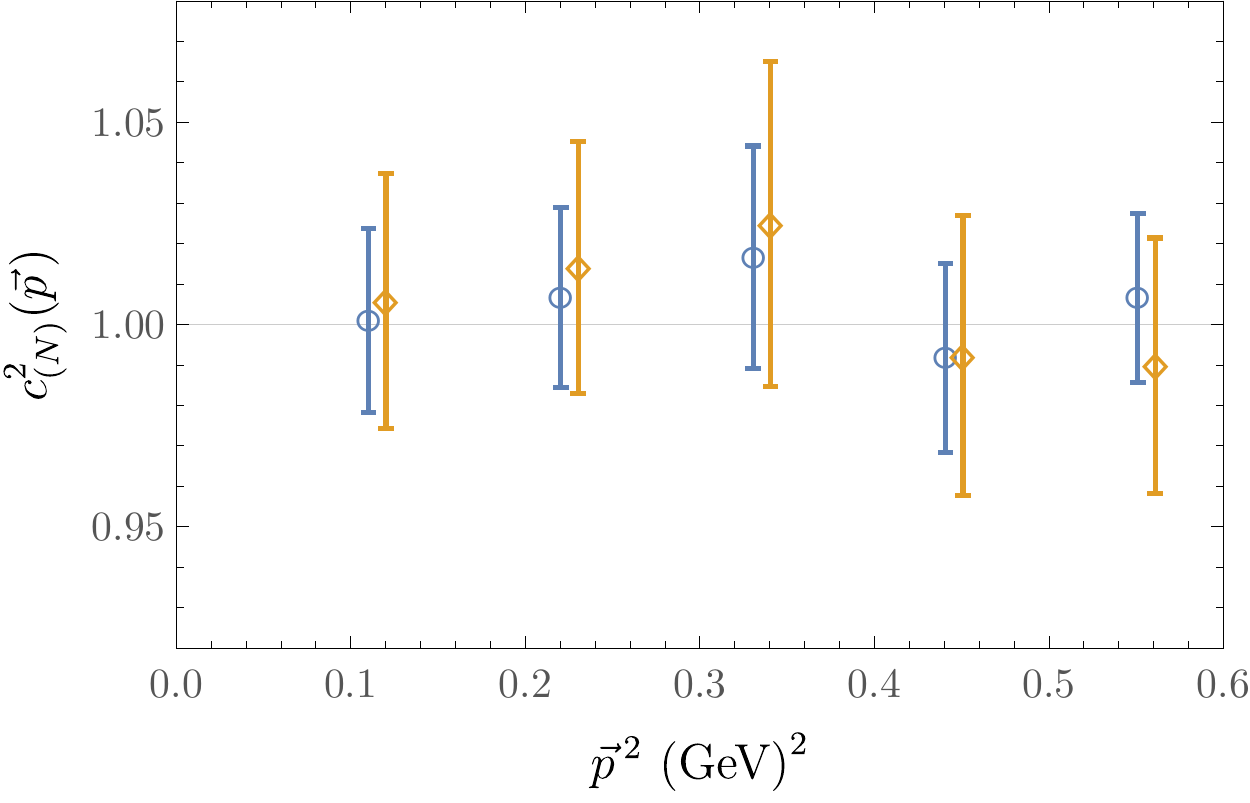} }
	\caption{The speed of light $c_{(N/\pi)}^2(\vec{p}\,)=(E^{(N/\pi)2}_{\vec{p}}-M_{(N/\pi)}^2)/\vec{p}\,^2)$ entering the dispersion relation for the pion and nucleon as a function of the squared momentum of the hadron. The blue circles and orange diamonds show results obtained using SP and SS correlation functions respectively (the SS results are slightly offset on the horizontal axis for clarity).}
	\label{fig:energies}
\end{figure}

Nucleon and pion two-point functions are evaluated for all three-momenta such that $|\vec{p}\,|^2 \leq 5 (2\pi/L)^2$, and for both spin components of the nucleon.
Effective mass functions defined as 
\begin{equation}\label{eq:effmass}
E^{(N/\pi)}(t_f-t_0)=\ln\left[C_{(N/\pi)}^{\rm{2pt}}(t_f-t_0)/C_{(N/\pi)}^{\rm{2pt}}(t_f-t_0+1)\right],
\end{equation}
constructed from the two-point functions of the nucleon (averaged over spins) and pion, are shown in Fig.~\ref{fig:2pts} for the SP correlators. As the momentum increases, the signal quality degrades for both the nucleon and pion. Energies are extracted from constant fits to the effective masses over the longest time region with $\chi^2/\text{d.o.f}\le1$, accounting for the correlations in the data. These time windows define the range of sink times $t_f$ where excited state contamination is small in comparison with the statistical uncertainties of the data.
The energies extracted as a function of momentum using both SS and SP two-point correlators are used to construct  the effective speed of light (in units of $c$) shown for both hadrons in Fig.~\ref{fig:energies}; comparison of these quantities to unity provides a measure of discretisation errors in this calculation. On this ensemble, discretisation effects on the speed of light are at the percent level for all momenta considered. Consistent values for $c^2_{(N/\pi)}$ were found on this ensemble in Ref.~\cite{PhysRevD.92.114512}.

For the gluon operators $O_i^\mathfrak{R}$ defined in Eqs.~\eqref{eq:oprep} and \eqref{eq:oprep2}, nucleon three-point correlation functions are defined by
\begin{eqnarray}
C_{s;\mathfrak{R},i}^{\rm{3pt}}(\vec{p},\vec{p}\,' = \vec{p} + \vec{\Delta},t_f,\tau; \vec{x}_0, t_0) &=& \sum_{\vec{x},\vec{y}} e^{-i \vec{p}\cdot\vec{x}} e^{i \vec{\Delta}\cdot\vec{y}} (\Gamma_s)_{\alpha\beta} \langle 0 | \chi_\beta(\vec{x},t_f) {\cal O}^\mathfrak{R}_{i}(\vec{y},\tau)\overline\chi_\alpha(\vec{x}_0,t_0) | 0 \rangle \nonumber \\
&\stackrel{t_f\gg\tau\gg t_0}{\longrightarrow}&
\frac{\sqrt{Z^\ast(p) \tilde{Z}(p')}}{4E^{(N)}_{\vec{p}}E^{(N)}_{\vec{p}\,'}} 
e^{-E^{(N)}_{\vec{p}} (t_f-\tau)} e^{-E^{(N)}_{\vec{p}\,'} (\tau -t_0)}\nonumber \\
&&\times    {\rm Tr}\left[ \Gamma_s (p\!\!\!/ +M_N) {\cal F}^\mathfrak{R}_i[A_g,B_g,D_g] ( p\!\!\!/' +M_N) \right]   + \ldots\,,
\label{eq:c3ptprot}
\end{eqnarray}
where ${\cal F}^\mathfrak{R}_i$ for $i=\{1,2,3\}$ denotes the linear combination of ${\cal F}_{\mu\nu}$ (defined in Eq.~\eqref{eq:Fdef}) with indices matching the structure of the corresponding operator $\mathcal{O}^\mathfrak{R}_i$.
Similarly, the three-point correlation functions of the pion are defined by
\begin{eqnarray}
C_{(\pi); \mathfrak{R},i}^{\rm{3pt}}(\vec{p},\vec{p}\,'= \vec{p} + \vec{\Delta},t_f,\tau;\vec{x}_0, t_0) &=&\sum_{\vec{x},\vec{y}} e^{-i \vec{p}\cdot\vec{x}} e^{i \vec{\Delta}\cdot\vec{y}}  \langle 0 | \chi_{(\pi)}(\vec{x},t_f) {\cal O}^\mathfrak{R}_{i}(\vec{y},\tau) \chi_{(\pi)}^\dagger(\vec{x}_0,t_0) | 0 \rangle
\nonumber \\
&\stackrel{t_f\gg\tau\gg t_0}{\longrightarrow}&
\frac{\sqrt{Z^\ast_\pi(p) \tilde{Z}_{\pi}(p')}}{4E^{(\pi)}_{\vec{p}}E^{(\pi)}_{\vec{p}\,'}} 
e^{-E^{(\pi)}_{\vec{p}} (t_f-\tau)} e^{-E^{(\pi)}_{\vec{p}\,'} (\tau -t_0)}{\cal K}^\mathfrak{R}_i[A_g^{(\pi)},D_g^{(\pi)}]+\ldots\,,
\label{eq:c3ptpi}
\end{eqnarray}
where again the representation and subscript labels $\{\mathfrak{R},i\}$ correspond to the operators defined in Eqs.~\eqref{eq:oprep} and \eqref{eq:oprep2}, as discussed for the nucleon. 
For both the nucleon and the pion, three-point functions are constructed with all possible sink three-momenta that satisfy $|\vec{p}\,|^2 \leq 5 (2\pi/L)^2$, with operator three-momenta $|\vec{\Delta}|^2 \leq 18 (2\pi/L)^2$.

The two- and three-point correlation functions are evaluated on an average of $N_{\rm src}=203$ randomly placed sources on each of the $N_{\rm cfg}= 2821$ configurations of the ensemble detailed in Table~\ref{tab:configs}. At the first stage of analysis,  results on each configuration are averaged (after translation such that all sources coincide at the origin). In the discussion of further analysis, the $x_0$ and $t_0$ labels are thus omitted. 
A bootstrap resampling procedure over the $N_{\rm cfg}$ independent samples is used to propagate the statistical uncertainties of the two- and three- point functions. In this procedure, $N_{\rm boot}=200$ bootstrap ensembles each with $N_{\rm cfg}$ elements are randomly drawn (allowing replacement). Repeating the analysis with $N_{\rm boot}=100$ or 1000 yields consistent values and uncertainties. To test the assumption that the configurations, each separated by 10 hybrid Monte-Carlo trajectories, are independent, the analysis is also undertaken with correlation functions calculated on sets of $N_{\rm block}=10$ successive configurations (still spaced by 10 hybrid Monte-Carlo trajectories) averaged before bootstrap resampling. This blocking process does not modify the results at a statistically significant level.
%

\begin{figure}
	\centering
	\subfigure[Pion]{
		\includegraphics[width=0.45\linewidth]{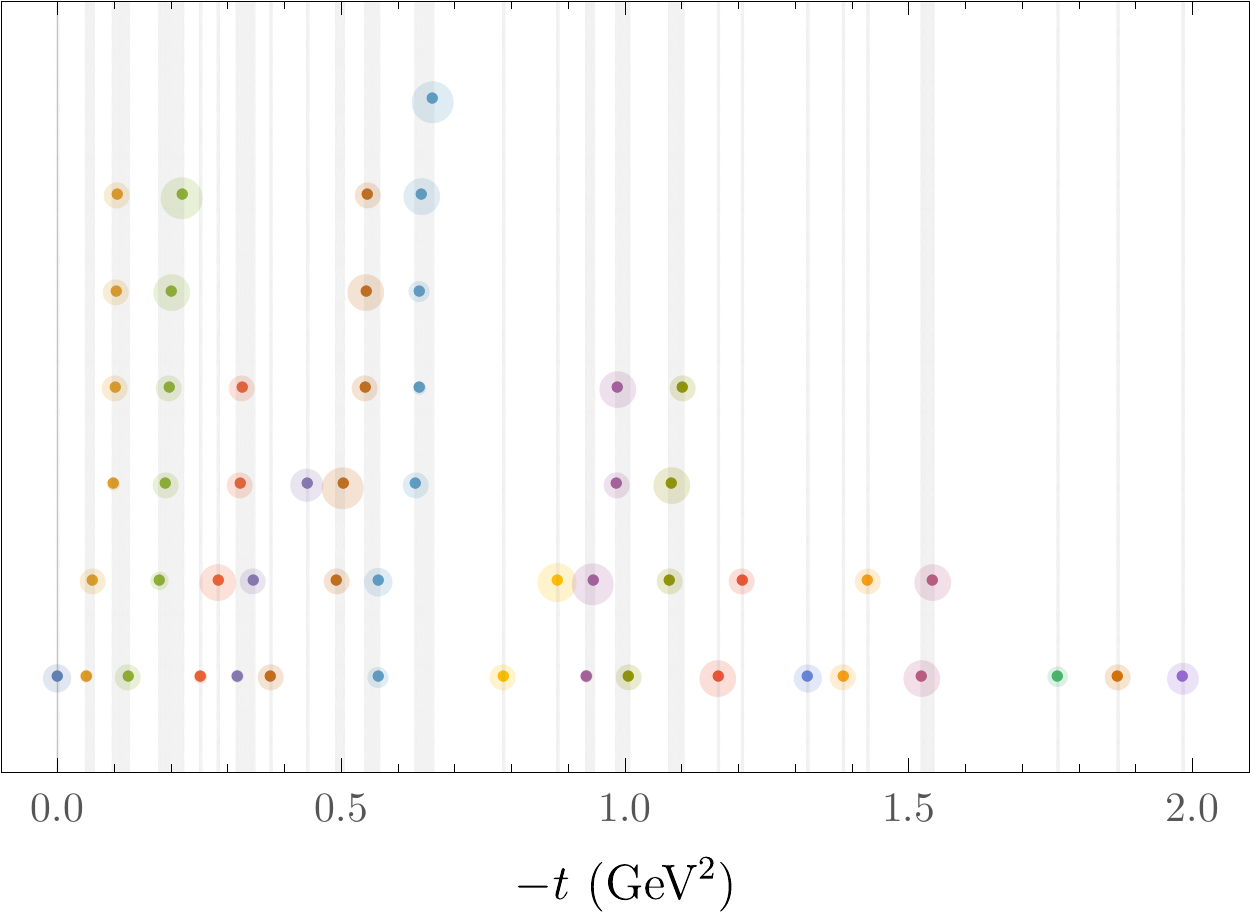}}\hspace{4mm}
	\subfigure[Nucleon]{
		\includegraphics[width=0.45\linewidth]{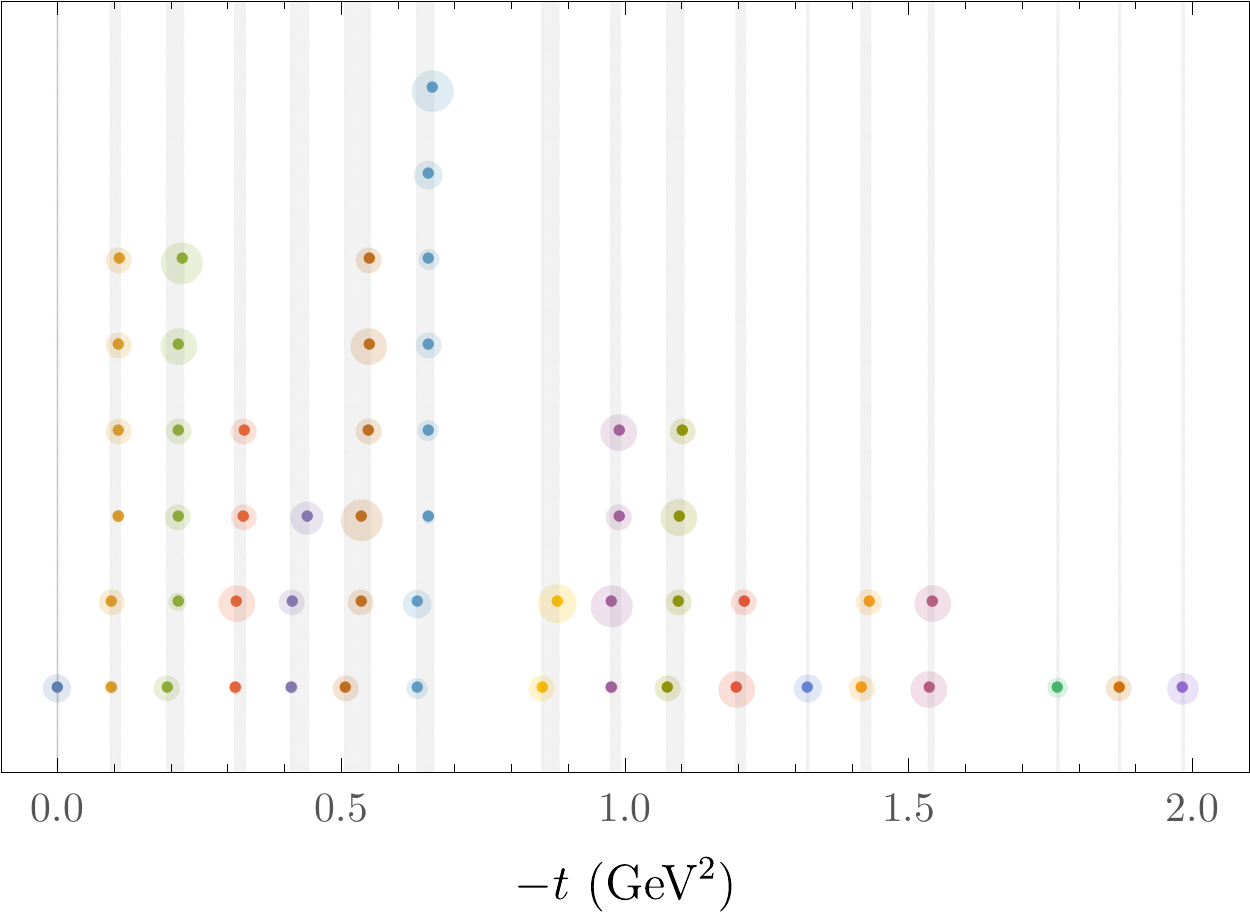} }
	\caption{Accessible $t=(p'-p)^2$ values using all possible sink three-momenta $\vec{p}\,'$ with $|\vec{p'}\,|^2 \leq 5 (2\pi/L)^2$, and all operator momenta with $|\vec{\Delta}|^2 \leq 18 (2\pi/L)^2$, for both the nucleon and pion. Each colour denotes $t$ values corresponding to a single choice of squared three-momentum transfer $\vec{\Delta}^2=(\vec{p}\,'-\vec{p})^2$. The size of each point is proportional to the square root of the number of three-momenta at that $t$. The grey vertical bands highlight the range of each binning; data at $t$-values within each bin are analysed as a single system.}
	\label{fig:groupfigs}
\end{figure}

To extract the GFFs, ratios of the nucleon and pion three-point and two-point functions are formed for each of the $N_{\rm boot}$ bootstrap resamplings: 
\begin{eqnarray}
R_{s;\mathfrak{R},i}(\vec{p},\vec{p}\,',t_f,\tau) &=& \frac{C^{3\text{pt}}_{s ;\mathfrak{R},i}(\vec{p},\vec{p}\,',t_f,\tau)}{C_{s}^{2\text{pt}}(\vec{p}\,',t_f)}\sqrt{\frac{C_{s}^{2\text{pt}}(\vec{p},t_f-\tau)C_{s}^{2\text{pt}}(\vec{p}\,',t_f)C_{s}^{2\text{pt}}(\vec{p}\,',\tau)}{C_{s}^{2\text{pt}}(\vec{p}\,',t_f-\tau)C_{s}^{2\text{pt}}(\vec{p},t_f)C_{s}^{2\text{pt}}(\vec{p},\tau)}}
\nonumber \\ \label{eq:rat}
&\stackrel{t_f\gg\tau\gg0}{\longrightarrow}& \frac{  {\rm Tr}\left[ \Gamma_s (p\!\!\!/ +M_N) {\cal F}_i[A_g,B_g,D_g] (p\!\!\!/' +M_N) \right]   }{8\sqrt{E^{(N)}_{\vec{p}} E^{(N)}_{\vec{p}\,'}  (E^{(N)}_{\vec{p}}+M_N)(E^{(N)}_{\vec{p}'}+M_N)  }}+ \ldots ,
\end{eqnarray}
\begin{eqnarray}
R^{(\pi)}_{\mathfrak{R},i}(\vec{p},\vec{p}\,',t_f,\tau) &=& \frac{C^{3\text{pt}}_{(\pi) ;\mathfrak{R},i}(\vec{p},\vec{p}\,',t_f,\tau)}{C_{(\pi)}^{2\text{pt}}(\vec{p}\,',t_f)}\sqrt{\frac{C_{(\pi)}^{2\text{pt}}(\vec{p},t_f-\tau)C_{(\pi)}^{2\text{pt}}(\vec{p}\,',t_f)C_{(\pi)}^{2\text{pt}}(\vec{p}\,',\tau)}{C_{(\pi)}^{2\text{pt}}(\vec{p}\,',t_f-\tau)C_{(\pi)}^{2\text{pt}}(\vec{p},t_f)C_{(\pi)}^{2\text{pt}}(\vec{p},\tau)}}
\nonumber \\ \label{eq:pirat}
&\stackrel{t_f\gg\tau\gg0}{\longrightarrow}& \frac{  {\cal K}_i[A_g^{(\pi)},D_g^{(\pi)}]    }{2\sqrt{E^{(\pi)}_{\vec{p}} E^{(\pi)}_{\vec{p}\,'}  }}+ \ldots,
\end{eqnarray}
where the nucleon and pion energies are constructed as $E^{(N/\pi)}_{\vec{p}} =\sqrt{M_{N/\pi}^2 + |\vec{p}\,|^2}$, rather than determined from the two-point functions at each three-momentum.
In these ratios, the exponential time-dependence and overlap factors cancel for the ground state contribution for $0\ll\tau\ll t_f$. The decompositions of the nucleon and pion matrix elements in Eqs.~\eqref{Eq:EMT-FFs-spin-12} and \eqref{Eq:EMT-FFs-spin-0} thus allow the ratios in this limit to be constructed in terms of only the unknown GFFs, which are functions of the squared momentum transfer $t=(p'-p)^2$, and known kinematic factors. At each value of the squared momentum transfer $t$, the various consistent choices of $\vec{p}$, $\vec{p}\,'$, and operator index $i$ for a given representation $\mathfrak{R}$ (and spin $s$ for the nucleon) thus provide a system of equations that can be solved to isolate the GFFs at that $t$. Of the large number of $t$ values accessible using the three-momenta considered here, many are very close together (in comparison with the overall scale of momenta). To provide the best determination of the GFFs, nearby $t$ values are thus binned as illustrated graphically in Fig.~\ref{fig:groupfigs}, and their constraints are treated as a single system of equations at each average $t$. The bins are defined such that no adjacent accessible $t$ values that differ by 0.03 GeV$^2$ or more will be in the same bin.

To reduce the dimensionality of the highly overdetermined systems of equations that must be solved to isolate the GFFs for each momentum transfer bin, ratios that give the same linear combination of GFFs (up to a sign) in the limit $0\ll\tau\ll t_f$ are averaged (including the appropriate signs). The averaged ratios are denoted $\overline{R}^{(\pi/N)}_{\mathfrak{R};k}(t_f,\tau)$, where the subscript $k$ now enumerates the different averaged ratios, rather than indexing specific operators. This averaging is performed separately for ratios constructed with SS and SP correlation functions, and separately for each representation $\mathfrak{R}$. For the nucleon there are between 4 and 101 averaged ratios for each smearing at different momenta, and for the pion there are between 2 and 48. Subsequently, constant fits to the $t_f$ and $\tau$ dependence of the averaged ratios are performed to extract the ground state contribution. 
For each averaged ratio, fits are performed over windows of timeslices in the two-dimensional $\{t_f,\tau\}$ plane with $t_f^\text{min}<t_f<t_f^\text{max}$ and $\Delta \tau < \tau < (t_f-\Delta_\tau +1)$. Fit windows are constrained to have the earliest sink time $t_f^\text{min}$ no earlier than the time at which the hadron two-point functions are consistent with a single state. This constraint is imposed despite the fact that the ratios are typically noisier than the two-point functions and are consistent with a single state considerably earlier than the two-point correlators. For a given $t_f^\text{min}$, the maximum sink time $t_f^\text{max}$ in the window is constrained to be at least 4 timeslices larger, and no later than than the latest time at which the hadron two-point functions are consistent with a single state (under correlated fits). The gap $\Delta \tau$ is also chosen to be at least 4. Subject to these constraints, all possible constant fits are performed to a given average ratio by minimizing the total $\chi^2$ function including both SS and SP ratios for $\{t_f,\tau\}$ pairs in each window. The mean of the bootstrap results for the best fit is taken as the central value, while the uncertainty is formed by taking half of the variation in central values of all fits with $\chi^2/\text{d.o.f.}\le1$ in quadrature with the standard deviation of the bootstrap results for the best fit. For the pion ratios, the variation over acceptable fits is typically small compared with the statistical uncertainty, while for the nucleon, whose signal degenerates more quickly with sink time $t_f$, this systematic uncertainty dominates. An alternative approach to the analysis using the summation method~\cite{Capitani:2012gj} yields fits for each averaged ratio that are consistent within uncertainties.

The GFFs $A_g(t)$, $B_g(t)$, and $D_g(t)$ for the nucleon, and $A_g^{(\pi)}(t)$ and $D_g^{(\pi)}(t)$ for the pion, are determined at each binned value of the momentum transfer $t$ by solving the overdetermined system of equations defined by fits to the averaged ratios at that $t$. The averaged ratios are first renormalised with the appropriate $Z_\mathfrak{R}^{\overline{\text{MS}}}(\mu=2~\text{GeV})$, allowing results from both representations to be combined in a simultaneous fit. To propagate the uncertainties on the renormalisation, this fit is performed $N_\text{sample}=250$ times at each $t$, sampling from the distributions of $Z_\mathfrak{R}^{\overline{\text{MS}}}(\mu=2~\text{GeV})$ in Eqs.~\eqref{eq:Z1} and \eqref{eq:Z2} for the two representations $\mathfrak{R}$. The standard deviation of the variation of the best fit values over the samplings is taken in quadrature with the statistical uncertainty of the fit with the central values of the renormalisation constants to define the total uncertainty of the GFFs at that $t$. Choices of $N_\text{sample}$ between 150 and 1000 give consistent uncertainty determinations.

Appendix~\ref{app:plateaufits} presents examples of the fits to the averaged ratios $\overline{R}^{(\pi/N)}_{\mathfrak{R};k}(t_f,\tau)$ at a number of momenta $t$. Also shown are graphical representations of examples of the systems of equations that are solved to determine the GFFs.

\section{Results}
\label{sec:results}

\begin{figure}[!p]
	\centering
	\subfigure[]{
		\includegraphics[width=0.47\textwidth]{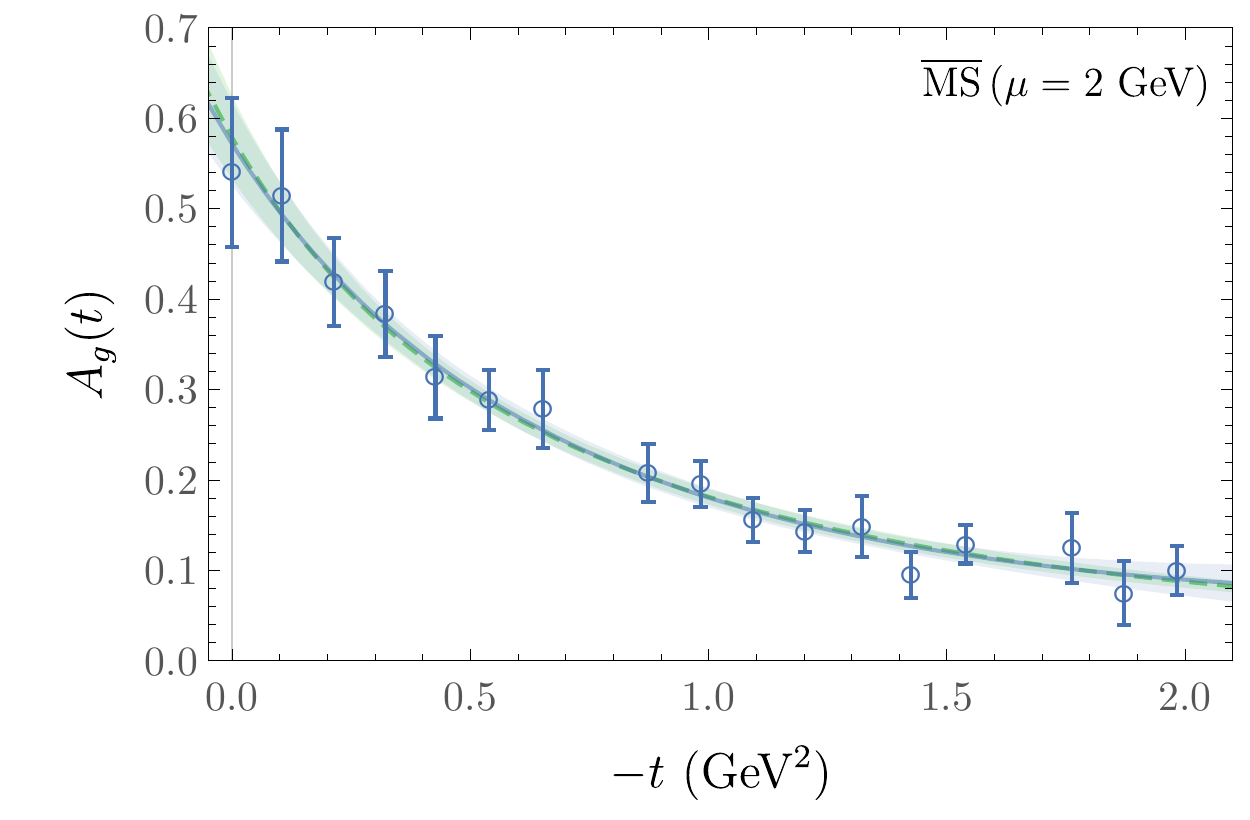} }
	\subfigure[]{
		\includegraphics[width=0.47\textwidth]{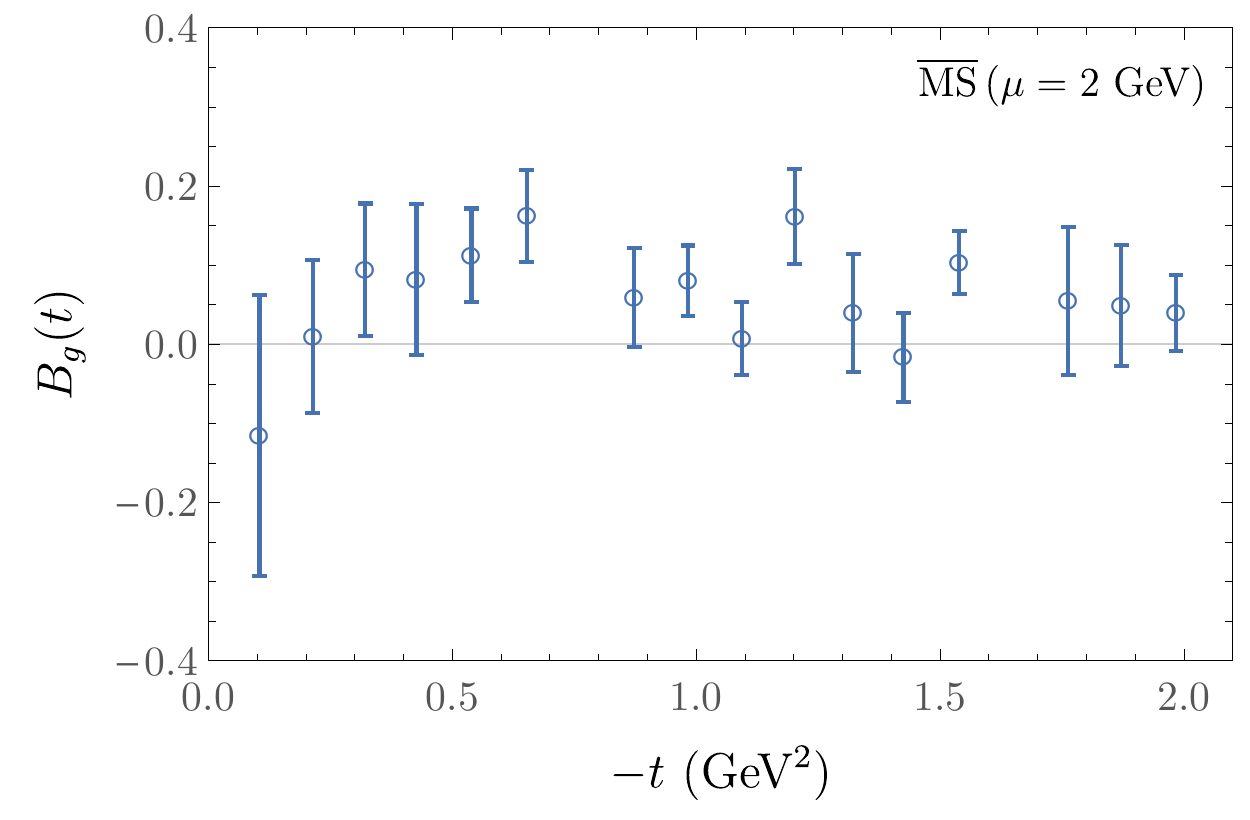}}
	\subfigure[]{
		\includegraphics[width=0.47\textwidth]{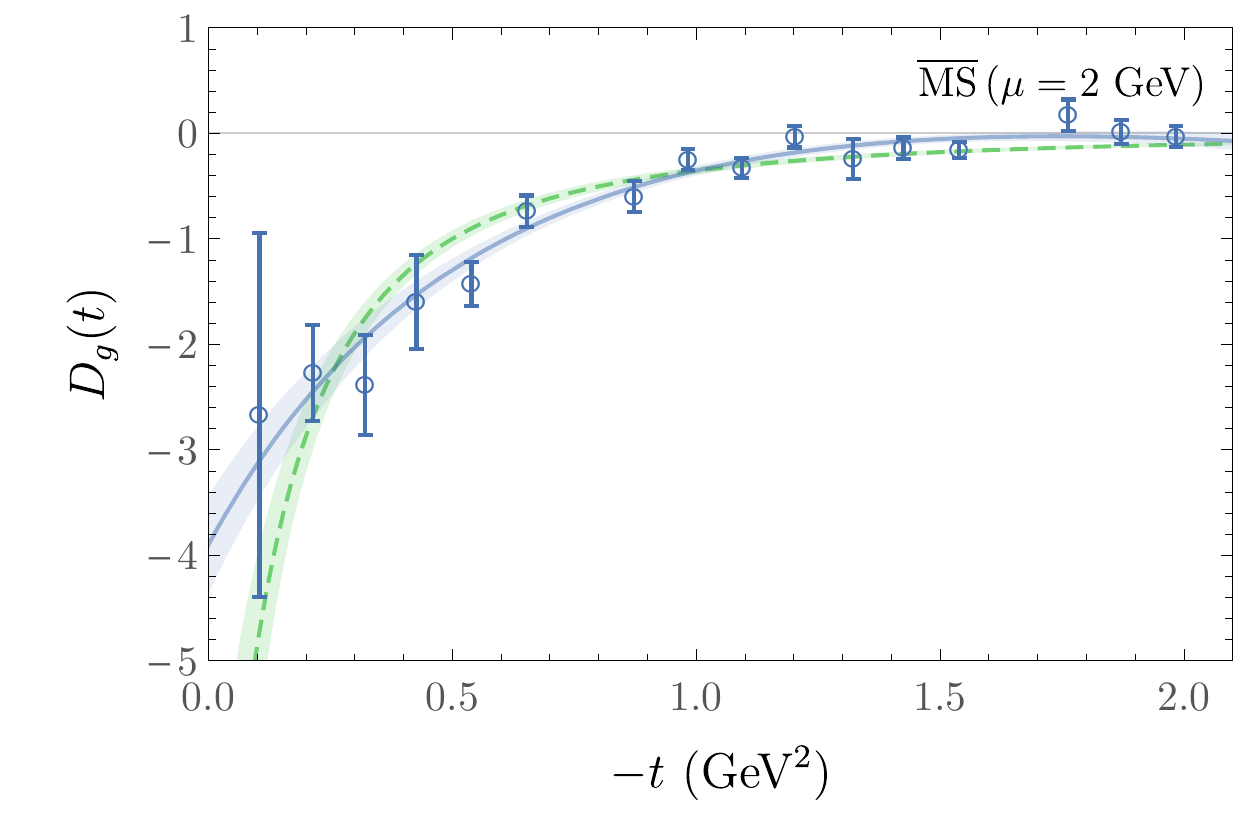}}
	\caption{Gluon GFFs of the nucleon, renormalised in the $\overline{\text{MS}}$ scheme at a scale of $\mu=2$~GeV. The solid blue bands illustrate $z$-expansion fits as described in the text, while the dashed green bands show dipole fits to the data. Horizontal error bars denoting the non-zero widths of the bins in $t$ (described in the text) are omitted as they are comparable to the sizes of the point markers, or smaller.}
	\label{fig:protonGFFs}
\end{figure}

\begin{figure}[!p]
	\centering
	\subfigure[]{
		\includegraphics[width=0.47\textwidth]{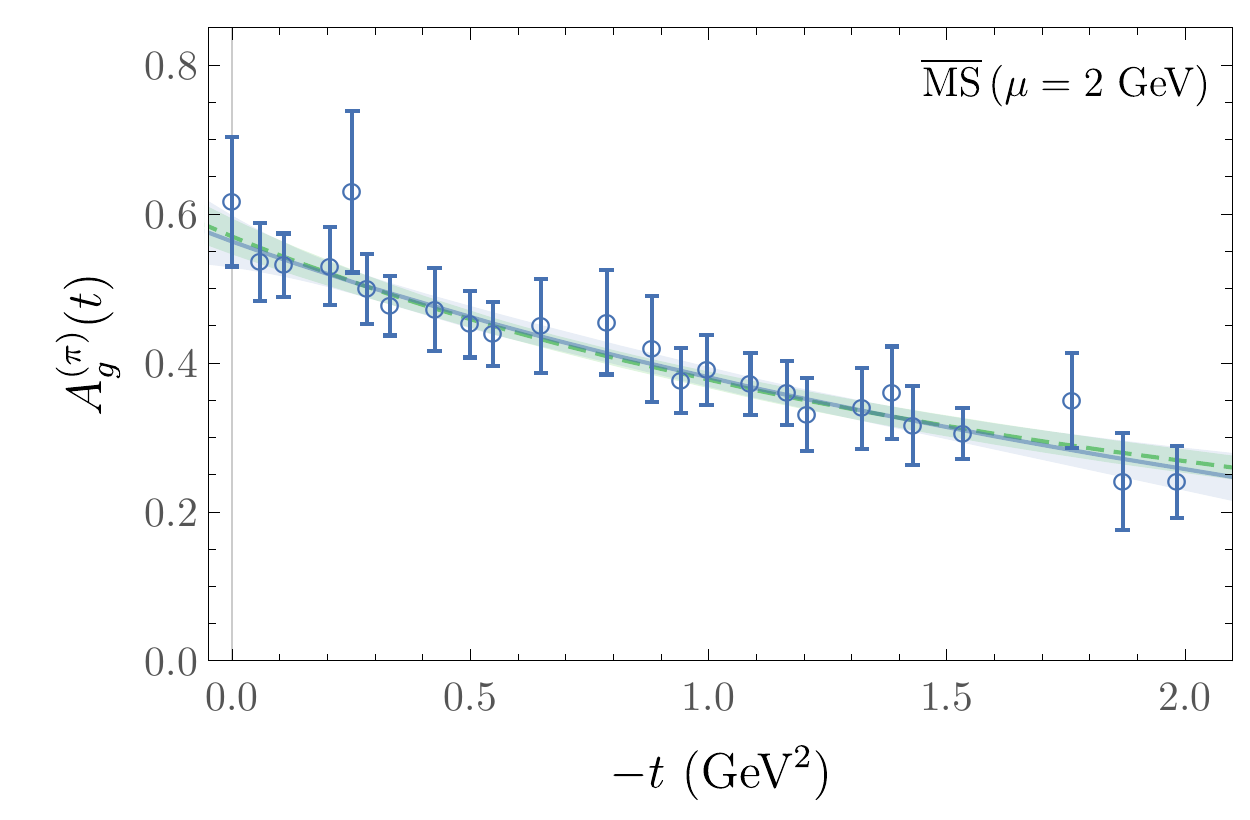}} \hspace*{5mm}
	\subfigure[]{
		\includegraphics[width=0.47\linewidth]{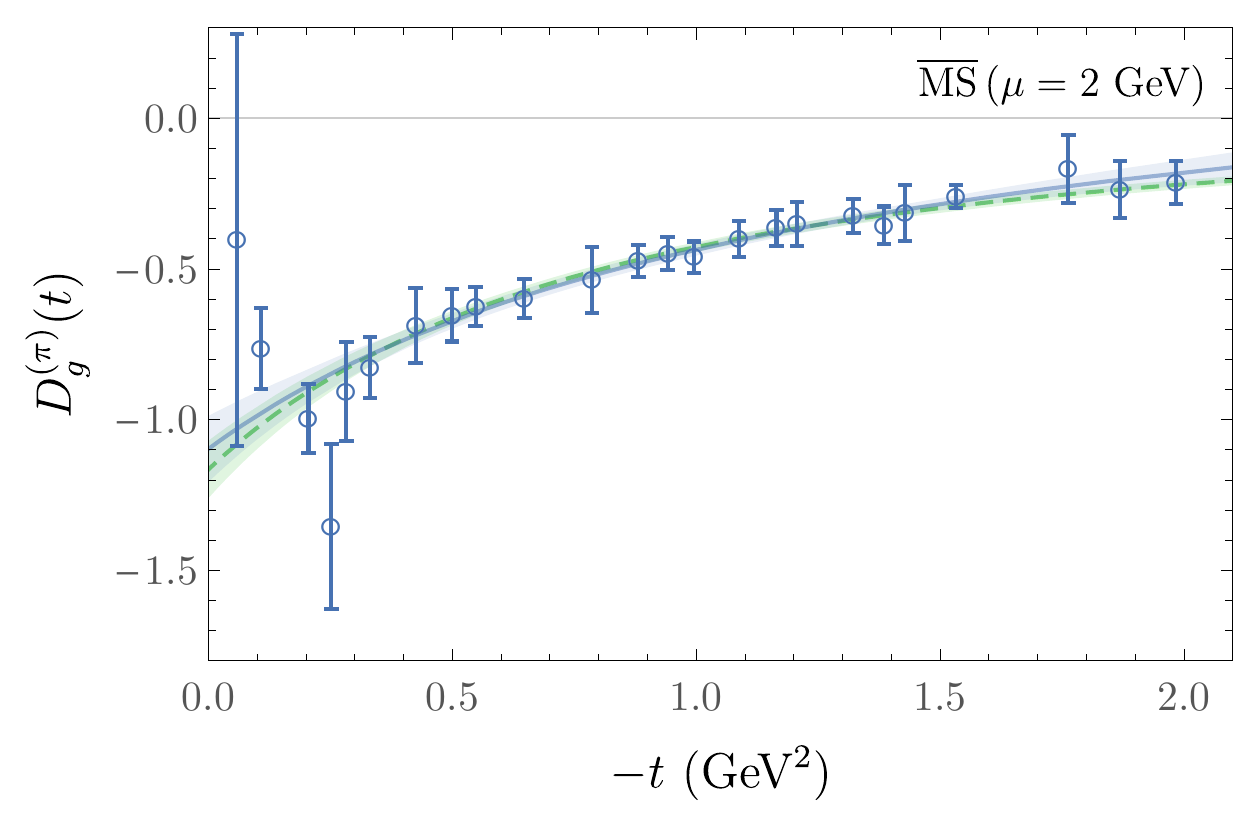}}
	\caption{Gluon GFFs of the pion, renormalised in the $\overline{\text{MS}}$ scheme at a scale of $\mu=2$~GeV. As in Fig.~\ref{fig:protonGFFs} for the nucleon, the solid blue bands illustrate $z$-expansion fits as described in the text, while the dashed green bands show dipole fits to the data. As in Fig.~\ref{fig:protonGFFs}, horizontal error bars denoting the non-zero widths of the bins in $t$ (described in the text) are omitted as they are comparable to the sizes of the point markers, or smaller.}
	\label{fig:pionGFFs}
\end{figure}

The three gluon GFFs of the nucleon and two gluon GFFs of the pion that are extracted from the LQCD calculations detailed in Sec.~\ref{sec:latt} are shown in Figs.~\ref{fig:protonGFFs} and \ref{fig:pionGFFs}, respectively. Results are shown in the $\overline{\text{MS}}$ scheme at $\mu=2$~GeV, where the renormalisation procedure is as described in Sec.~\ref{sec:NPR}. The $A(t)$ and $D(t)$ GFFs for the  nucleon and pion fall off as $|t|$ increases and are well-described by dipole forms as well as the more general $z$-expansion~\cite{Hill:2010yb} parametrisation:
\begin{equation}
\label{eq:fitformdipole}
X_\text{dipole}(t) = \frac{\alpha}{(1-t/m^2)^2}, \hspace{6mm} X_\text{$z$-exp.}(t)=\sum_{k=0}^{k_\text{max}}a_k \left[z(t)\right]^k, \hspace{2mm} z(t) = \frac{\sqrt{t_\text{cut}-t}-\sqrt{t_\text{cut}}}{\sqrt{t_\text{cut}-t}+\sqrt{t_\text{cut}}},
\end{equation}
where $k_\text{max}=2$ and for both pion and nucleon $t_\text{cut}=4m_\pi^2$~\cite{Hill:2010yb}. Higher-order $z$-expansion fits yield a $\chi^2/\text{d.o.f.}\ll1$, overfitting statistical fluctuations in the data. Fit parameters for each parametrisation are tabulated in Table~\ref{tab:dipolemasses}. Notably, the dipole masses are a factor of two larger for the pion than for the nucleon for both $A(t)$ and $D(t)$. Correspondingly, the gluon radii defined from either form factor are smaller for the pion than the nucleon, as is found in experiment for the respective charge radii~\cite{Tanabashi:2018oca}. The nucleon $B_g(t)$ GFF is consistent with zero over the entire range of $t$ of this study.
\begin{table}
	\begin{tabular}{c|cc|ccc}\toprule
		& $m$ (GeV) & $\alpha$  &  $a_0$  & $a_1$ & $a_2$ \\ \hline
		$A_g$ & 1.13(6) & 0.58(5) & 0.57(5) & -2.7(5) & 4(1)  \\
		$D_g$ & 0.48(5) & -10(3) & -3.9(5) & 28(4) & 50(10) \\
		$A_g^{(\pi)}$  & 2.1(2) & 0.56(3) & 0.55(4) & -0.7(5) & -0.9(1.6) \\
		$D_g^{(\pi)}$ & 1.24(7) & -1.2(1) & -1.1(1) & 3.9(1.2) & -3(3)  \\\toprule
	\end{tabular}
	\caption{\label{tab:dipolemasses} 
		Fit parameters of dipole and $z$-expansion fits (Eq.~\eqref{eq:fitformdipole}) to the $t$-dependence of the nucleon and pion gluon GFFs. The fits are displayed as bands in Figs.~\ref{fig:protonGFFs} and \ref{fig:pionGFFs}.  }
\end{table}


\begin{figure}[!th]
	\centering
	\subfigure[\label{fig:pirat}]{
		\includegraphics[width=0.47\textwidth]{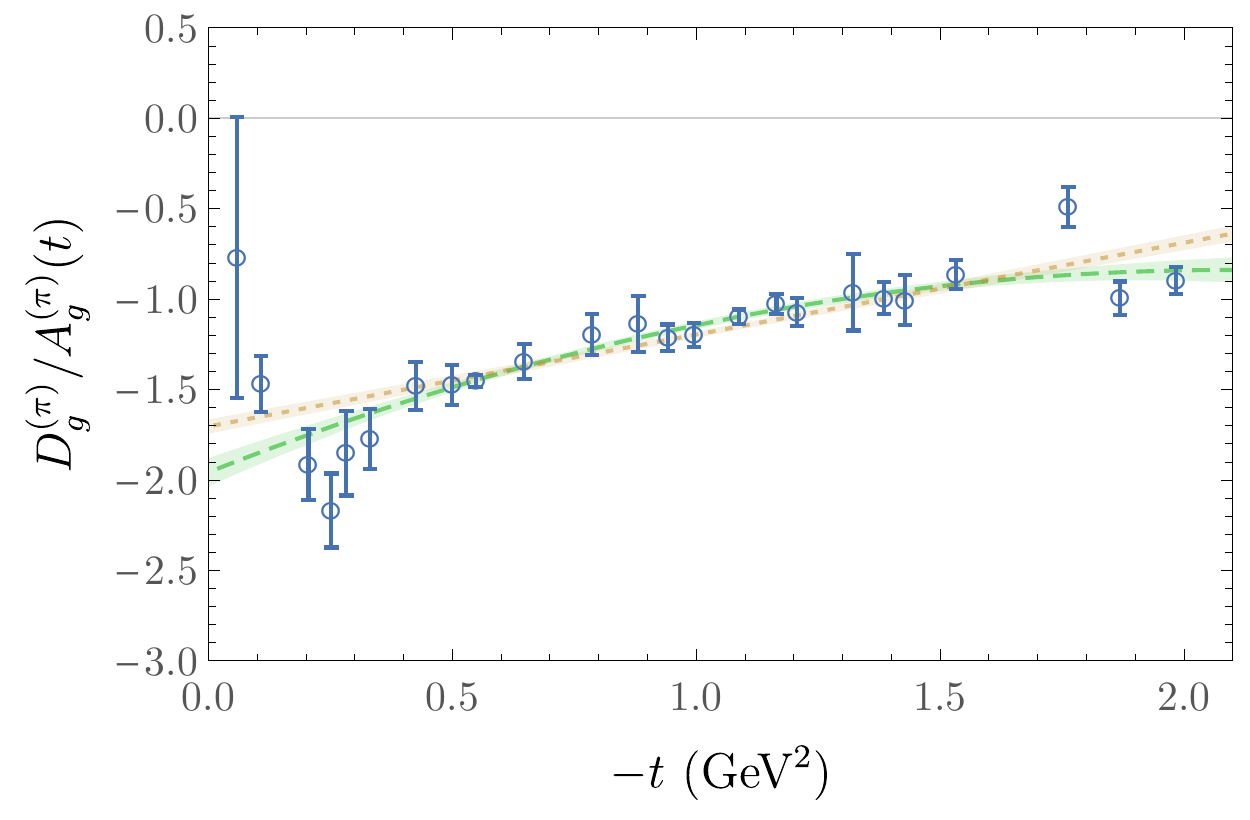}}\\
	\subfigure[]{
		\includegraphics[width=0.47\textwidth]{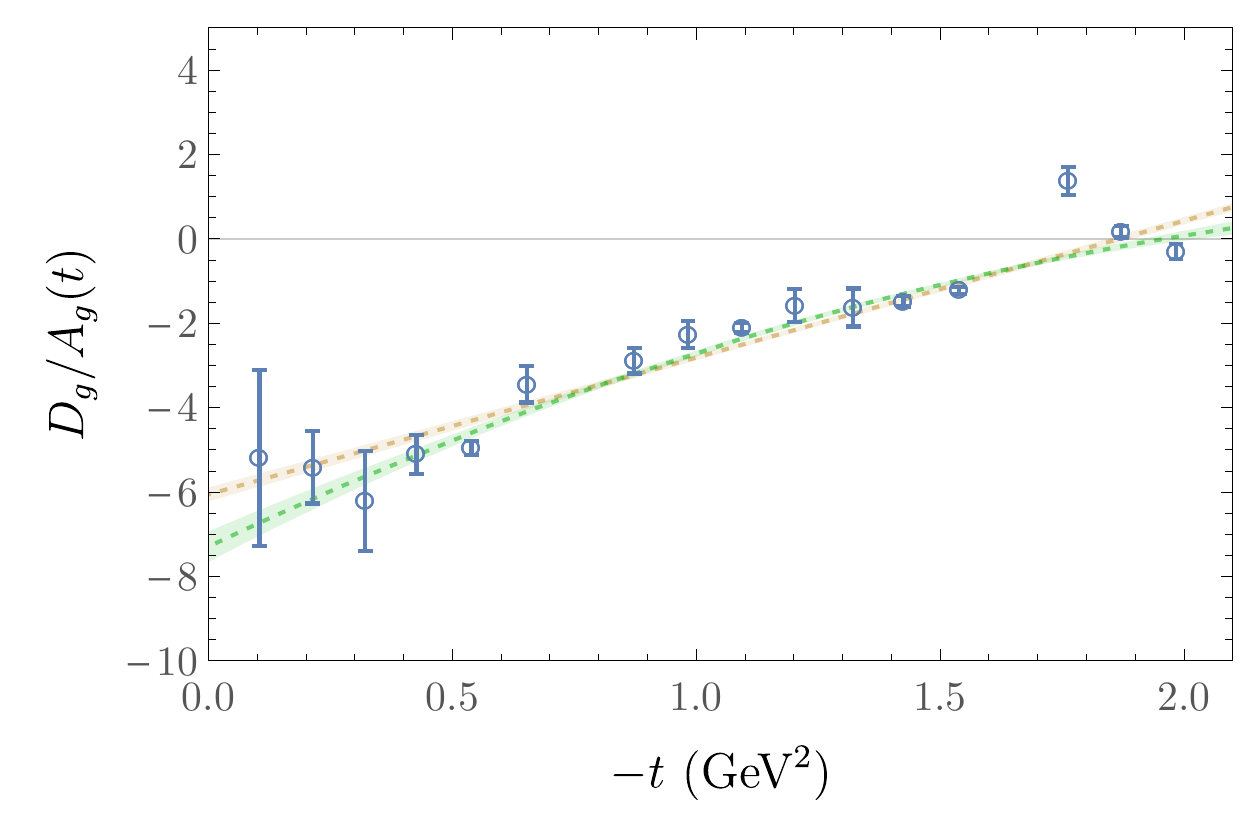}}
	\subfigure[]{
		\includegraphics[width=0.47\textwidth]{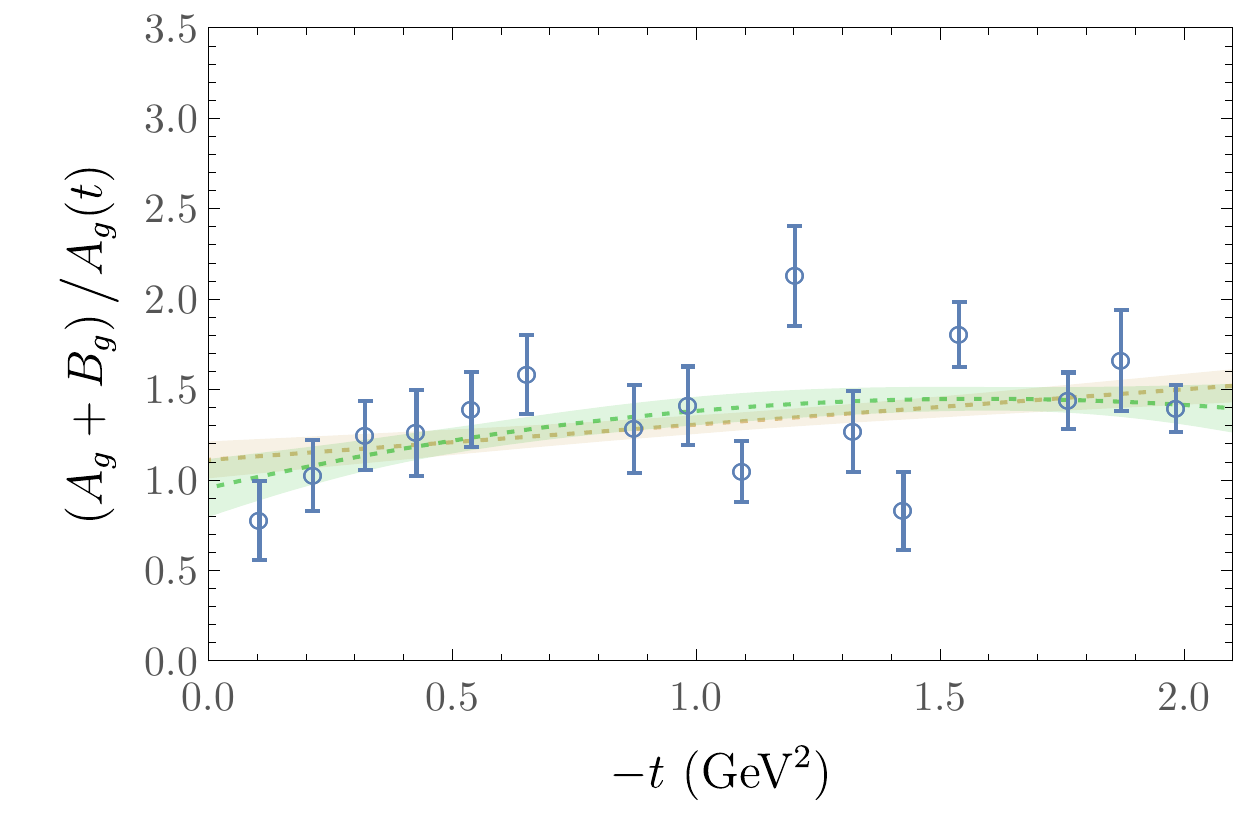}}
	\caption{Ratios of the extracted gluon GFFs of the pion and nucleon, which are independent of renormalisation under the assumption of negligible mixing with the corresponding quark GFFs. The green dashed and orange dotted fit bands show constant and linear fits to the data, respectively. Horizontal error bars denoting the non-zero widths of the bins in $t$ (described in the text) are omitted as they are comparable to the sizes of the point markers, or smaller.}
	\label{fig:DonA}
\end{figure}

Ratios of the GFFs are renormalisation-independent if mixing with the corresponding quark GFFs is negligible.  Fig.~\ref{fig:DonA} displays these ratios for both the nucleon and the pion, along with linear and quadratic fits to their $t$-dependence. Several notable features are apparent.  First, the ratios of $D_g^{(\pi)}/A_g^{(\pi)}$ and  $D_g/A_g$ are approximately linear over a wide range of $t$, and the nucleon ratio is larger than the pion ratio at $t=0$.  Second, since $B_q(t)\sim0$ , the ratio $(A(t)_g+B(t)_g)/A(t)_g$ is approximately unity, and, correspondingly, the fractional contributions of gluons to the nucleon angular momentum and momentum are similar. 

\begin{figure}
	\centering
	\subfigure[Nucleon	\label{fig:nucleonxvsmpi2}]{
		\includegraphics[width=0.47\textwidth]{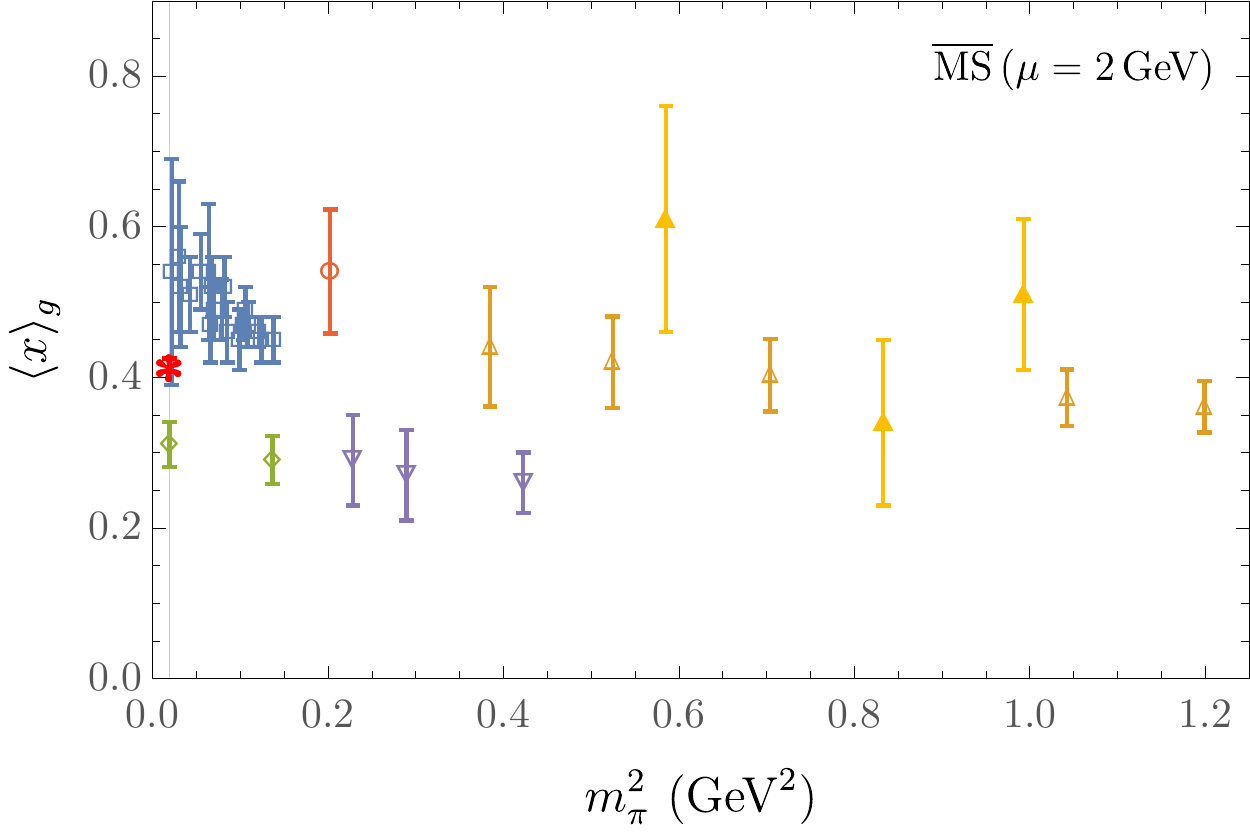}} \hspace*{5mm}
	\subfigure[Pion	\label{fig:pionxvsmpi2}]{
		\includegraphics[width=0.47\textwidth]{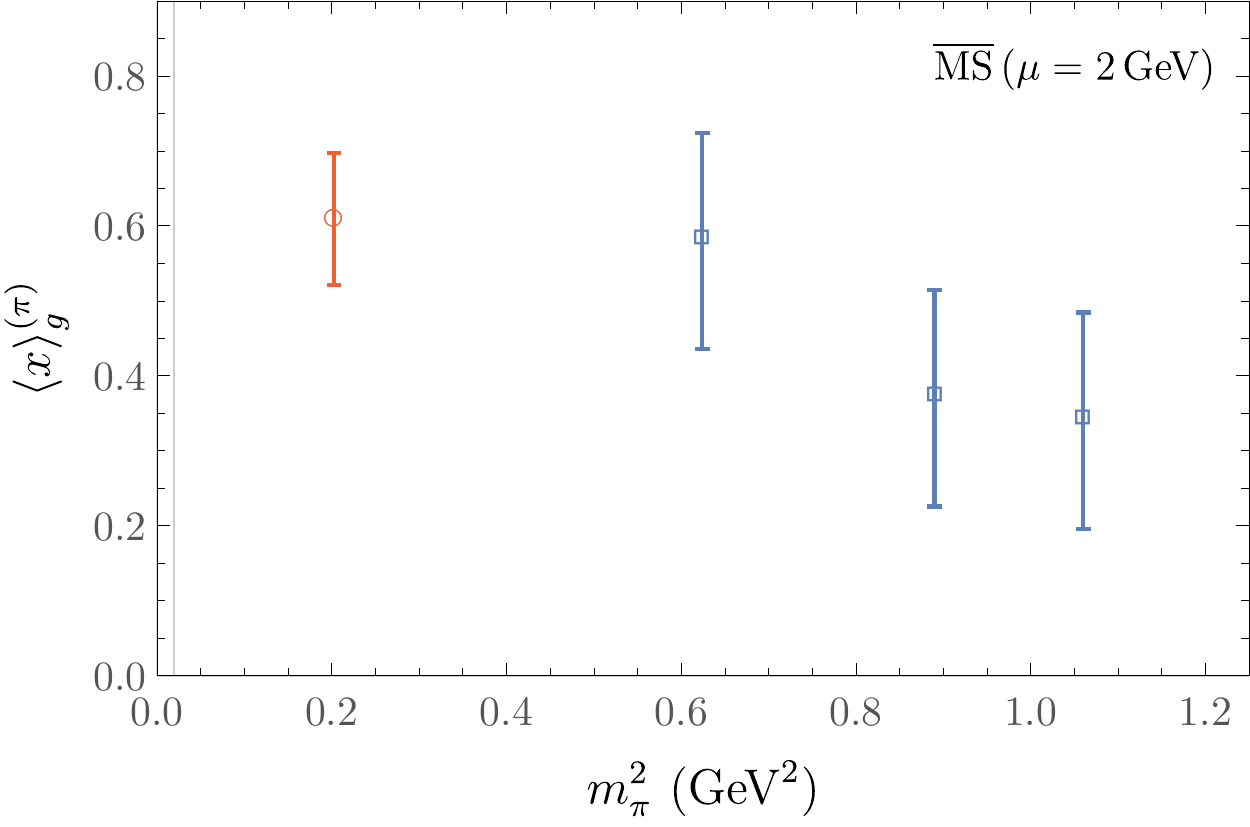}}
	\caption{
		Comparison of the gluon momentum fractions of the nucleon and pion determined in this study, denoted by red circles on each subfigure, to the results of previous calculations at different values of the pion mass. In subfigure (a) the blue squares show data taken from Ref.~\cite{Yang:2018nqn} ($\chi$QCD collaboration) which were computed using various ensembles of domain wall fermion configurations, and the green diamonds show results from Ref.~ \cite{Alexandrou:2016ekb} (ETM collaboration) obtained using twisted-mass fermions. Results from quenched QCD are also shown: the purple inverted triangles show the results of Ref.~\cite{Deka:2013zha} ($\chi$QCD collaboration) determined using quenched QCD, the orange triangles show those from Ref.~\cite{Horsley:2012pz} (QCDSF collaboration), and the yellow filled triangles denote those from Ref.~\cite{Gockeler:1996zg}  (QCDSF collaboration). 
		The experimental value for the proton is shown as the red star and is taken from the CT14 PDF parametrisation~\cite{Dulat:2015mca}. In subfigure (b), the blue squares show data from the quenched QCD calculation reported in Ref.~\cite{Meyer:2007tm}.
	}
\end{figure}

While there are no previous QCD calculations of the $t$-dependence of the three gluon GFFs of the nucleon, a quenched QCD calculation of $A_g(t)$ and $B_g(t)$ was presented in Ref.~\cite{Deka:2013zha} at larger than physical quark masses. The $A_g(t)$ form factor determined in that study is considerably smaller in magnitude although with a similar $t$-dependence to that calculated here, while the $B_g(t)$  form factor is consistent with zero.
The gluon momentum fraction of the nucleon, which corresponds to the forward limit $A_g(0)=\langle x \rangle_g$, is found in the present study to be $\langle x\rangle_g(\mu= 2\ {\rm GeV})=0.54(8)$ at $m_\pi=450$ MeV in the $\overline{\rm MS}$ scheme. This quantity has previously been determined in a number of LQCD calculations~\cite{Horsley:2012pz,Deka:2013zha,Alexandrou:2016ekb,Yang:2018nqn,Gockeler:1996zg}. These results are collated in Fig.~\ref{fig:nucleonxvsmpi2}, which also includes the phenomenological result for the momentum fraction from the CT14 PDF parametrisation~\cite{Dulat:2015mca}. It is notable that while there is some scatter in the LQCD results, likely a result of systematic uncertainties that are as-yet uncontrolled, the gluon momentum fraction is approximately constant with changes in quark masses within each study (which one could expect to have correlated systematic effects at different masses). It is expected that the GFFs at nonzero $t$ will also be approximately independent of quark mass. 
The gluon momentum fraction of the pion from this work is $\langle x\rangle^{(\pi)}_g(\mu= 2\ {\rm GeV})=0.61(9)$ at $m_\pi=450$ MeV in the $\overline{\rm MS}$ scheme.  A phenomenological estimate of $\langle x\rangle^{(\pi)}_g\sim 0.3$  in reported in Ref.~\cite{Gluck:1999xe} but without an uncertainty. A quenched QCD calculation of this quantity has been presented in Ref.~\cite{Meyer:2007tm}, and a comparison with the results of this study is shown in Fig.~\ref{fig:pionxvsmpi2}. As was found for the nucleon, no significant quark-mass dependence is evident in this quantity.

\begin{figure}[!hp]
	\centering
	\subfigure[]{
		\includegraphics[width=0.47\textwidth]{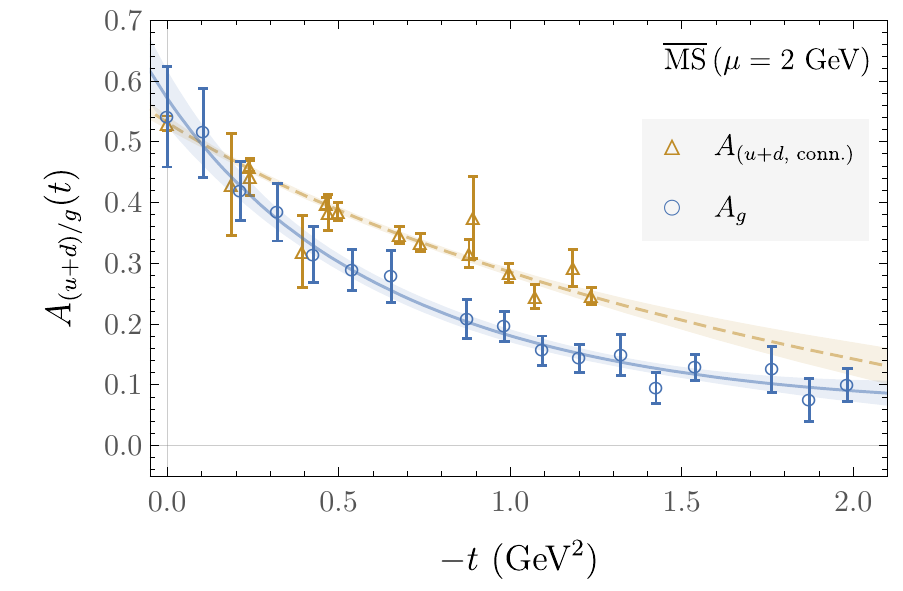}}
	\subfigure[]{
		\includegraphics[width=0.47\textwidth]{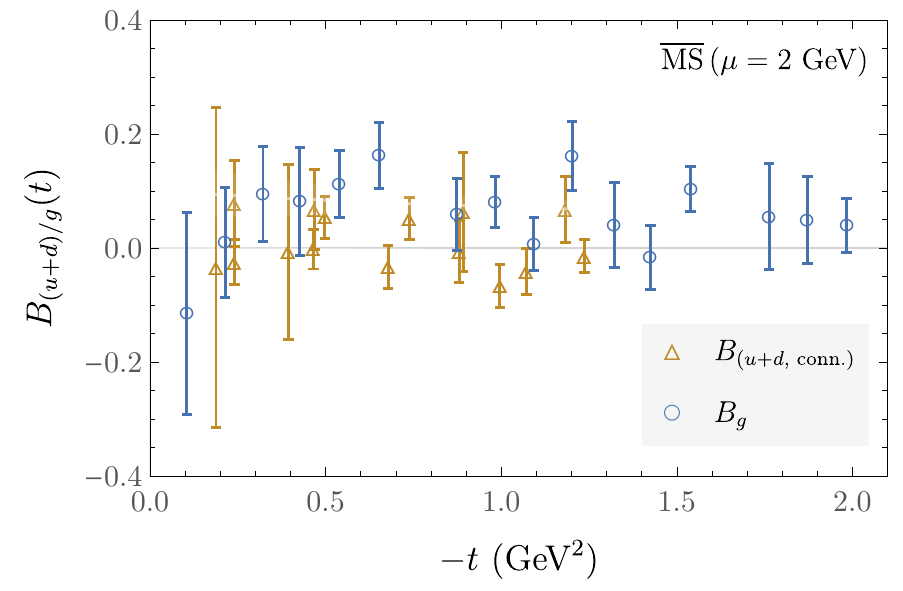}}
	\subfigure[]{
		\includegraphics[width=0.47\textwidth]{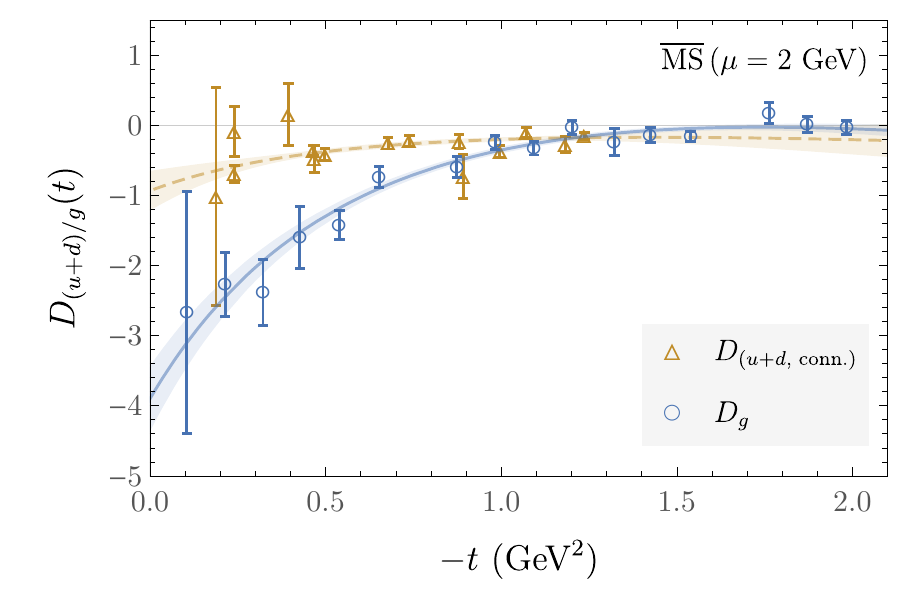}}
	\caption{The renormalised  gluon GFFs of the nucleon (blue circles) compared with the corresponding connected isoscalar quark GFFs (orange triangles) from Ref.~\cite{Hagler:2007xi} that are calculated using a similar light quark mass (corresponding to $m_\pi=496$ MeV).	Results are presented at a 
		renormalisation scale of $\mu=2$~GeV in the $\overline{\text{MS}}$ scheme.  For the $A_a(t)$ and $D_a(t)$ form factors, the shaded bands show $z$-expansion fits as described in the text. 
	}
	\label{fig:QvsGNucleon}
\end{figure}
\begin{figure}[!hp]
	\centering
	\subfigure[]{
		\includegraphics[width=0.47\textwidth]{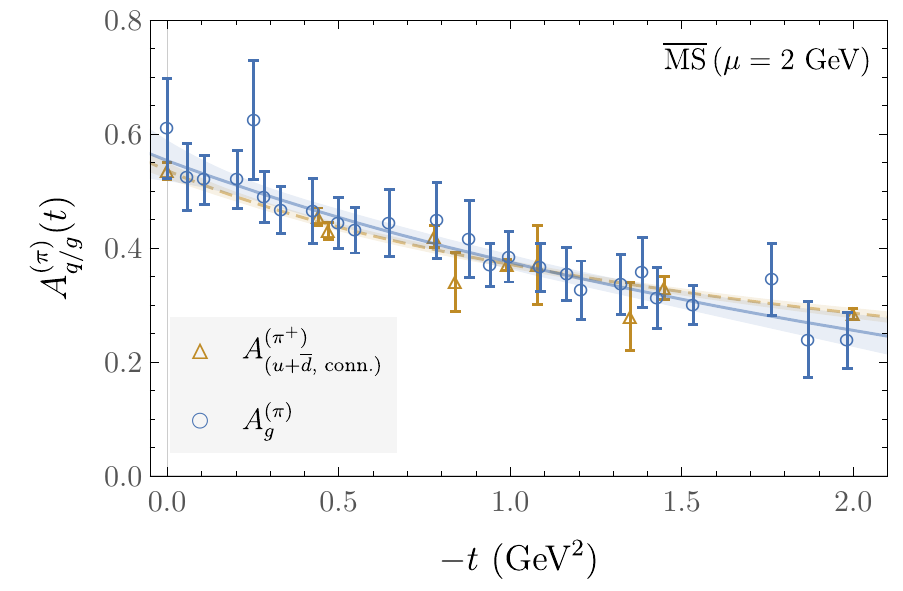} }
	\subfigure[]{
		\includegraphics[width=0.47\textwidth]{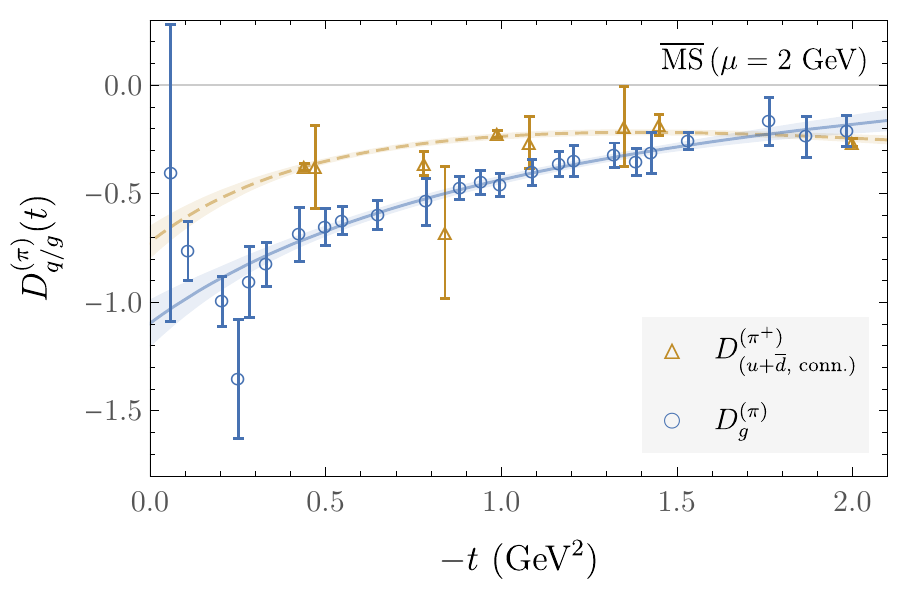}}
	\caption{The renormalised gluon GFFs of the pion (blue circles) compared with the corresponding connected contributions to the quark GFFs (orange triangles) computed in Ref.~\cite{Brommel:2007zz} (taken from Fig. 7.6 in that reference) at a quark mass 
		corresponding to $m_\pi=842$ MeV using non-perturbatively improved clover fermions.  Results are presented at a renormalisation scale of $\mu=2$~GeV in the $\overline{\text{MS}}$ scheme.  The shaded bands show $z$-expansion fits as described in the text for the quark and gluon GFFs.
	}
	\label{fig:QvsGPion}
\end{figure}

The quark GFFs of the nucleon and pion have been previously computed at various quark masses in LQCD using a variety of lattice actions and quark masses. Only the connected quark-line contributions have been determined in most cases. Where they have been computed at similar quark masses to those in this study, disconnected contributions are found to be at the percent level~\cite{Deka:2013zha}.   Figure~\ref{fig:QvsGNucleon} shows a comparison of the three gluon GFFs of the nucleon to the corresponding connected isoscalar quark GFFs computed at a similar quark mass (corresponding to $m_\pi\sim496$ MeV) to that used in this work using domain wall valence quarks on configurations generated with an ASqTad staggered quark action at a similar lattice spacing ($a=0.125$ fm)~\cite{Hagler:2007xi} to that used here. Fits using the $z$-parametrisation of Eq.~\eqref{eq:fitformdipole} are shown to both quark and gluon of GFFs. The $A_g(t)$ and $D_g(t)$ GFFs fall off more quickly with $t$ than the corresponding quark GFFs, and thus the generalised nucleon radii defined from the gluon GFFs are larger than those from the corresponding quark quantities. 
In the forward limit, $A_g(0)\approx A_{u+d,\text{conn.}}(0)$, indicating that quarks and gluons each contribute approximately half of the  nucleon momentum  at this unphysically heavy value of the quark mass (the sum of the gluon and connected quark momentum fraction is $\sim1.07(9)$, indicating that undetermined systematic effects from disconnected contributions and lattice artefacts are likely small).  The quark and gluon $D$-terms are both negative, with $D_g(t)\sim 2 D_{u+d, ({\rm conn.})}(t)$. The $B_g$ and $B_{u+d, ({\rm conn.})}$ GFFs are both  consistent with zero.

The connected quark GFFs of the pion have previously been calculated using  a different formulation of the clover quark action at somewhat heavier quark masses corresponding to a  pion mass of 842 MeV and a lattice spacing of $a=0.073$ fm~\cite{Brommel:2007zz}. These results are shown alongside the gluon GFFs of the pion computed here in Fig. \ref{fig:QvsGPion}.  Fits using the $z$-parametrisation of Eq.~\eqref{eq:fitformdipole} are shown for both quark and gluon GFFs. As for the nucleon, the gluon and quark contributions to the lightcone momentum of the pion, defined by the forward limits $A_g(0)$ and $A_{u+\overline{d},\text{conn.}}(0)$ respectively, are similar. Unlike for the nucleon, the corresponding pion quark and gluon GFFs $A_a(t)$ remain similar over the entire range of $t$ that is investigated. The quark and gluon $D$-term GFFs in the pion are also similar in magnitude, relative to the uncertainties of each calculation.

\section{Conclusion}
\label{sec:conc}

In this article, the first determination of the complete set of gluon generalised gravitational form factors of the nucleon and pion from lattice QCD is presented. All GFFs are found to have dipole-like dependence on the squared momentum transfer $t$, with the exception of the $B_g(t)$ GFF of the nucleon that is consistent with zero over the entire rage of $t$ that is investigated. For the nucleon, the gluon GFFs fall off faster in $|t|$ and can be parametrised with larger dipole masses than the corresponding quark GFFs computed using similar lattice discretisations and at a similar value of the quark masses, indicating the gluon distributions have a smaller spatial size than those of the quarks. In contrast, the quark and gluon GFFs of the pion have very similar $t$-dependences. For both the pion and the nucleon, the gluon momentum fraction, corresponding to the forward limit of one of the GFFs, is found to be approximately 0.5--0.6, somewhat larger the phenomenological value in both cases. The gluon contributions to the nucleon momentum and angular momentum are of similar relative size.

All calculations presented here have been performed at a single lattice spacing and volume and at a single unphysical  value of the light quark masses, and mixing of the isoscalar quark GFFs with the gluon GFFs has been neglected based on expectations from lattice perturbation theory \cite{Alexandrou:2016ekb} that these effects are small. The as-yet-unquantified systematic uncertainties that result from the lattice spacing and finite volume effects are expected to be considerably smaller than the uncertainties reported on the renormalised GFFs. Since the gluon GFFs are determined from purely gluonic operators (up to effects of mixing), the quark-mass--dependence is also expected to be mild, and extrapolation to the physical quark masses will likely not shift the GFFs outside their uncertainties.
Future calculations will control these remaining systematic uncertainties and thereby allow more precise comparisons with phenomenology and also controlled predictions for the gluon contributions to the shear and pressure distributions of the nucleon and pion that are determined by the $D$-term GFFs \cite{Polyakov:1998ze,Polyakov:2018zvc}.

\section*{Acknowledgements}

This work used the facilities of the Extreme Science and Engineering Discovery Environment (XSEDE), which is supported by National Science Foundation grant number ACI-1548562, under allocation TG-PHY170018, as well as facilities of the USQCD Collaboration, which are funded by the Office of Science of the U.S. Department of Energy. Resources of the National Energy Research Scientific Computing Center (NERSC), a U.S. Department of Energy Office of Science User Facility operated under Contract No. DE-AC02-05CH11231, were also used, as were computing facilities at the College of William and Mary which were provided by contributions from the National Science Foundation, the Commonwealth of Virginia Equipment Trust Fund and the Office of Naval Research.
PES is supported by the National Science Foundation under CAREER Award 1841699 and in part by Perimeter Institute for Theoretical Physics. Research at Perimeter Institute is supported by the Government of Canada through the Department of Innovation, Science and Economic Development and by the Province of Ontario through the Ministry of Research and Innovation. WD is supported by the U.S. Department of Energy under Early Career Research Award DE-SC0010495 and grant DE-SC0011090. WD is also supported  within the framework of the TMD Topical Collaboration of  the U.S. Department of Energy, Office of Science, Office of Nuclear Physics, and  by the SciDAC4 award DE-SC0018121. The Chroma software library~\cite{Edwards:2004sx} was used in the data analysis. The authors thank Robert Edwards, Balint Joo, Kostas Orginos, Dimitra Pefkou, and the NPLQCD collaboration for generating the ensembles used in this study, and Ross Young and James Zanotti for helpful discussions.

\appendix
\section{Plateau fits and GFF extractions}
\label{app:plateaufits}

\begin{figure}
	\centering
	\subfigure[$t_f=12$ (dark points), $t_f$=14 (pale points)]{
	\includegraphics[width=\linewidth]{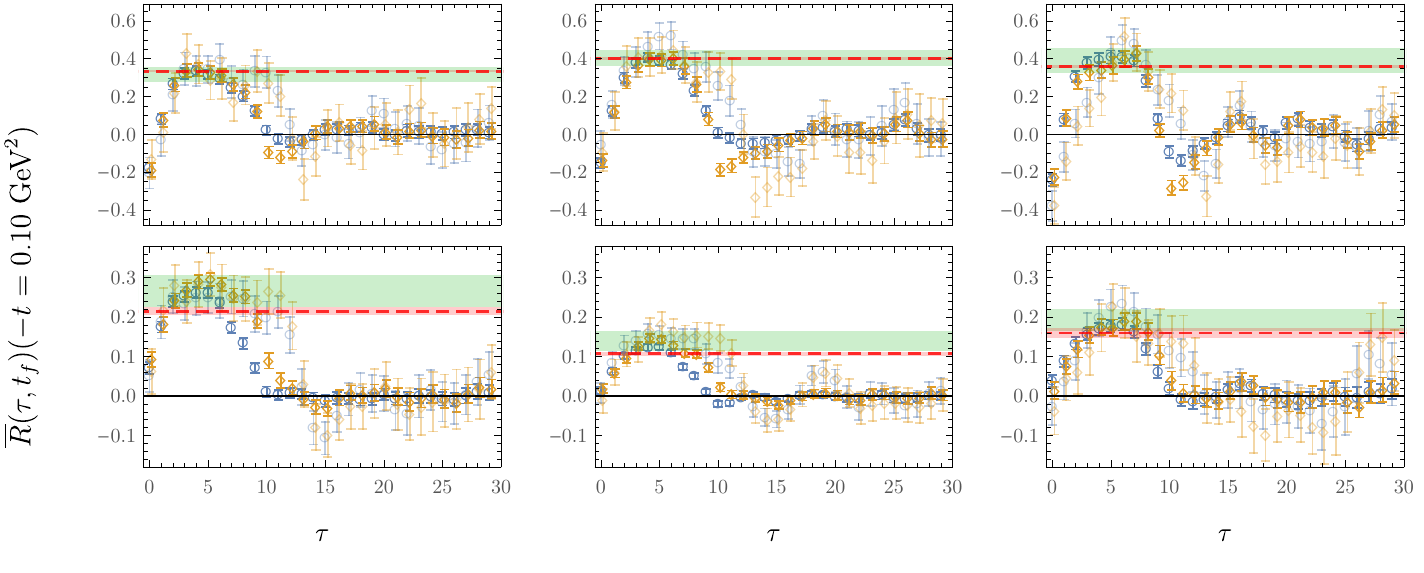}}
	\subfigure[$\tau=4$ (dark points), $\tau$=6 (pale points)]{
	\includegraphics[width=\linewidth]{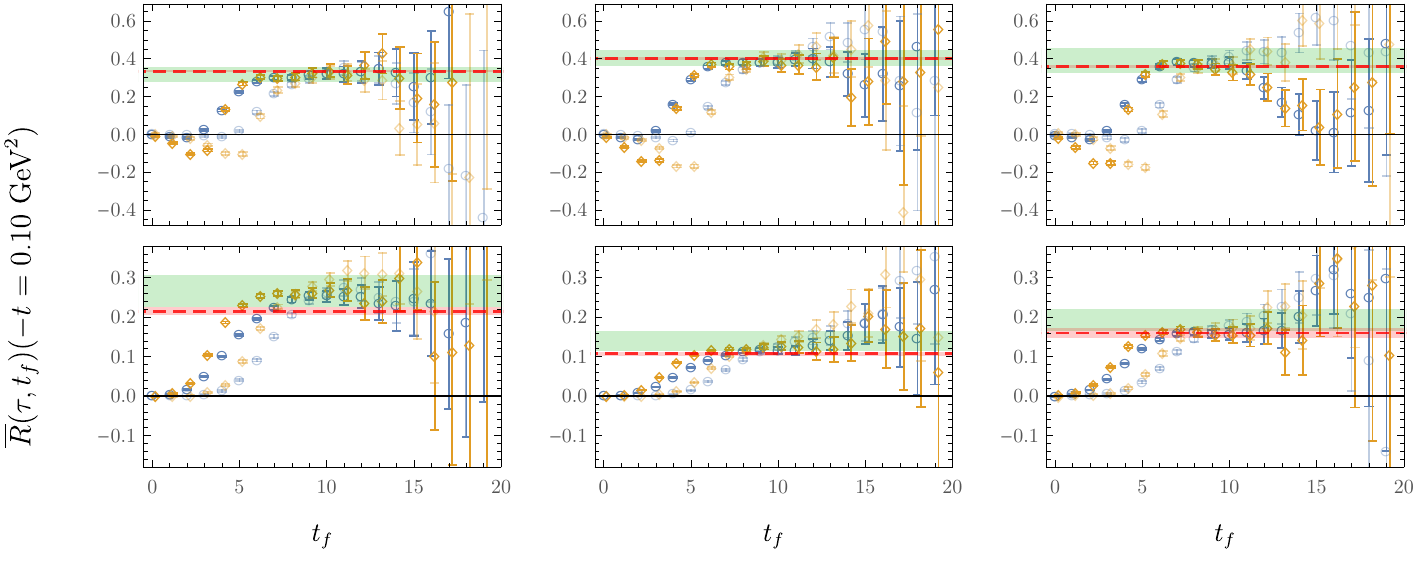}}
	\caption{Examples of the raw lattice data for averaged ratios of three and two point functions $\overline{R}^{(N)}_{\mathfrak{R};k}(t_f,\tau)$ for the nucleon formed from SP (blue circles) and SS (orange diamonds) correlation functions, as well as plateau fits to their $t_f$ and $\tau$ dependence (green bands). The red dashed bands show the central value and uncertainty of the final fitted result for the GFFs projected back to the linear combination of GFFs corresponding to each ratio. The upper (lower) row of figures in each panel  shows examples of ratios determined using operators in representation $\mathfrak{R}=\tau_1^{(3)}$ ($\tau_3^{(6)}$). }
	\label{fig:pbump4}
\end{figure}

\begin{figure}
	\centering
	\subfigure[$t_f=12$ (dark points), $t_f$=14 (pale points)]{
	\includegraphics[width=\linewidth]{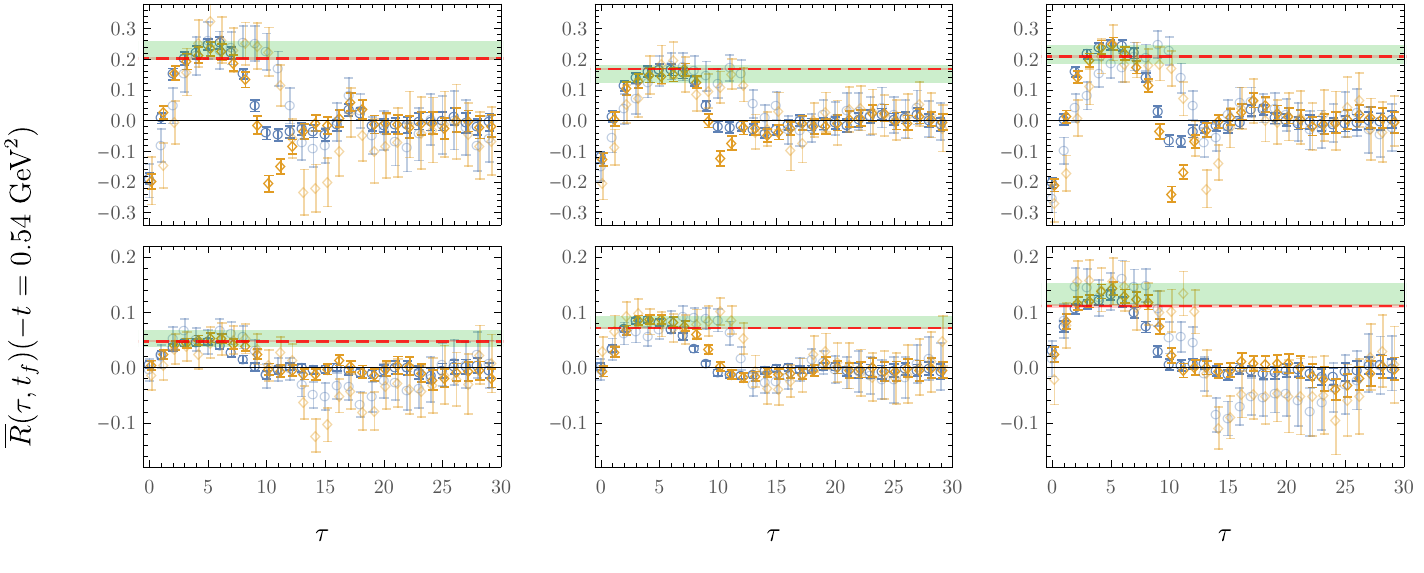}}
	\subfigure[$\tau=4$ (dark points), $\tau$=6 (pale points)]{
	\includegraphics[width=\linewidth]{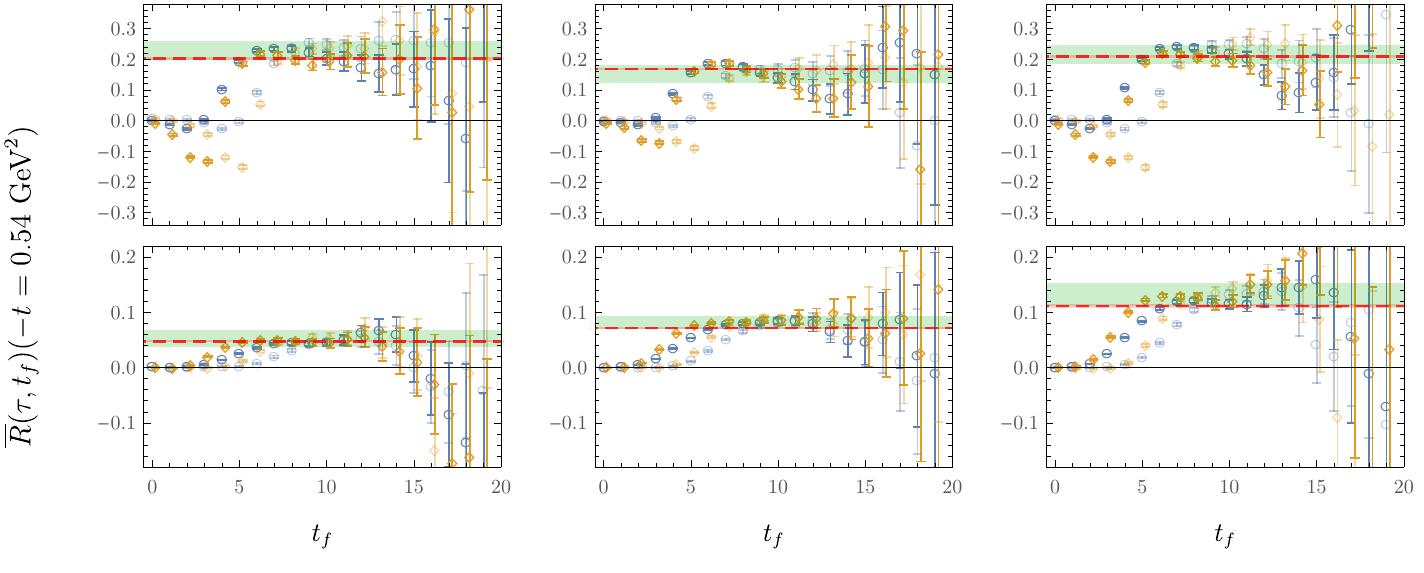}}
	\caption{As in Fig.~\ref{fig:pbump4}, for a different value of the squared momentum transfer $t$.}
	\label{fig:pbump9}
\end{figure}

\begin{figure}
	\centering
	\subfigure[$t_f=12$ (dark points), $t_f$=14 (pale points)]{
	\includegraphics[width=\linewidth]{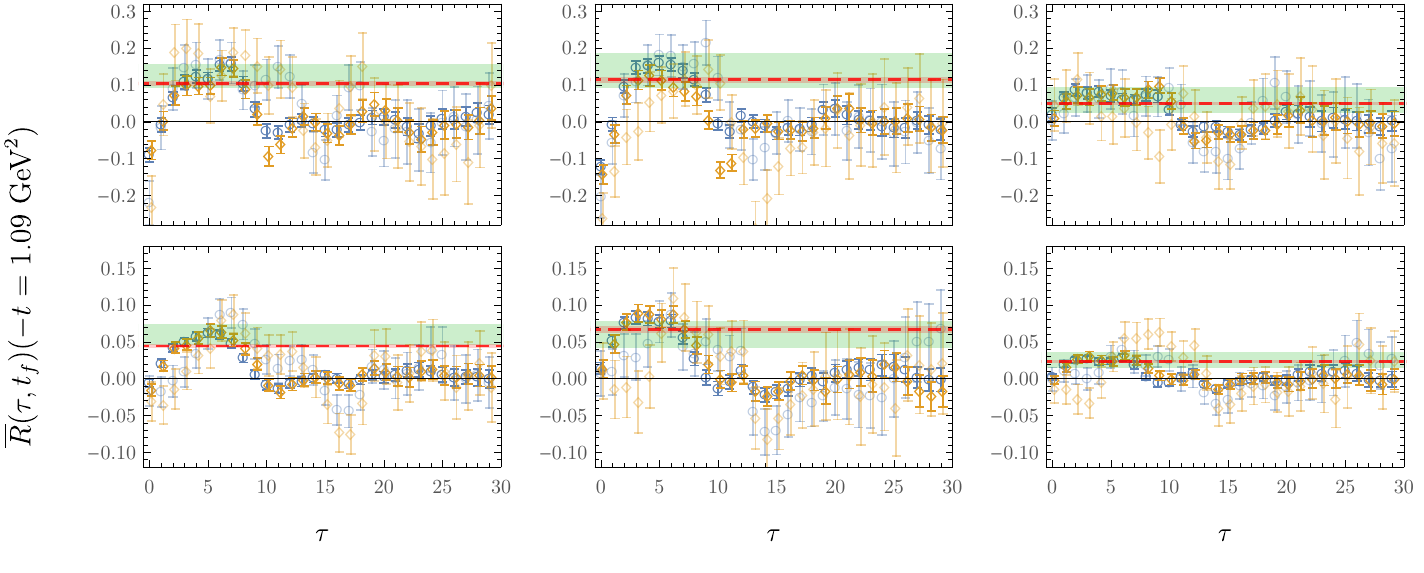}}
	\subfigure[$\tau=4$ (dark points), $\tau$=6 (pale points)]{
	\includegraphics[width=\linewidth]{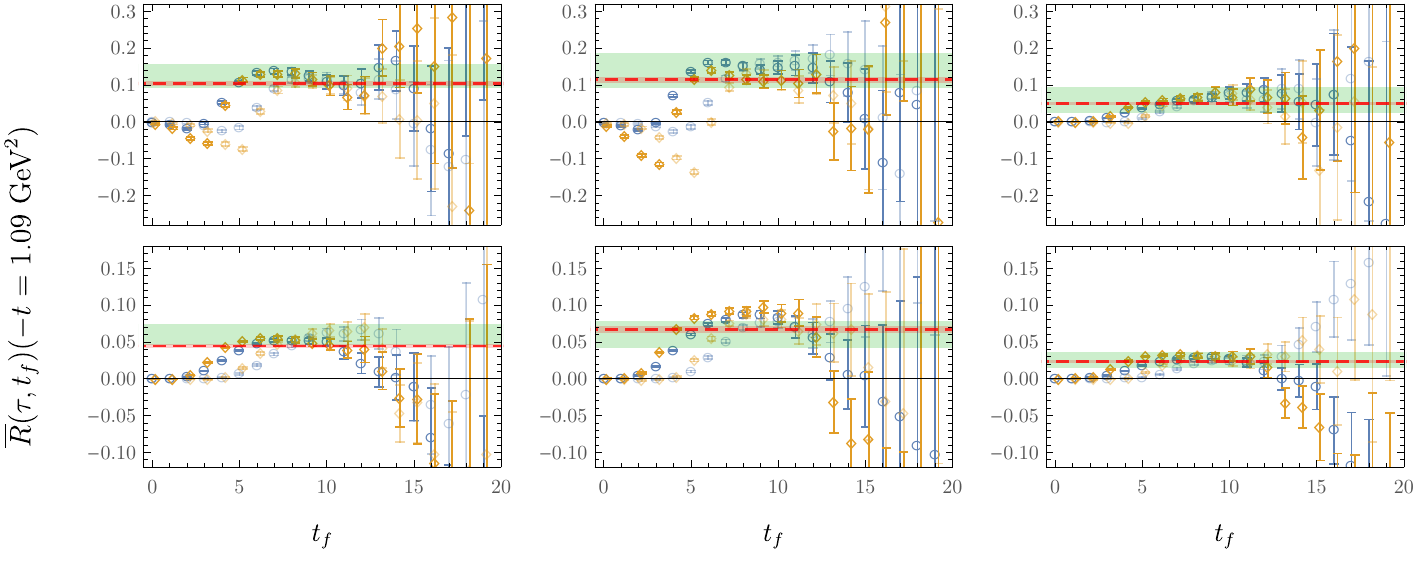}}
	\caption{As in Fig.~\ref{fig:pbump4}, for a different value of the squared momentum transfer $t$.}
	\label{fig:pbump14}
\end{figure}

\begin{figure}
	\centering
	\subfigure[$t_f=12$ (dark points), $t_f$=14 (pale points)]{
	\includegraphics[width=\linewidth]{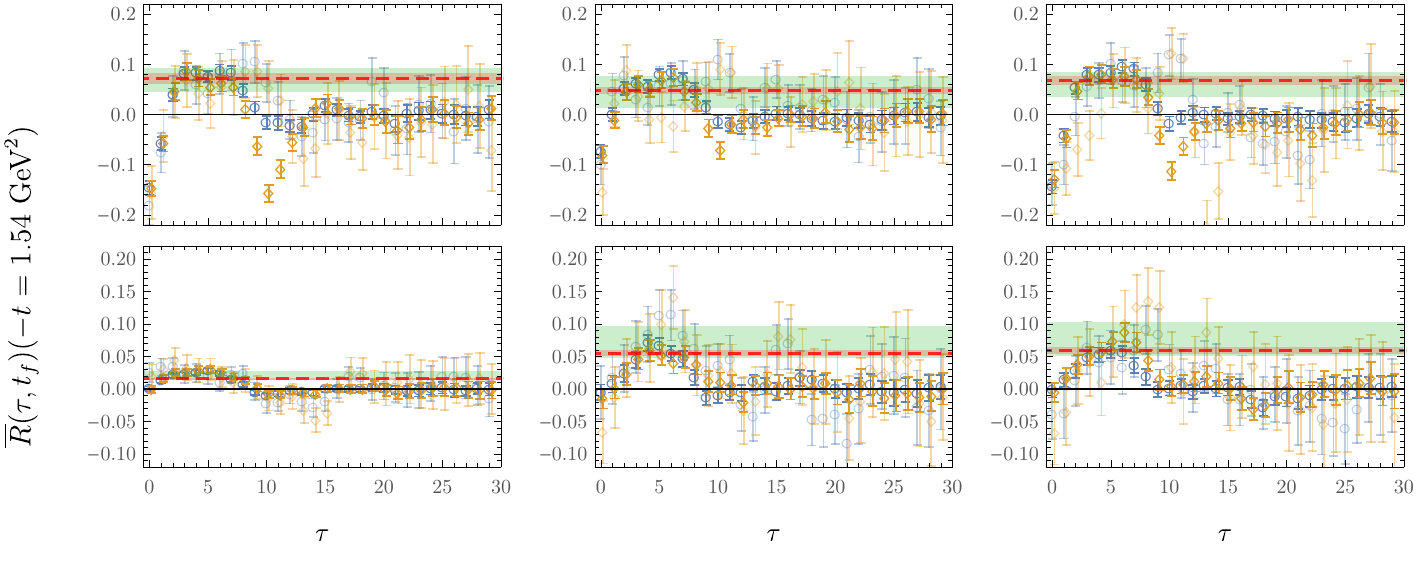}}
	\subfigure[$\tau=4$ (dark points), $\tau$=6 (pale points)]{
	\includegraphics[width=\linewidth]{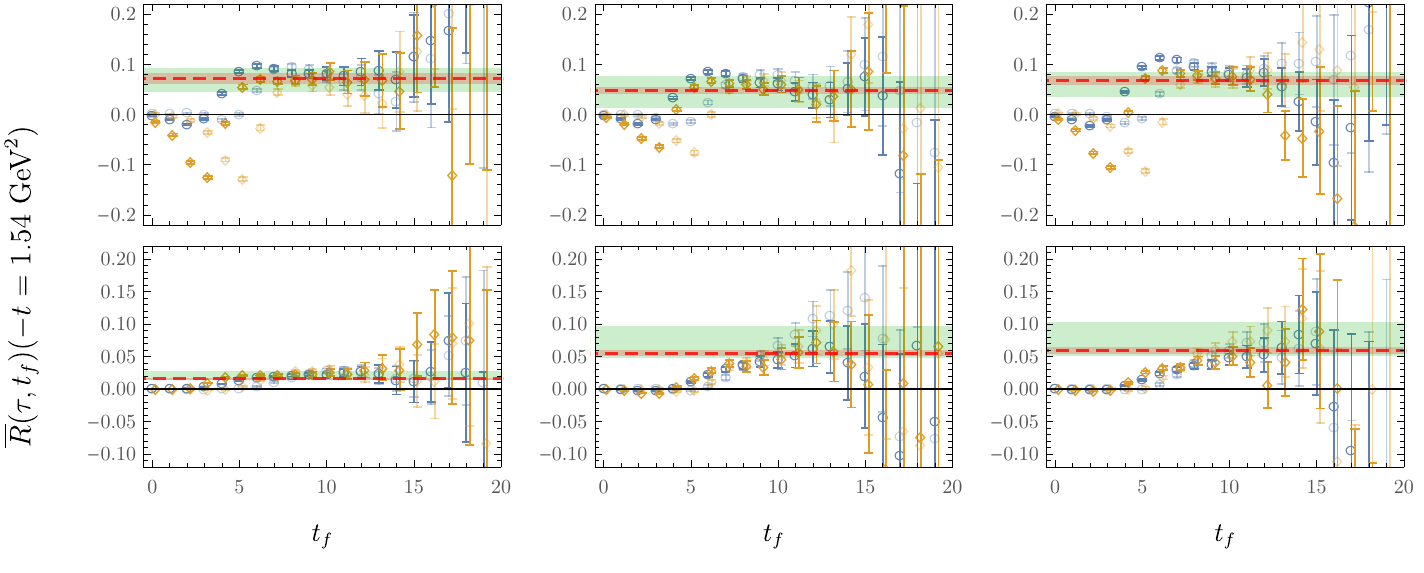}}
	\caption{As in Fig.~\ref{fig:pbump4}, for a different value of the squared momentum transfer $t$.}
	\label{fig:pbump18}
\end{figure}

\begin{figure}
	\centering
	\subfigure[$t_f=13$ (dark points), $t_f$=18 (pale points)]{
	\includegraphics[width=\linewidth]{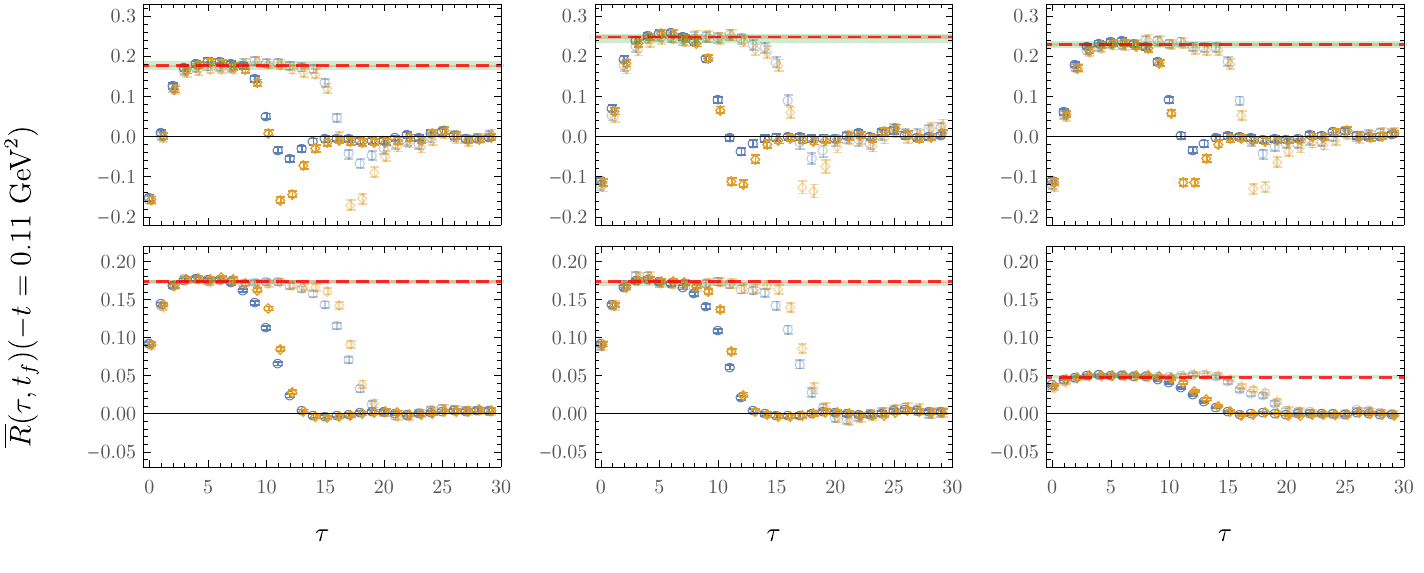}}
	\subfigure[$\tau=5$ (dark points), $\tau$=7 (pale points)]{
	\includegraphics[width=\linewidth]{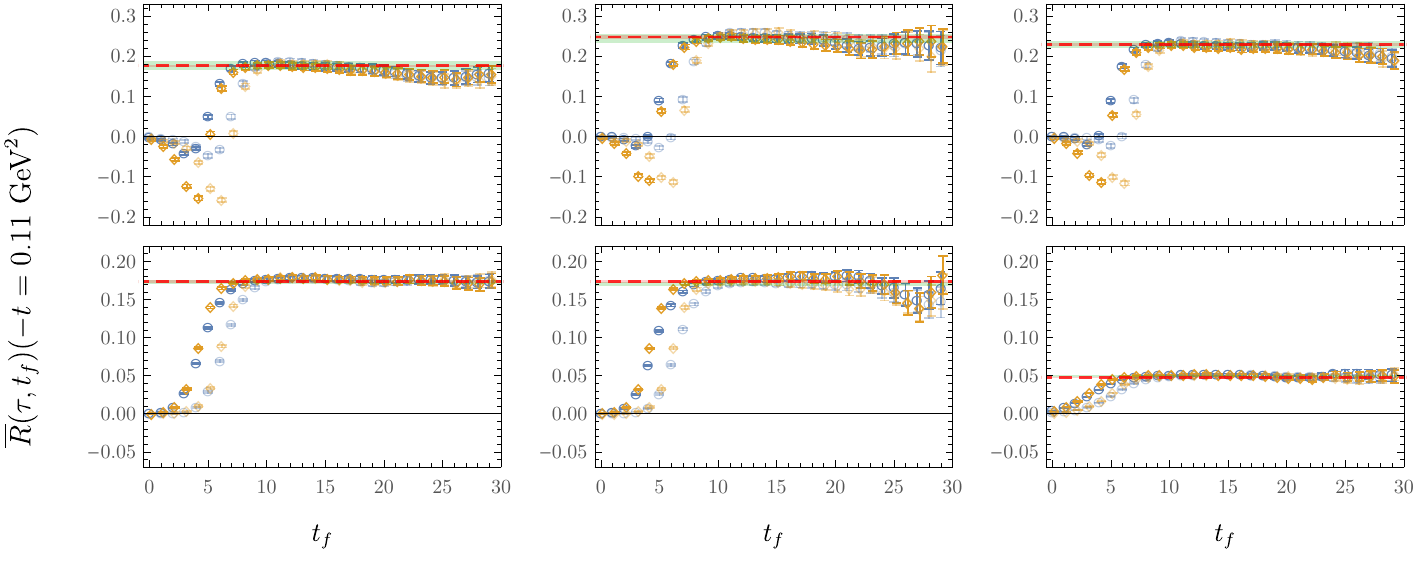}}
	\caption{As in Fig.~\ref{fig:pbump4}, for averaged ratios of three and two point functions for the pion, $\overline{R}^{(\pi)}_{\mathfrak{R};k}(t_f,\tau)$, formed from SP (blue circles) and SS (orange diamonds) correlation functions, as well as plateau fits to their $t_f$ and $\tau$ dependence (green bands). The red dashed bands show the central value and uncertainty of the final fitted result for the GFFs projected back to the linear combination of GFFs corresponding to each ratio. The upper (lower) row of figures in each panel  shows examples of ratios determined using operators in representation $\mathfrak{R}=\tau_1^{(3)}$ ($\tau_3^{(6)}$). }
	\label{fig:pibump4}
\end{figure}

\begin{figure}
	\centering
	\subfigure[$t_f=13$ (dark points), $t_f$=18 (pale points)]{
	\includegraphics[width=\linewidth]{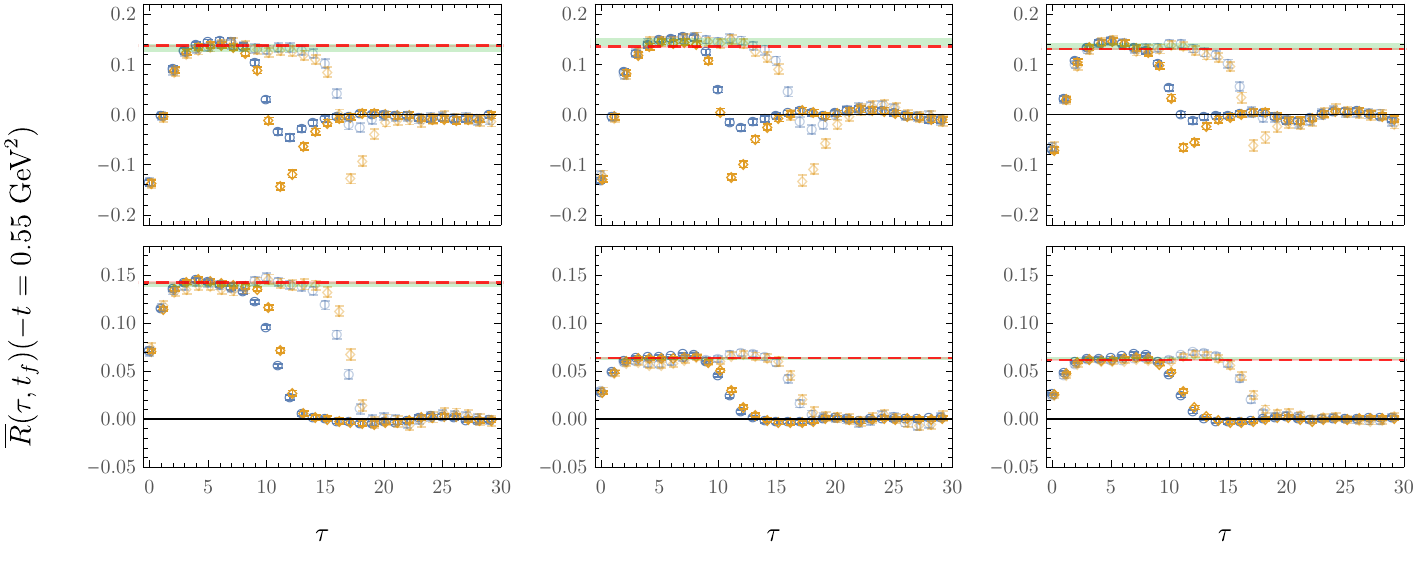}}
	\subfigure[$\tau=5$ (dark points), $\tau$=7 (pale points)]{
	\includegraphics[width=\linewidth]{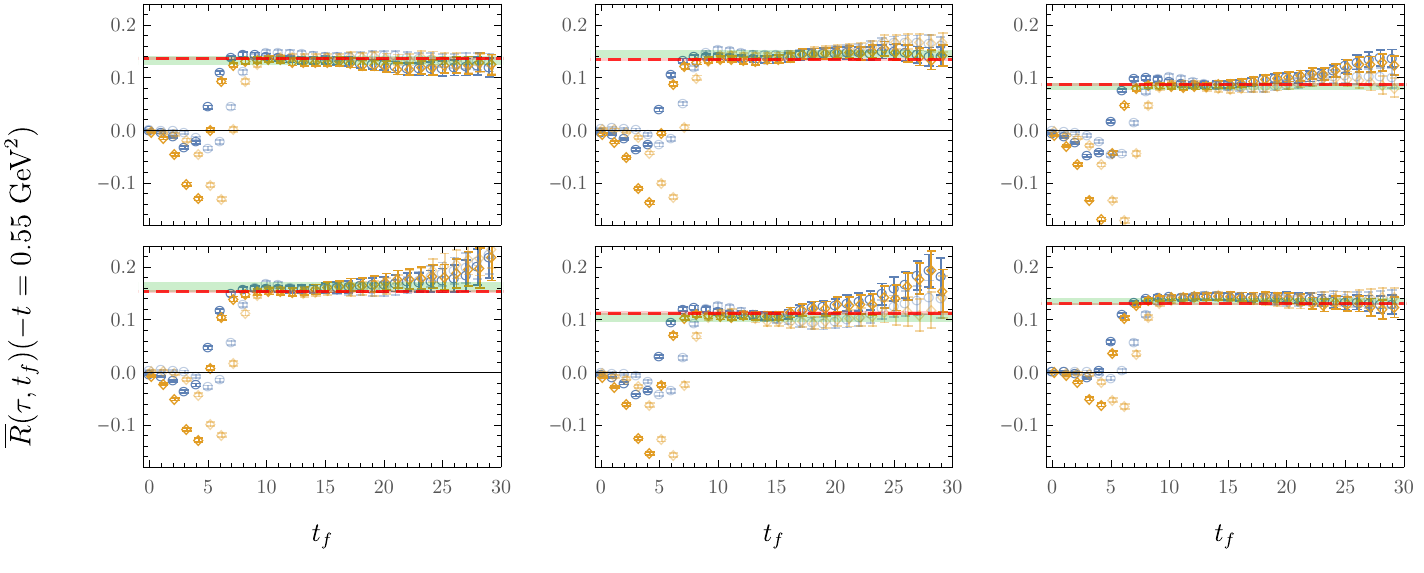}}
	\caption{As in Fig.~\ref{fig:pibump4}, for a different value of the squared momentum transfer $t$.}
	\label{fig:pibump9}
\end{figure}

\begin{figure}
	\centering
	\subfigure[$t_f=13$ (dark points), $t_f$=18 (pale points)]{
	\includegraphics[width=\linewidth]{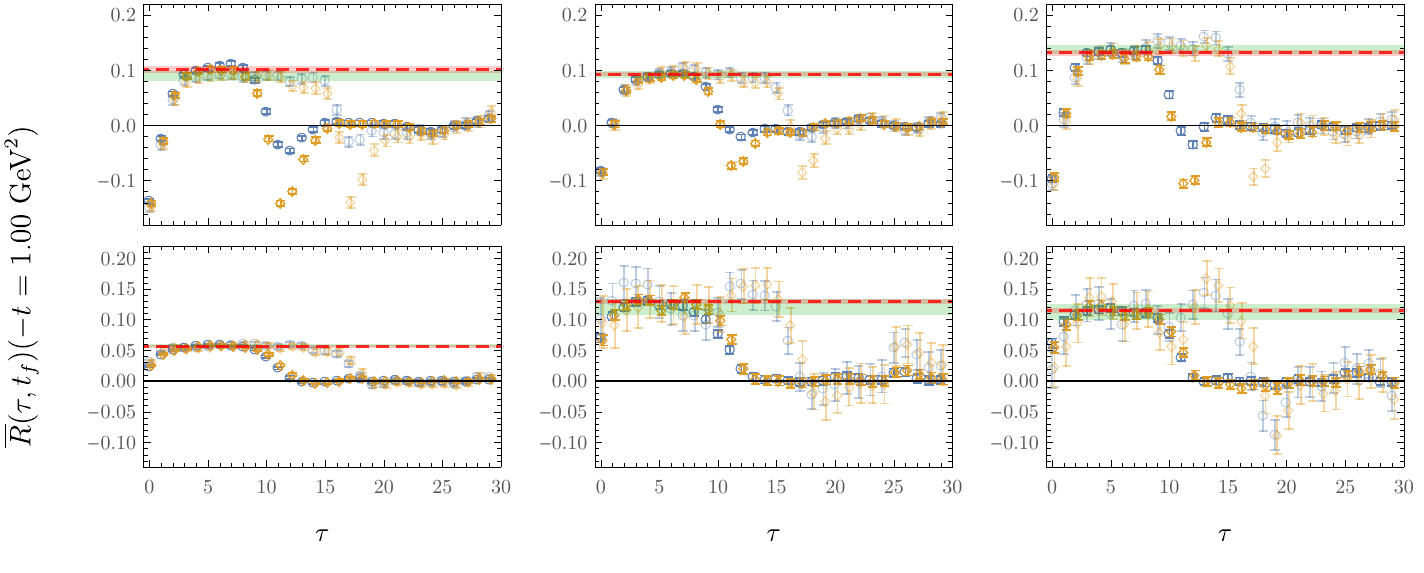}}
	\subfigure[$\tau=5$ (dark points), $\tau$=7 (pale points)]{
	\includegraphics[width=\linewidth]{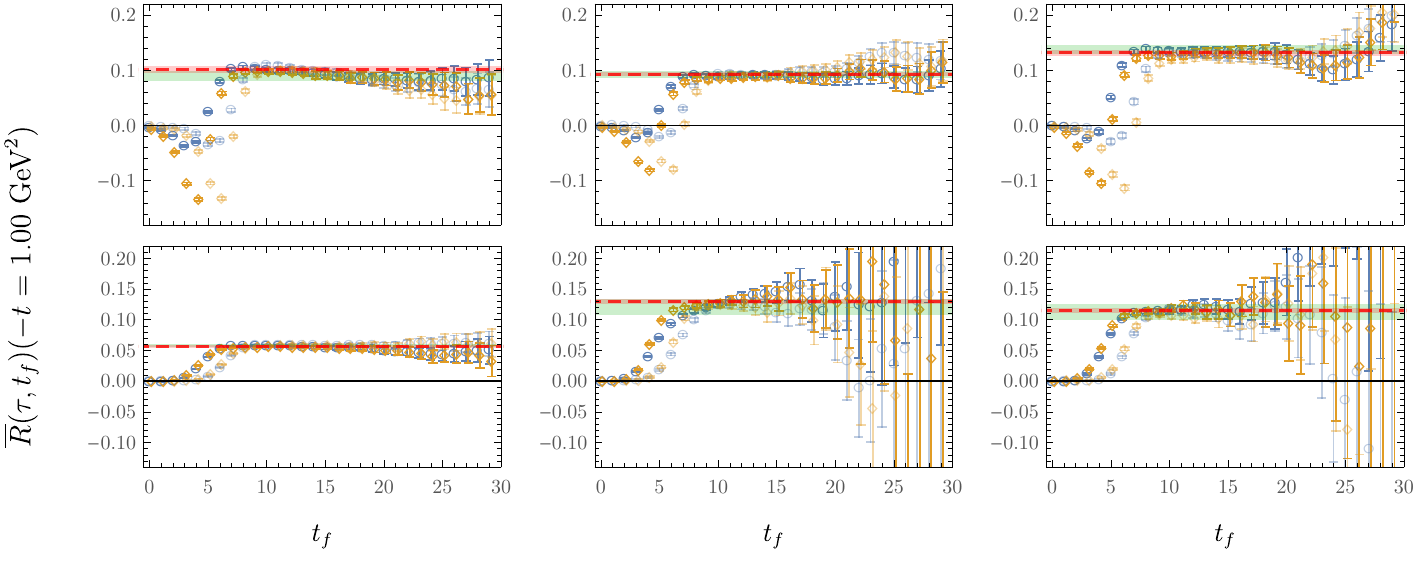}}
	\caption{As in Fig.~\ref{fig:pibump4}, for a different value of the squared momentum transfer $t$.}
	\label{fig:pibump14}
\end{figure}

\begin{figure}
	\centering
	\subfigure[$t_f=13$ (dark points), $t_f$=18 (pale points)]{
	\includegraphics[width=\linewidth]{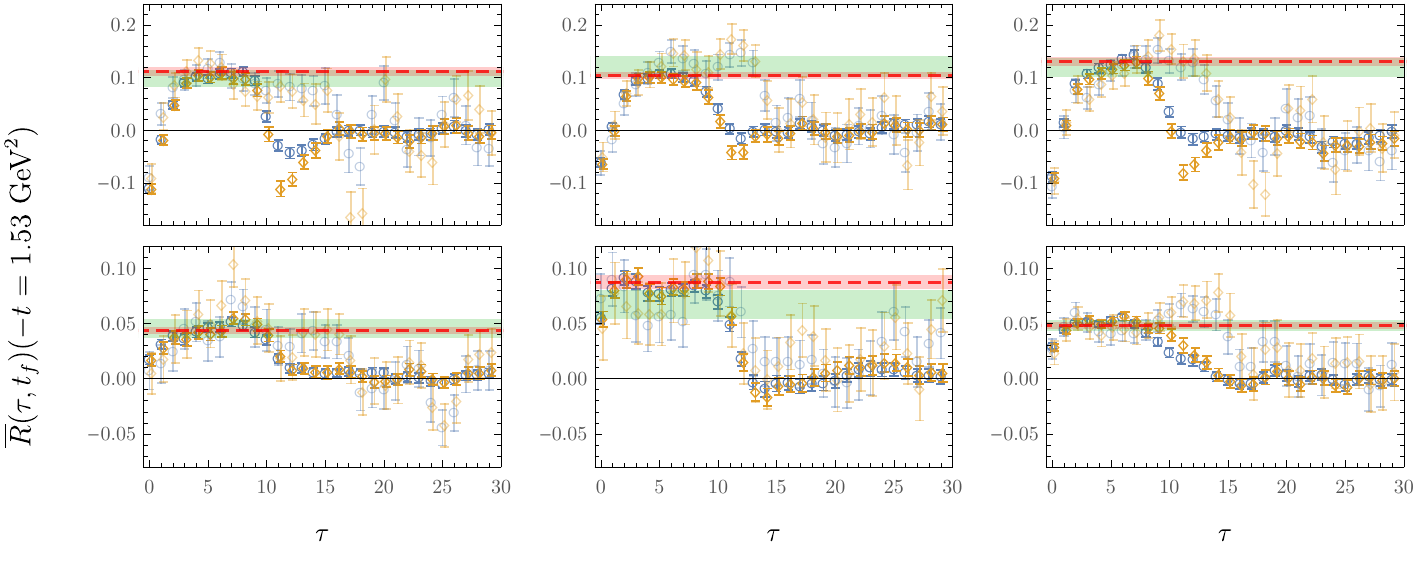}}
	\subfigure[$\tau=5$ (dark points), $\tau$=7 (pale points)]{
	\includegraphics[width=\linewidth]{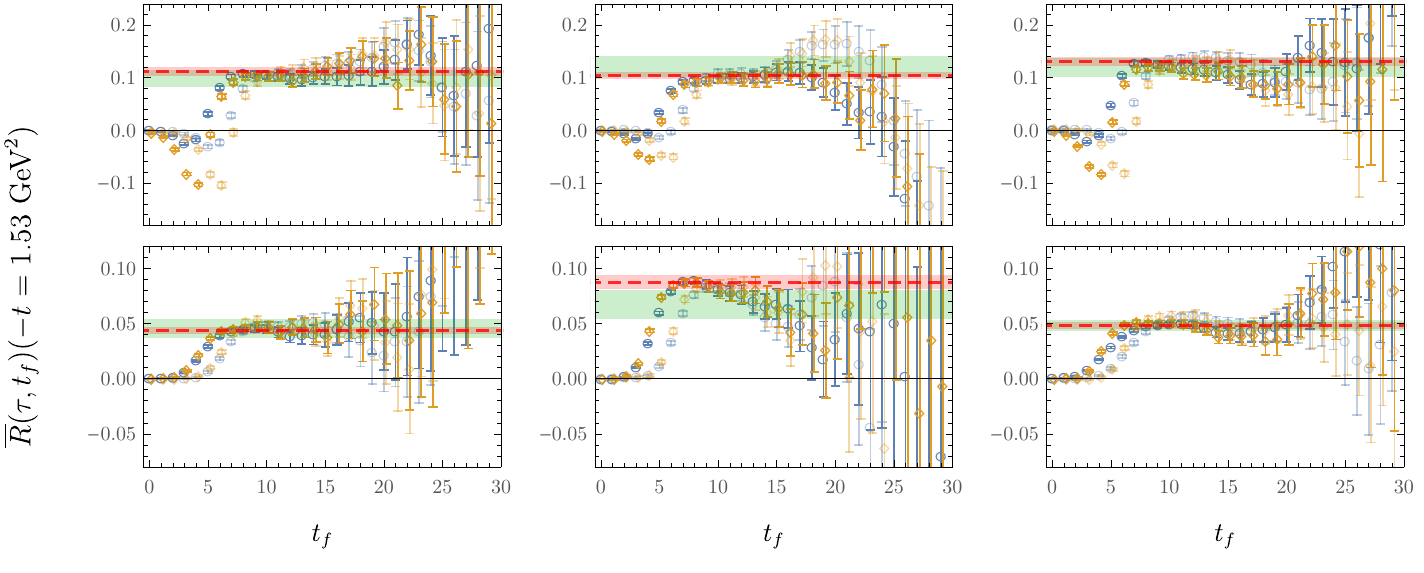}}
	\caption{As in Fig.~\ref{fig:pibump4}, for a different value of the squared momentum transfer $t$.}
	\label{fig:pibump18}
\end{figure}

\begin{figure}
	\centering
	\includegraphics[width=0.31\linewidth]{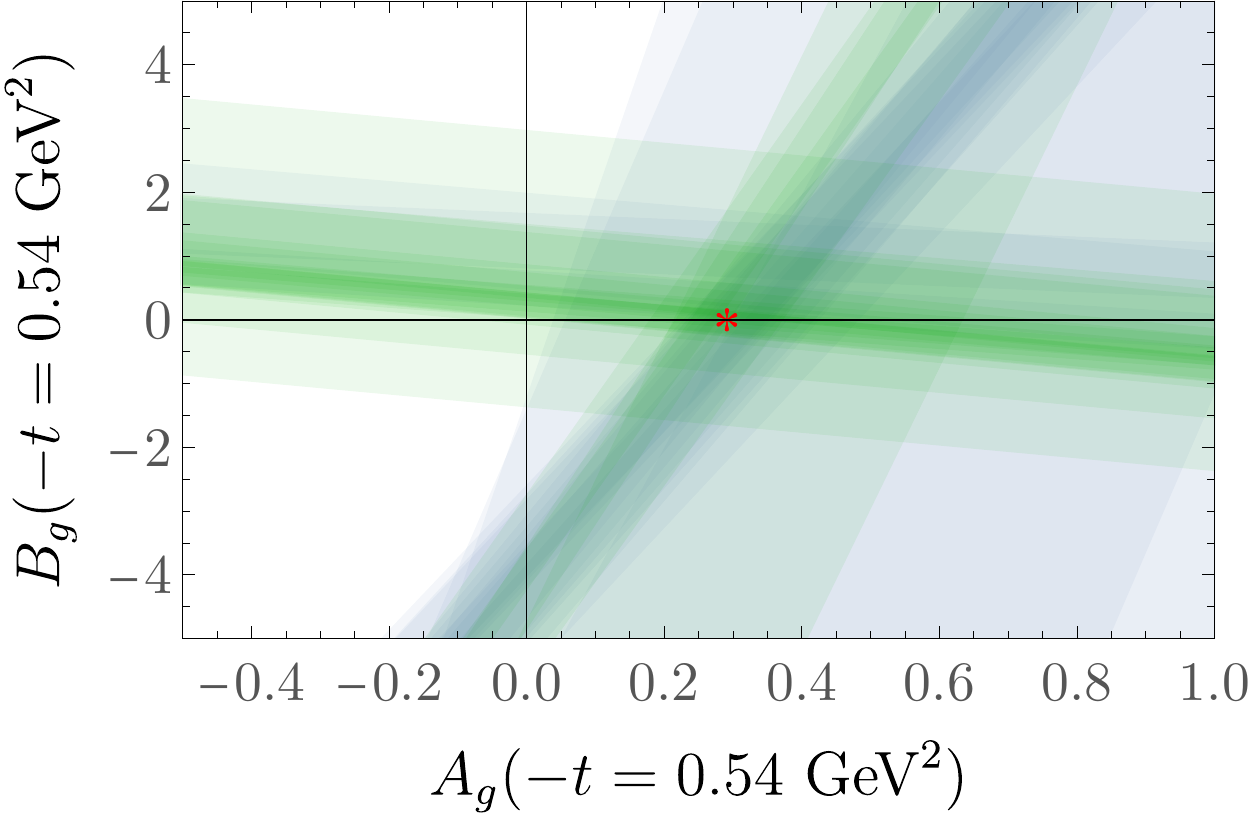}\, \includegraphics[width=0.31\linewidth]{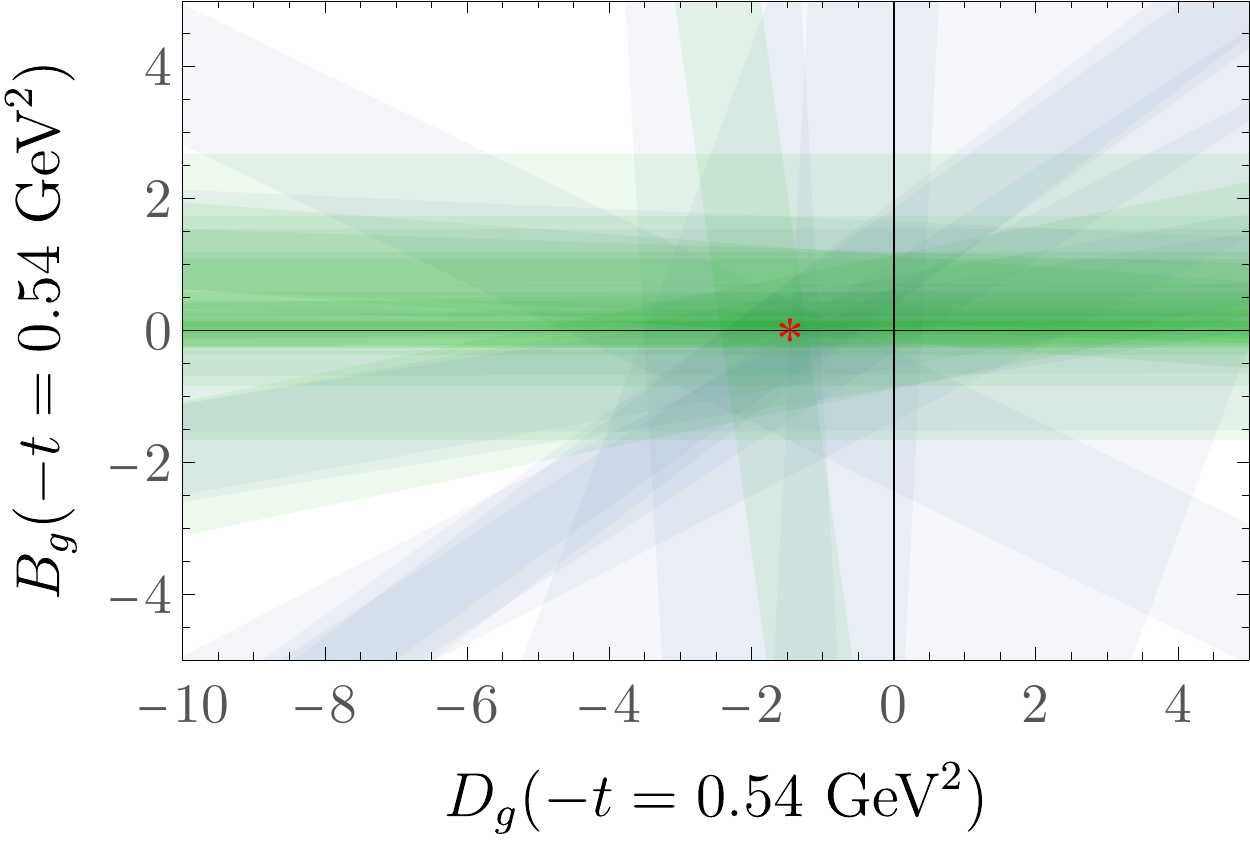}\,	\includegraphics[width=0.31\linewidth]{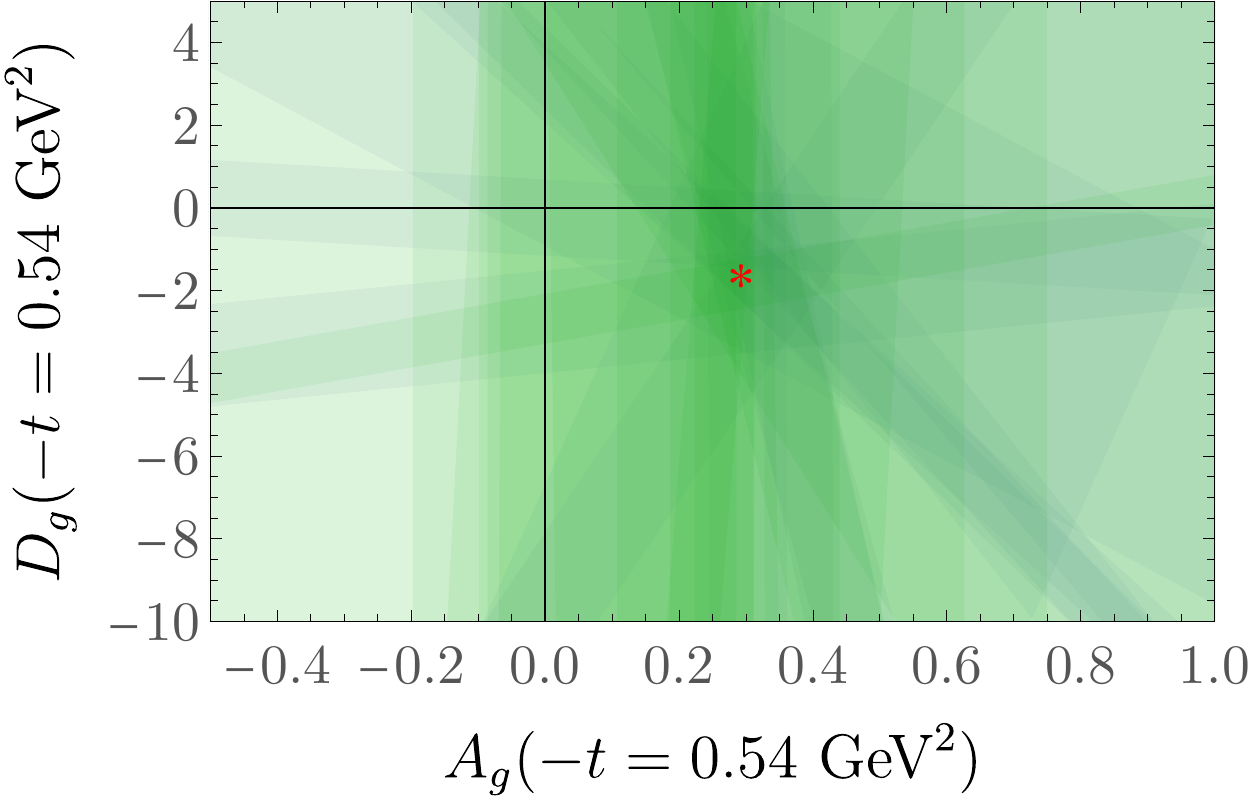} \\ \vspace*{3mm}
	\includegraphics[width=0.31\linewidth]{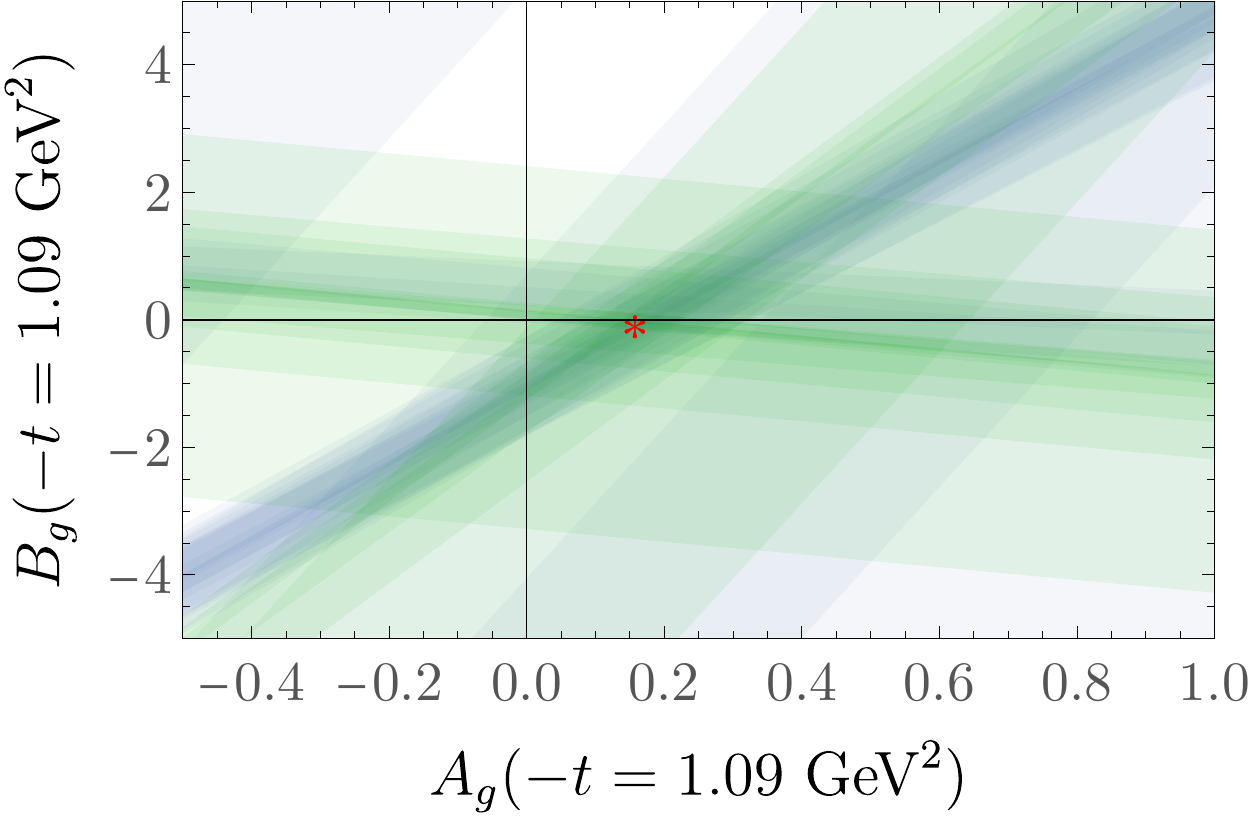}\, \includegraphics[width=0.31\linewidth]{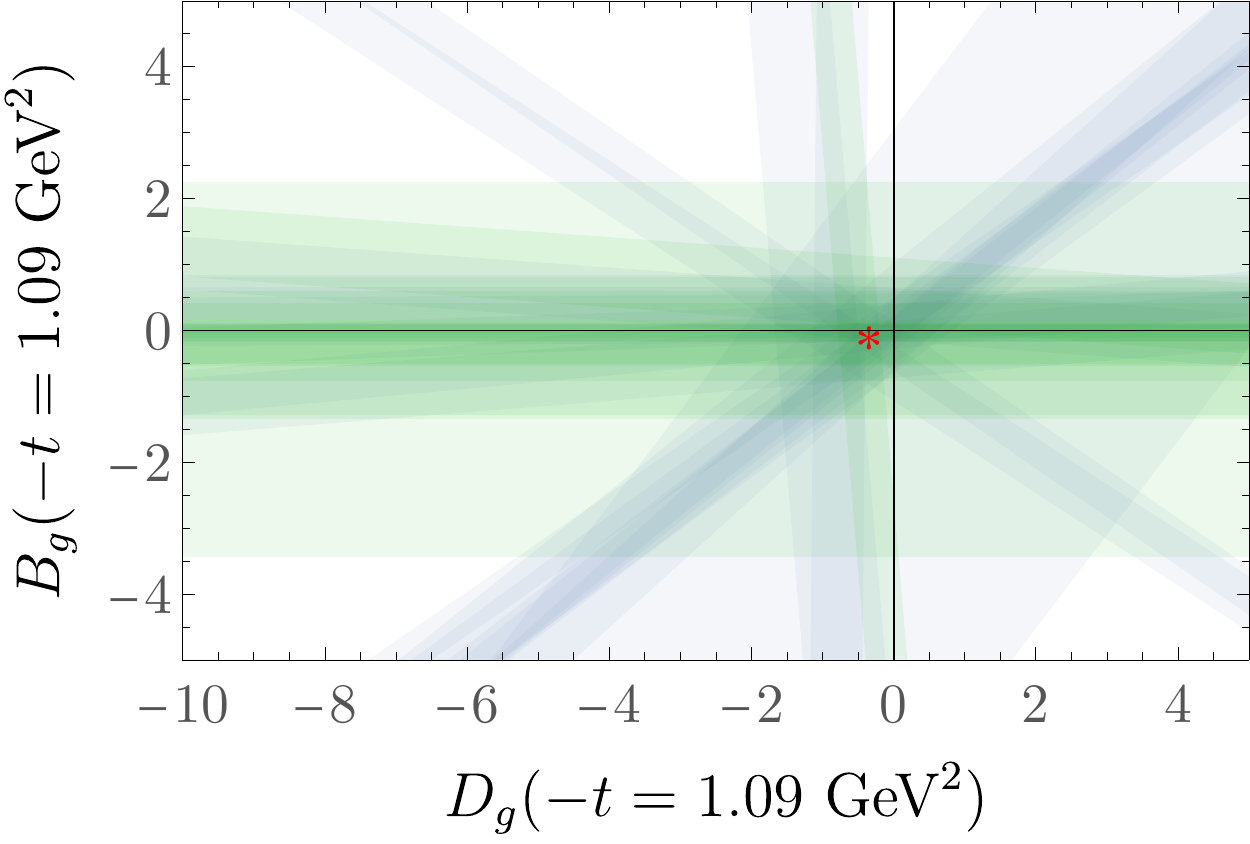}\,	\includegraphics[width=0.31\linewidth]{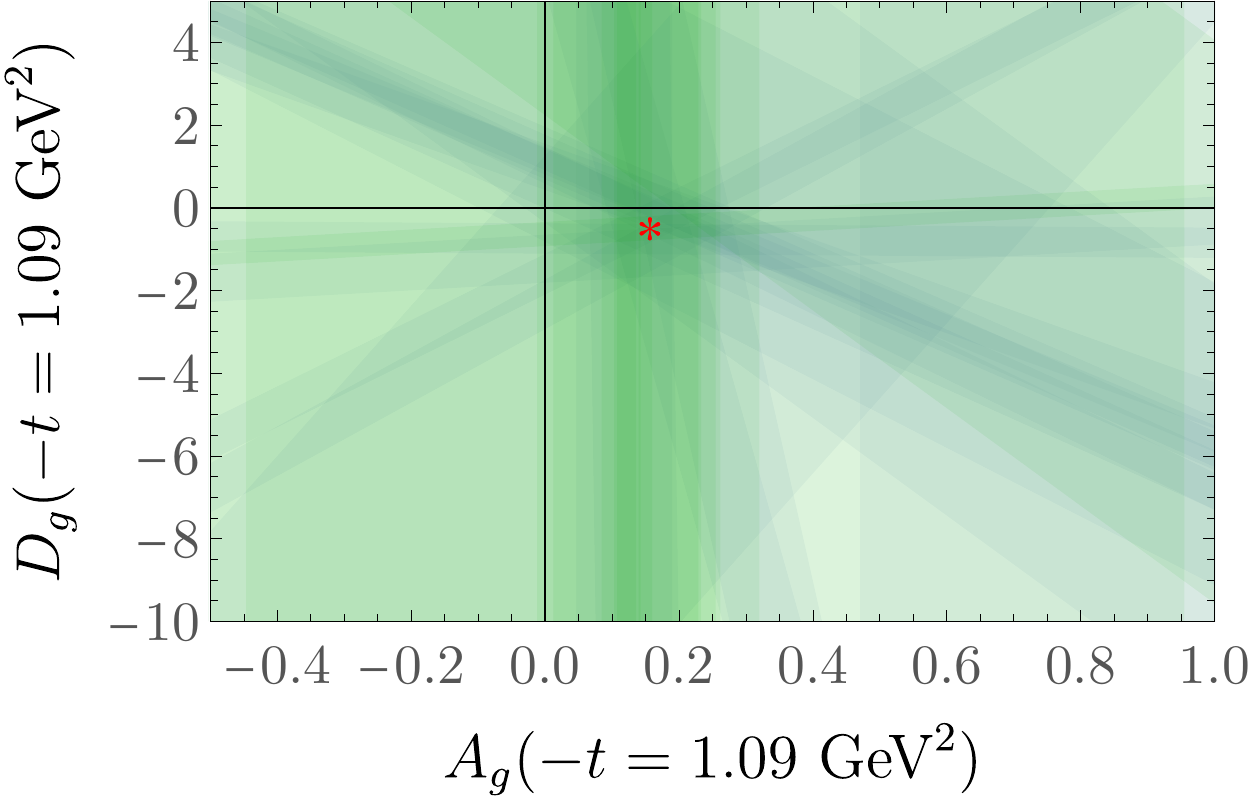} \\ \vspace*{3mm}
	\includegraphics[width=0.31\linewidth]{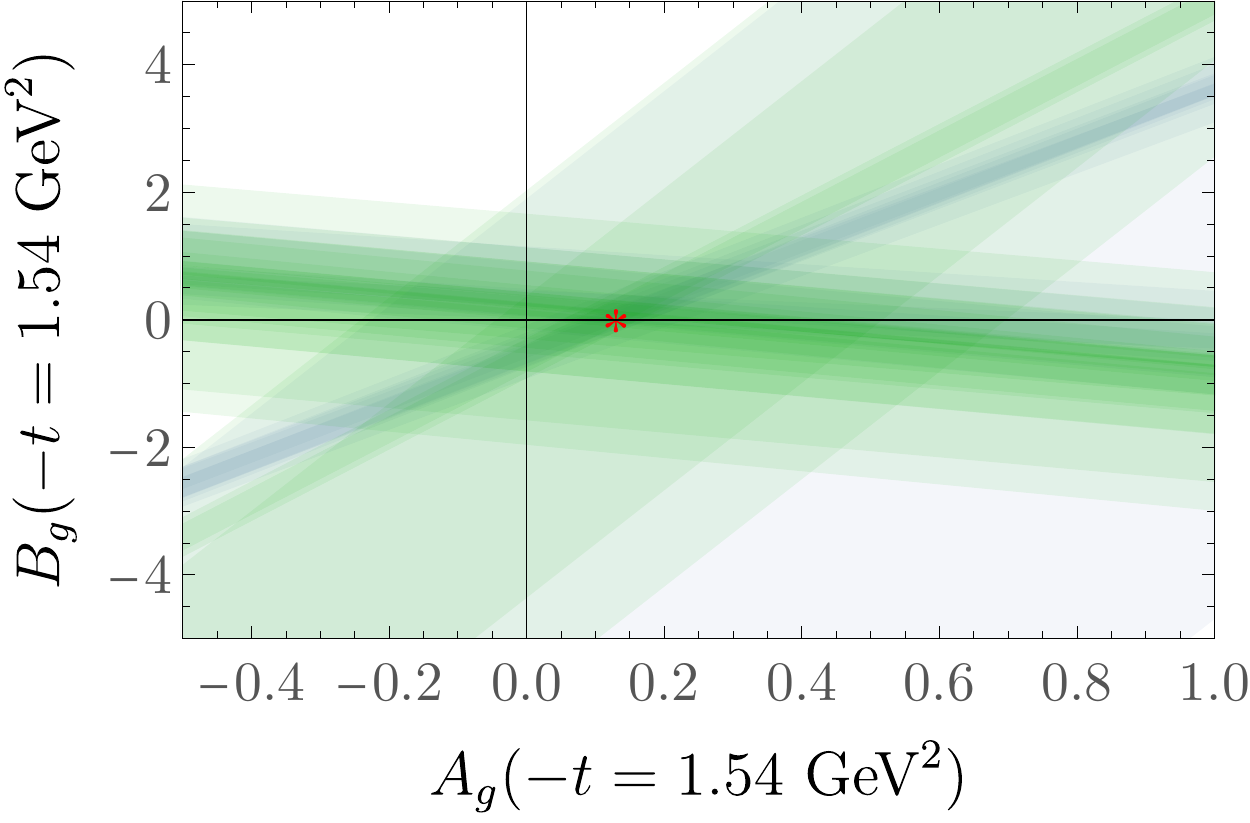}\, \includegraphics[width=0.31\linewidth]{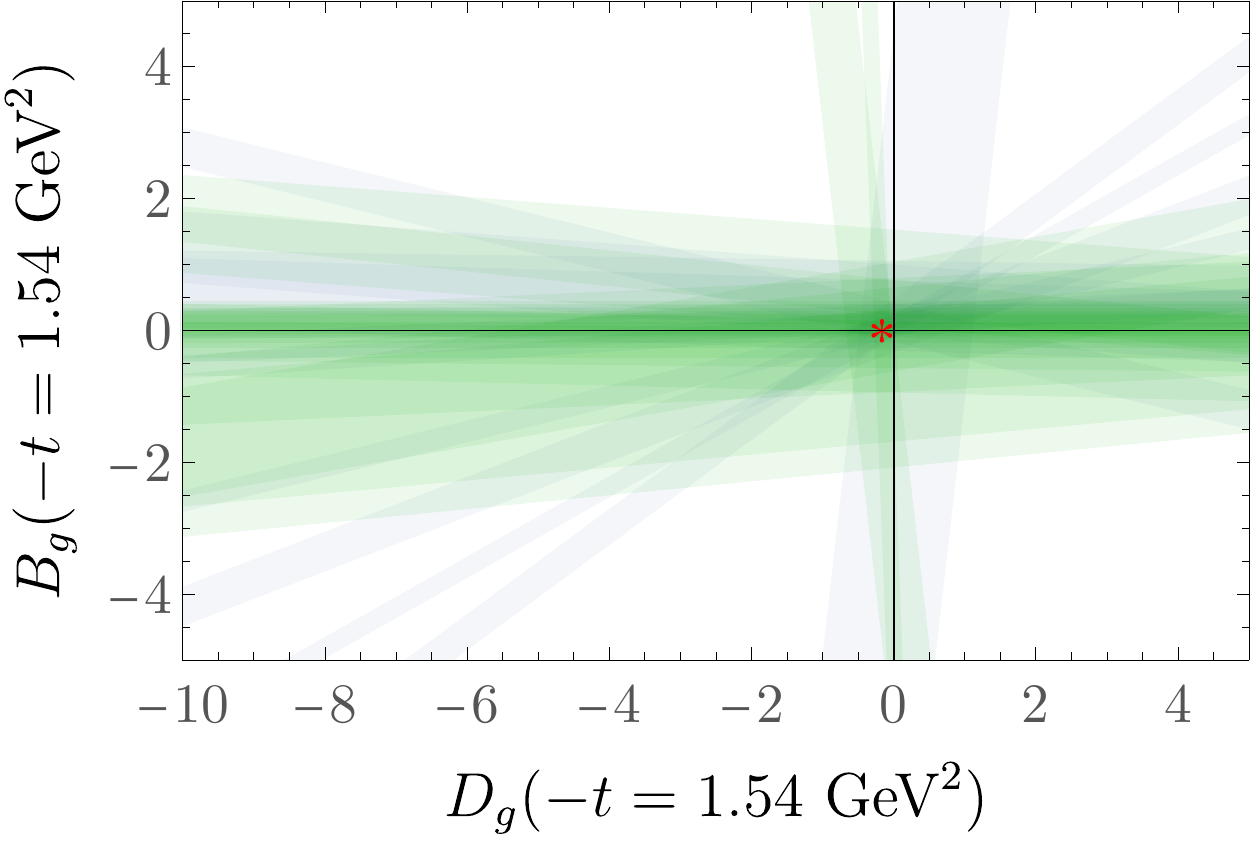}\,	\includegraphics[width=0.31\linewidth]{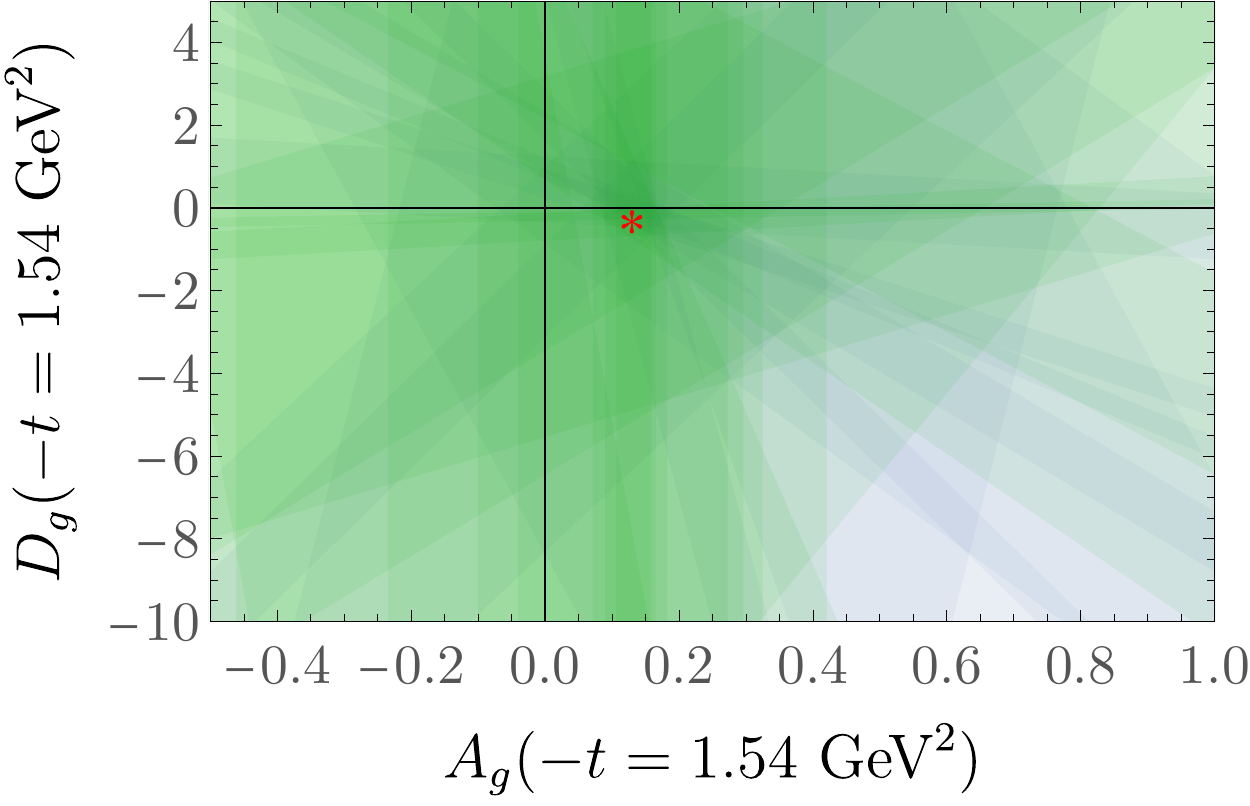} \\ \vspace*{3mm}
	\caption{Constraints on the  renormalised gluon GFFs of the nucleon at various values of the squared momentum transfer $t$. The three columns show the projections onto the $A_g(t)$--$B_g(t)$, $D_g(t)$--$B_g(t)$ and $A_g(t)$--$D_g(t)$-planes, with the GFF not shown in each projection taken to its central value. On each figure, every shaded band shows the 1-standard-deviation uncertainty arising from the plateau fit to a single averaged ratio, described in Sec.~\ref{sec:latt}. Blue and green colours denote constraints from operators in the $\tau_1^{(3)}$ and $\tau_3^{(6)}$ representations respectively.  For clarity, only the 30 most important constraints (as defined by their contribution to the fit $\chi^2$) are shown, although all constraints are used in the analysis. Uncertainties associated with the renormalisation constants are not shown. The stars correspond to the central value of the fits to the form factors at each $t$.}
	\label{fig:band1}
\end{figure}

\begin{figure}
	\centering
	\includegraphics[width=0.42\linewidth]{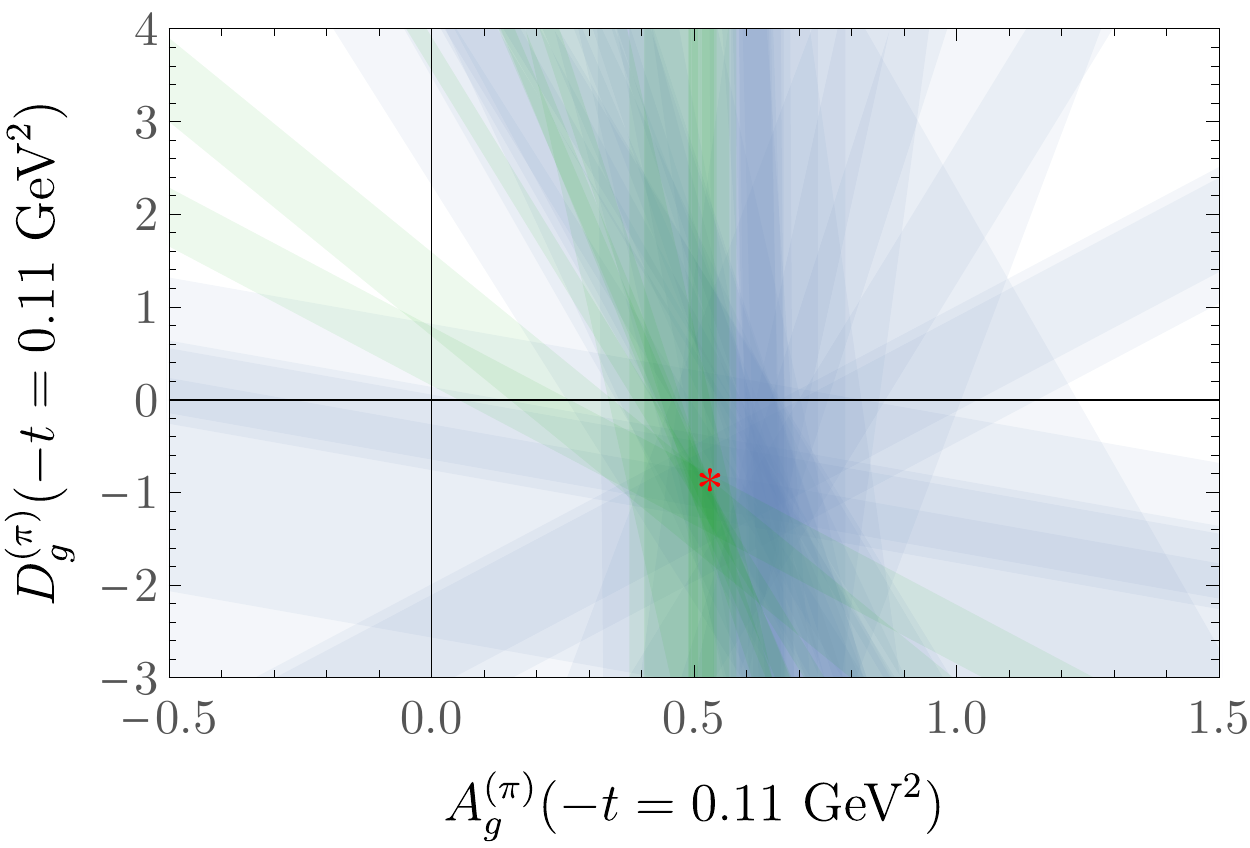}\qquad	\includegraphics[width=0.42\linewidth]{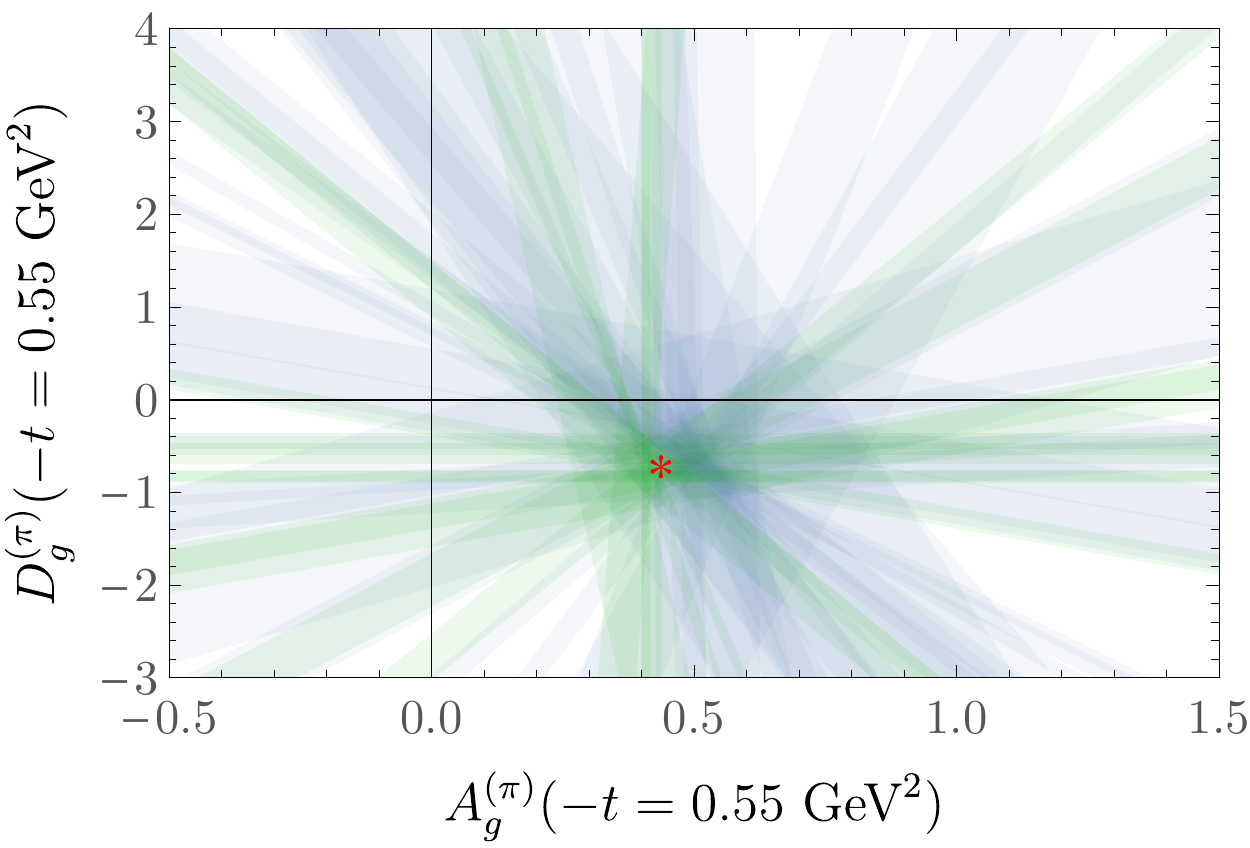} \\ \vspace*{8mm}
	\includegraphics[width=0.42\linewidth]{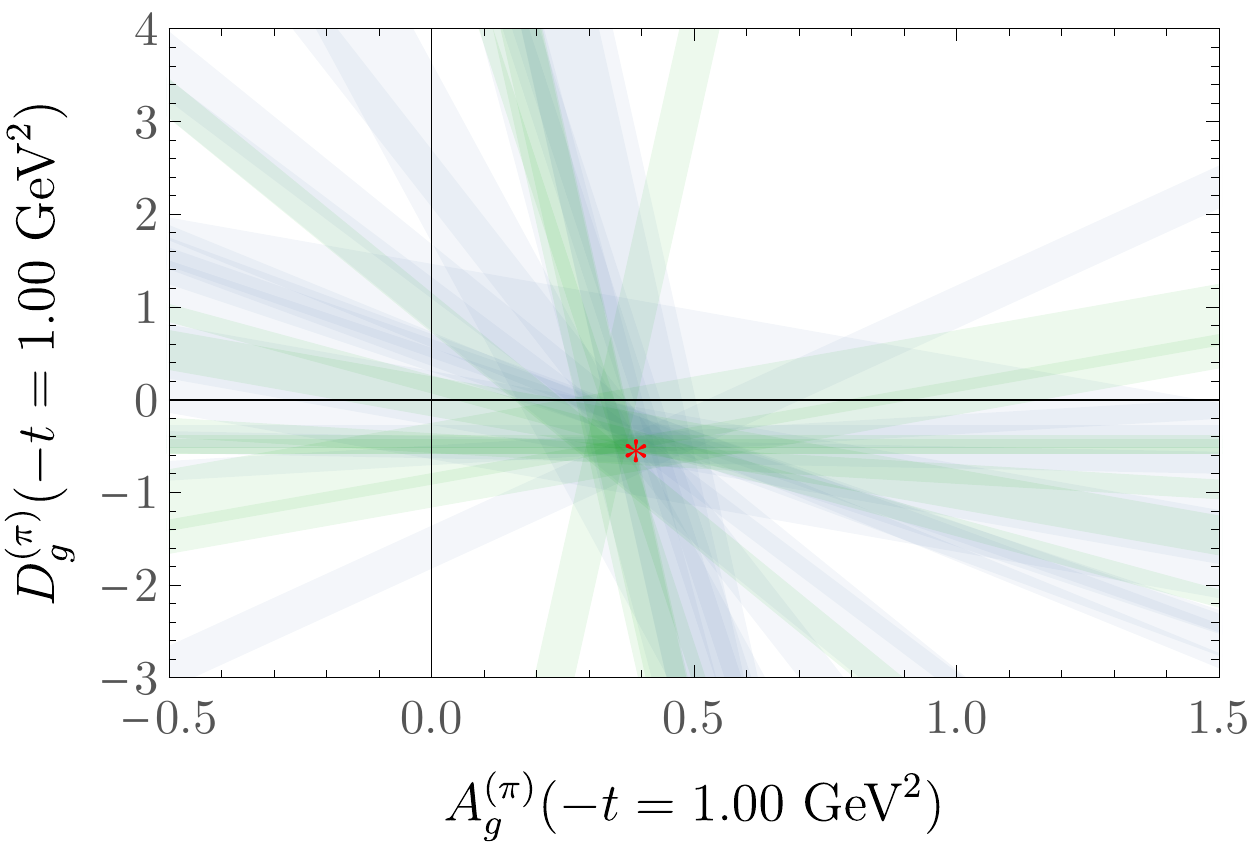}\qquad	\includegraphics[width=0.42\linewidth]{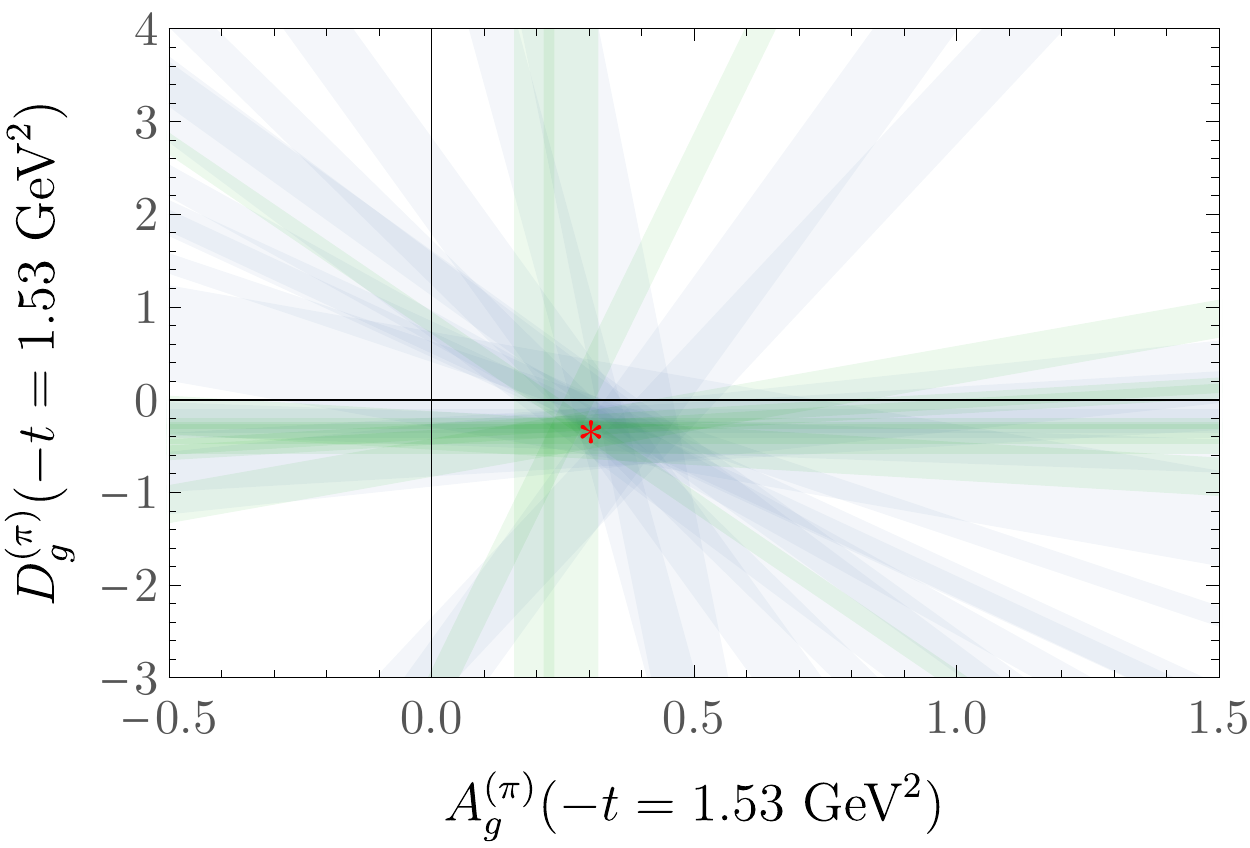} \\ \vspace*{8mm}
	\caption{Constraints on the  renormalised gluon GFFs of the pion at various values of the squared momentum transfer $t$, as in Fig.~\ref{fig:band1} for the nucleon. All constraints are shown, however the uncertainties associated with the renormalisation constants are not displayed. The red stars show the central values of the fits to the form factors at each $t$. }
	\label{fig:piband1}
\end{figure}

This Appendix displays examples of raw LQCD data for the averaged ratios of three- and two-point functions $\overline{R}^{(\pi/N)}_{\mathfrak{R};k}(t_f,\tau)$ (defined in Section~\ref{sec:latt}), and illustrates the results of plateau fits to the $t_f$ and $\tau$ dependence of these ratios.
Figures~\ref{fig:pbump4}--\ref{fig:pbump18} show data for the nucleon at a selection of values of the squared momentum transfer $t$. In each case, ratios are plotted vs $\tau$ at two different sink times for both SP and SS correlation functions, and vs sink time $t_f$ for two choices of operator insertion time. Also shown on each figure are both the fit band resulting from a simultaneous fit to the $t_f$ and $\tau$ dependence of the ratios formed with both SS and SP three-point functions within the plateau region (discussed in Section~\ref{sec:latt}), and the central value and uncertainty of the final fitted result for the GFFs at that momentum, projected back to the linear combination of that particular ratio. 
Analogous figures for the pion are shown in Figs.~\ref{fig:pibump4}--\ref{fig:pibump18} (a later sink time is shown for the pion than for the nucleon since the signal in the statistically cleaner pion data continues to later times). Figures~\ref{fig:band1} and \ref{fig:piband1} show the constraints from the fits to each averaged ratio on the GFFs graphically at a selection of values of the squared momentum transfer $t$.



\bibliography{bob}

\begin{thebibliography}{67}
\expandafter\ifx\csname natexlab\endcsname\relax\def\natexlab#1{#1}\fi
\expandafter\ifx\csname bibnamefont\endcsname\relax
  \def\bibnamefont#1{#1}\fi
\expandafter\ifx\csname bibfnamefont\endcsname\relax
  \def\bibfnamefont#1{#1}\fi
\expandafter\ifx\csname citenamefont\endcsname\relax
  \def\citenamefont#1{#1}\fi
\expandafter\ifx\csname url\endcsname\relax
  \def\url#1{\texttt{#1}}\fi
\expandafter\ifx\csname urlprefix\endcsname\relax\def\urlprefix{URL }\fi
\providecommand{\bibinfo}[2]{#2}
\providecommand{\eprint}[2][]{\url{#2}}

\bibitem[{\citenamefont{M{\"u}ller et~al.}(1994)\citenamefont{M{\"u}ller,
  Robaschik, Geyer, Dittes, and Ho{\v r}ej{\v s}i}}]{Mueller:1998fv}
\bibinfo{author}{\bibfnamefont{D.}~\bibnamefont{M{\"u}ller}},
  \bibinfo{author}{\bibfnamefont{D.}~\bibnamefont{Robaschik}},
  \bibinfo{author}{\bibfnamefont{B.}~\bibnamefont{Geyer}},
  \bibinfo{author}{\bibfnamefont{F.~M.} \bibnamefont{Dittes}},
  \bibnamefont{and} \bibinfo{author}{\bibfnamefont{J.}~\bibnamefont{Ho{\v
  r}ej{\v s}i}}, \bibinfo{journal}{Fortsch. Phys.}
  \textbf{\bibinfo{volume}{42}}, \bibinfo{pages}{101} (\bibinfo{year}{1994}),
  \eprint{hep-ph/9812448}.

\bibitem[{\citenamefont{Ji}(1997)}]{Ji:1996nm}
\bibinfo{author}{\bibfnamefont{X.-D.} \bibnamefont{Ji}},
  \bibinfo{journal}{Phys. Rev.} \textbf{\bibinfo{volume}{D55}},
  \bibinfo{pages}{7114} (\bibinfo{year}{1997}), \eprint{hep-ph/9609381}.

\bibitem[{\citenamefont{Radyushkin}(1997)}]{Radyushkin:1997ki}
\bibinfo{author}{\bibfnamefont{A.~V.} \bibnamefont{Radyushkin}},
  \bibinfo{journal}{Phys. Rev.} \textbf{\bibinfo{volume}{D56}},
  \bibinfo{pages}{5524} (\bibinfo{year}{1997}), \eprint{hep-ph/9704207}.

\bibitem[{\citenamefont{Ji}(1998)}]{Ji:1998pc}
\bibinfo{author}{\bibfnamefont{X.-D.} \bibnamefont{Ji}}, \bibinfo{journal}{J.
  Phys.} \textbf{\bibinfo{volume}{G24}}, \bibinfo{pages}{1181}
  (\bibinfo{year}{1998}), \eprint{hep-ph/9807358}.

\bibitem[{\citenamefont{Diehl}(2003)}]{Diehl:2003ny}
\bibinfo{author}{\bibfnamefont{M.}~\bibnamefont{Diehl}},
  \bibinfo{journal}{Phys. Rept.} \textbf{\bibinfo{volume}{388}},
  \bibinfo{pages}{41} (\bibinfo{year}{2003}), \eprint{hep-ph/0307382}.

\bibitem[{\citenamefont{Belitsky and Radyushkin}(2005)}]{Belitsky:2005qn}
\bibinfo{author}{\bibfnamefont{A.~V.} \bibnamefont{Belitsky}} \bibnamefont{and}
  \bibinfo{author}{\bibfnamefont{A.~V.} \bibnamefont{Radyushkin}},
  \bibinfo{journal}{Phys. Rept.} \textbf{\bibinfo{volume}{418}},
  \bibinfo{pages}{1} (\bibinfo{year}{2005}), \eprint{hep-ph/0504030}.

\bibitem[{\citenamefont{Polyakov and Shuvaev}(2002)}]{Polyakov:2002wz}
\bibinfo{author}{\bibfnamefont{M.~V.} \bibnamefont{Polyakov}} \bibnamefont{and}
  \bibinfo{author}{\bibfnamefont{A.~G.} \bibnamefont{Shuvaev}}
  (\bibinfo{year}{2002}), \eprint{hep-ph/0207153}.

\bibitem[{\citenamefont{Polyakov}(2003)}]{Polyakov:2002yz}
\bibinfo{author}{\bibfnamefont{M.~V.} \bibnamefont{Polyakov}},
  \bibinfo{journal}{Phys. Lett.} \textbf{\bibinfo{volume}{B555}},
  \bibinfo{pages}{57} (\bibinfo{year}{2003}), \eprint{hep-ph/0210165}.

\bibitem[{\citenamefont{Polyakov and Schweitzer}(2018)}]{Polyakov:2018zvc}
\bibinfo{author}{\bibfnamefont{M.~V.} \bibnamefont{Polyakov}} \bibnamefont{and}
  \bibinfo{author}{\bibfnamefont{P.}~\bibnamefont{Schweitzer}}
  (\bibinfo{year}{2018}), \eprint{1805.06596}.

\bibitem[{\citenamefont{Niccolai}(2016)}]{Niccolai:2016sdf}
\bibinfo{author}{\bibfnamefont{S.}~\bibnamefont{Niccolai}}
  (\bibinfo{collaboration}{CLAS}), \bibinfo{journal}{PoS}
  \textbf{\bibinfo{volume}{DIS2016}}, \bibinfo{pages}{233}
  (\bibinfo{year}{2016}).

\bibitem[{\citenamefont{Capua}(2012)}]{Capua:2012sd}
\bibinfo{author}{\bibfnamefont{M.}~\bibnamefont{Capua}}
  (\bibinfo{collaboration}{ZEUS, H1}), in
  \emph{\bibinfo{booktitle}{{Proceedings, 3rd International Workshop on
  Multiple Partonic Interactions at the LHC (MPI@LHC 2011): Hamburg, Germany,
  21-25 Nov 2011}}} (\bibinfo{year}{2012}), pp. \bibinfo{pages}{137--143},
  \eprint{1202.2828},
  \urlprefix\url{http://inspirehep.net/record/1088834/files/arXiv:1202.2828.pdf}.

\bibitem[{\citenamefont{Sandacz}(2017)}]{Sandacz:2018zsy}
\bibinfo{author}{\bibfnamefont{A.}~\bibnamefont{Sandacz}}
  (\bibinfo{collaboration}{COMPASS}), \bibinfo{journal}{J. Phys. Conf. Ser.}
  \textbf{\bibinfo{volume}{938}}, \bibinfo{pages}{012015}
  (\bibinfo{year}{2017}).

\bibitem[{\citenamefont{Favart et~al.}(2016)\citenamefont{Favart, Guidal, Horn,
  and Kroll}}]{Favart:2015umi}
\bibinfo{author}{\bibfnamefont{L.}~\bibnamefont{Favart}},
  \bibinfo{author}{\bibfnamefont{M.}~\bibnamefont{Guidal}},
  \bibinfo{author}{\bibfnamefont{T.}~\bibnamefont{Horn}}, \bibnamefont{and}
  \bibinfo{author}{\bibfnamefont{P.}~\bibnamefont{Kroll}},
  \bibinfo{journal}{Eur. Phys. J.} \textbf{\bibinfo{volume}{A52}},
  \bibinfo{pages}{158} (\bibinfo{year}{2016}), \eprint{1511.04535}.

\bibitem[{\citenamefont{d'Hose et~al.}(2016)\citenamefont{d'Hose, Niccolai, and
  Rostomyan}}]{dHose:2016mda}
\bibinfo{author}{\bibfnamefont{N.}~\bibnamefont{d'Hose}},
  \bibinfo{author}{\bibfnamefont{S.}~\bibnamefont{Niccolai}}, \bibnamefont{and}
  \bibinfo{author}{\bibfnamefont{A.}~\bibnamefont{Rostomyan}},
  \bibinfo{journal}{Eur. Phys. J.} \textbf{\bibinfo{volume}{A52}},
  \bibinfo{pages}{151} (\bibinfo{year}{2016}).

\bibitem[{\citenamefont{Kumeri{\v c}ki et~al.}(2016)\citenamefont{Kumeri{\v
  c}ki, Liuti, and Moutarde}}]{Kumericki:2016ehc}
\bibinfo{author}{\bibfnamefont{K.}~\bibnamefont{Kumeri{\v c}ki}},
  \bibinfo{author}{\bibfnamefont{S.}~\bibnamefont{Liuti}}, \bibnamefont{and}
  \bibinfo{author}{\bibfnamefont{H.}~\bibnamefont{Moutarde}},
  \bibinfo{journal}{Eur. Phys. J.} \textbf{\bibinfo{volume}{A52}},
  \bibinfo{pages}{157} (\bibinfo{year}{2016}), \eprint{1602.02763}.

\bibitem[{\citenamefont{Goeke et~al.}(2001)\citenamefont{Goeke, Polyakov, and
  Vanderhaeghen}}]{Goeke:2001tz}
\bibinfo{author}{\bibfnamefont{K.}~\bibnamefont{Goeke}},
  \bibinfo{author}{\bibfnamefont{M.~V.} \bibnamefont{Polyakov}},
  \bibnamefont{and}
  \bibinfo{author}{\bibfnamefont{M.}~\bibnamefont{Vanderhaeghen}},
  \bibinfo{journal}{Prog. Part. Nucl. Phys.} \textbf{\bibinfo{volume}{47}},
  \bibinfo{pages}{401} (\bibinfo{year}{2001}), \eprint{hep-ph/0106012}.

\bibitem[{\citenamefont{Brommel et~al.}(2008)}]{Brommel:2007xd}
\bibinfo{author}{\bibfnamefont{D.}~\bibnamefont{Brommel}} \bibnamefont{et~al.}
  (\bibinfo{collaboration}{UKQCD, QCDSF}), \bibinfo{journal}{Phys. Rev. Lett.}
  \textbf{\bibinfo{volume}{101}}, \bibinfo{pages}{122001}
  (\bibinfo{year}{2008}), \eprint{0708.2249}.

\bibitem[{\citenamefont{G{\"o}ckeler et~al.}(2004)\citenamefont{G{\"o}ckeler,
  Horsley, Pleiter, Rakow, Sch{\"a}fer, Schierholz, and
  Schroers}}]{Gockeler:2003jfa}
\bibinfo{author}{\bibfnamefont{M.}~\bibnamefont{G{\"o}ckeler}},
  \bibinfo{author}{\bibfnamefont{R.}~\bibnamefont{Horsley}},
  \bibinfo{author}{\bibfnamefont{D.}~\bibnamefont{Pleiter}},
  \bibinfo{author}{\bibfnamefont{P.~E.~L.} \bibnamefont{Rakow}},
  \bibinfo{author}{\bibfnamefont{A.}~\bibnamefont{Sch{\"a}fer}},
  \bibinfo{author}{\bibfnamefont{G.}~\bibnamefont{Schierholz}},
  \bibnamefont{and} \bibinfo{author}{\bibfnamefont{W.}~\bibnamefont{Schroers}}
  (\bibinfo{collaboration}{QCDSF}), \bibinfo{journal}{Phys. Rev. Lett.}
  \textbf{\bibinfo{volume}{92}}, \bibinfo{pages}{042002}
  (\bibinfo{year}{2004}), \eprint{hep-ph/0304249}.

\bibitem[{\citenamefont{H{\"a}gler et~al.}(2003)\citenamefont{H{\"a}gler,
  Negele, Renner, Schroers, Lippert, and Schilling}}]{Hagler:2003jd}
\bibinfo{author}{\bibfnamefont{P.}~\bibnamefont{H{\"a}gler}},
  \bibinfo{author}{\bibfnamefont{J.~W.} \bibnamefont{Negele}},
  \bibinfo{author}{\bibfnamefont{D.~B.} \bibnamefont{Renner}},
  \bibinfo{author}{\bibfnamefont{W.}~\bibnamefont{Schroers}},
  \bibinfo{author}{\bibfnamefont{T.}~\bibnamefont{Lippert}}, \bibnamefont{and}
  \bibinfo{author}{\bibfnamefont{K.}~\bibnamefont{Schilling}}
  (\bibinfo{collaboration}{LHPC, SESAM}), \bibinfo{journal}{Phys. Rev.}
  \textbf{\bibinfo{volume}{D68}}, \bibinfo{pages}{034505}
  (\bibinfo{year}{2003}), \eprint{hep-lat/0304018}.

\bibitem[{\citenamefont{Diehl and H{\"a}gler}(2005)}]{Diehl:2005jf}
\bibinfo{author}{\bibfnamefont{M.}~\bibnamefont{Diehl}} \bibnamefont{and}
  \bibinfo{author}{\bibfnamefont{P.}~\bibnamefont{H{\"a}gler}},
  \bibinfo{journal}{Eur. Phys. J.} \textbf{\bibinfo{volume}{C44}},
  \bibinfo{pages}{87} (\bibinfo{year}{2005}), \eprint{hep-ph/0504175}.

\bibitem[{\citenamefont{G{\"o}ckeler et~al.}(2005)\citenamefont{G{\"o}ckeler,
  H{\"a}gler, Horsley, Pleiter, Rakow, Sch{\"a}fer, Schierholz, and
  Zanotti}}]{Gockeler:2005cj}
\bibinfo{author}{\bibfnamefont{M.}~\bibnamefont{G{\"o}ckeler}},
  \bibinfo{author}{\bibfnamefont{P.}~\bibnamefont{H{\"a}gler}},
  \bibinfo{author}{\bibfnamefont{R.}~\bibnamefont{Horsley}},
  \bibinfo{author}{\bibfnamefont{D.}~\bibnamefont{Pleiter}},
  \bibinfo{author}{\bibfnamefont{P.~E.~L.} \bibnamefont{Rakow}},
  \bibinfo{author}{\bibfnamefont{A.}~\bibnamefont{Sch{\"a}fer}},
  \bibinfo{author}{\bibfnamefont{G.}~\bibnamefont{Schierholz}},
  \bibnamefont{and} \bibinfo{author}{\bibfnamefont{J.~M.}
  \bibnamefont{Zanotti}} (\bibinfo{collaboration}{UKQCD, QCDSF}),
  \bibinfo{journal}{Phys. Lett.} \textbf{\bibinfo{volume}{B627}},
  \bibinfo{pages}{113} (\bibinfo{year}{2005}), \eprint{hep-lat/0507001}.

\bibitem[{\citenamefont{G{\"o}ckeler et~al.}(2007)\citenamefont{G{\"o}ckeler,
  H{\"a}gler, Horsley, Nakamura, Pleiter, Rakow, Sch{\"a}fer, Schierholz,
  Stuben, and Zanotti}}]{Gockeler:2006zu}
\bibinfo{author}{\bibfnamefont{M.}~\bibnamefont{G{\"o}ckeler}},
  \bibinfo{author}{\bibfnamefont{P.}~\bibnamefont{H{\"a}gler}},
  \bibinfo{author}{\bibfnamefont{R.}~\bibnamefont{Horsley}},
  \bibinfo{author}{\bibfnamefont{Y.}~\bibnamefont{Nakamura}},
  \bibinfo{author}{\bibfnamefont{D.}~\bibnamefont{Pleiter}},
  \bibinfo{author}{\bibfnamefont{P.~E.~L.} \bibnamefont{Rakow}},
  \bibinfo{author}{\bibfnamefont{A.}~\bibnamefont{Sch{\"a}fer}},
  \bibinfo{author}{\bibfnamefont{G.}~\bibnamefont{Schierholz}},
  \bibinfo{author}{\bibfnamefont{H.}~\bibnamefont{Stuben}}, \bibnamefont{and}
  \bibinfo{author}{\bibfnamefont{J.~M.} \bibnamefont{Zanotti}}
  (\bibinfo{collaboration}{UKQCD, QCDSF}), \bibinfo{journal}{Phys. Rev. Lett.}
  \textbf{\bibinfo{volume}{98}}, \bibinfo{pages}{222001}
  (\bibinfo{year}{2007}), \eprint{hep-lat/0612032}.

\bibitem[{\citenamefont{H{\"a}gler et~al.}(2008)}]{Hagler:2007xi}
\bibinfo{author}{\bibfnamefont{P.}~\bibnamefont{H{\"a}gler}}
  \bibnamefont{et~al.} (\bibinfo{collaboration}{LHPC}), \bibinfo{journal}{Phys.
  Rev.} \textbf{\bibinfo{volume}{D77}}, \bibinfo{pages}{094502}
  (\bibinfo{year}{2008}), \eprint{0705.4295}.

\bibitem[{\citenamefont{Bratt et~al.}(2010)}]{Bratt:2010jn}
\bibinfo{author}{\bibfnamefont{J.~D.} \bibnamefont{Bratt}} \bibnamefont{et~al.}
  (\bibinfo{collaboration}{LHPC}), \bibinfo{journal}{Phys. Rev.}
  \textbf{\bibinfo{volume}{D82}}, \bibinfo{pages}{094502}
  (\bibinfo{year}{2010}), \eprint{1001.3620}.

\bibitem[{\citenamefont{Alexandrou et~al.}(2011)\citenamefont{Alexandrou,
  Carbonell, Constantinou, Harraud, Guichon, Jansen, Kallidonis, Korzec, and
  Papinutto}}]{Alexandrou:2011nr}
\bibinfo{author}{\bibfnamefont{C.}~\bibnamefont{Alexandrou}},
  \bibinfo{author}{\bibfnamefont{J.}~\bibnamefont{Carbonell}},
  \bibinfo{author}{\bibfnamefont{M.}~\bibnamefont{Constantinou}},
  \bibinfo{author}{\bibfnamefont{P.~A.} \bibnamefont{Harraud}},
  \bibinfo{author}{\bibfnamefont{P.}~\bibnamefont{Guichon}},
  \bibinfo{author}{\bibfnamefont{K.}~\bibnamefont{Jansen}},
  \bibinfo{author}{\bibfnamefont{C.}~\bibnamefont{Kallidonis}},
  \bibinfo{author}{\bibfnamefont{T.}~\bibnamefont{Korzec}}, \bibnamefont{and}
  \bibinfo{author}{\bibfnamefont{M.}~\bibnamefont{Papinutto}},
  \bibinfo{journal}{Phys. Rev.} \textbf{\bibinfo{volume}{D83}},
  \bibinfo{pages}{114513} (\bibinfo{year}{2011}), \eprint{1104.1600}.

\bibitem[{\citenamefont{H{\"a}gler}(2010)}]{Hagler:2009ni}
\bibinfo{author}{\bibfnamefont{P.}~\bibnamefont{H{\"a}gler}},
  \bibinfo{journal}{Phys. Rept.} \textbf{\bibinfo{volume}{490}},
  \bibinfo{pages}{49} (\bibinfo{year}{2010}), \eprint{0912.5483}.

\bibitem[{\citenamefont{Boer et~al.}(2011)}]{Boer:2011fh}
\bibinfo{author}{\bibfnamefont{D.}~\bibnamefont{Boer}} \bibnamefont{et~al.}
  (\bibinfo{year}{2011}), \eprint{1108.1713}.

\bibitem[{\citenamefont{Accardi et~al.}(2012)}]{Accardi:2012qut}
\bibinfo{author}{\bibfnamefont{A.}~\bibnamefont{Accardi}} \bibnamefont{et~al.}
  (\bibinfo{year}{2012}), \eprint{1212.1701}.

\bibitem[{\citenamefont{Kalantarians}(2014)}]{Kalantarians:2014eda}
\bibinfo{author}{\bibfnamefont{N.}~\bibnamefont{Kalantarians}},
  \bibinfo{journal}{J. Phys. Conf. Ser.} \textbf{\bibinfo{volume}{543}},
  \bibinfo{pages}{012008} (\bibinfo{year}{2014}).

\bibitem[{\citenamefont{Martinelli et~al.}(1995)\citenamefont{Martinelli,
  Pittori, Sachrajda, Testa, and Vladikas}}]{Martinelli:1994ty}
\bibinfo{author}{\bibfnamefont{G.}~\bibnamefont{Martinelli}},
  \bibinfo{author}{\bibfnamefont{C.}~\bibnamefont{Pittori}},
  \bibinfo{author}{\bibfnamefont{C.~T.} \bibnamefont{Sachrajda}},
  \bibinfo{author}{\bibfnamefont{M.}~\bibnamefont{Testa}}, \bibnamefont{and}
  \bibinfo{author}{\bibfnamefont{A.}~\bibnamefont{Vladikas}},
  \bibinfo{journal}{Nucl. Phys.} \textbf{\bibinfo{volume}{B445}},
  \bibinfo{pages}{81} (\bibinfo{year}{1995}), \eprint{hep-lat/9411010}.

\bibitem[{\citenamefont{Alexandrou et~al.}(2017)\citenamefont{Alexandrou,
  Constantinou, Hadjiyiannakou, Jansen, Panagopoulos, and
  Wiese}}]{Alexandrou:2016ekb}
\bibinfo{author}{\bibfnamefont{C.}~\bibnamefont{Alexandrou}},
  \bibinfo{author}{\bibfnamefont{M.}~\bibnamefont{Constantinou}},
  \bibinfo{author}{\bibfnamefont{K.}~\bibnamefont{Hadjiyiannakou}},
  \bibinfo{author}{\bibfnamefont{K.}~\bibnamefont{Jansen}},
  \bibinfo{author}{\bibfnamefont{H.}~\bibnamefont{Panagopoulos}},
  \bibnamefont{and} \bibinfo{author}{\bibfnamefont{C.}~\bibnamefont{Wiese}},
  \bibinfo{journal}{Phys. Rev.} \textbf{\bibinfo{volume}{D96}},
  \bibinfo{pages}{054503} (\bibinfo{year}{2017}), \eprint{1611.06901}.

\bibitem[{\citenamefont{Hoodbhoy and Ji}(1998)}]{Hoodbhoy:1998vm}
\bibinfo{author}{\bibfnamefont{P.}~\bibnamefont{Hoodbhoy}} \bibnamefont{and}
  \bibinfo{author}{\bibfnamefont{X.-D.} \bibnamefont{Ji}},
  \bibinfo{journal}{Phys. Rev.} \textbf{\bibinfo{volume}{D58}},
  \bibinfo{pages}{054006} (\bibinfo{year}{1998}), \eprint{hep-ph/9801369}.

\bibitem[{\citenamefont{Ji}(2013)}]{Ji:2013dva}
\bibinfo{author}{\bibfnamefont{X.}~\bibnamefont{Ji}}, \bibinfo{journal}{Phys.
  Rev. Lett.} \textbf{\bibinfo{volume}{110}}, \bibinfo{pages}{262002}
  (\bibinfo{year}{2013}), \eprint{1305.1539}.

\bibitem[{\citenamefont{Zhang et~al.}(2018)\citenamefont{Zhang, Ji,
  Sch{\"a}fer, Wang, and Zhao}}]{Zhang:2018diq}
\bibinfo{author}{\bibfnamefont{J.-H.} \bibnamefont{Zhang}},
  \bibinfo{author}{\bibfnamefont{X.}~\bibnamefont{Ji}},
  \bibinfo{author}{\bibfnamefont{A.}~\bibnamefont{Sch{\"a}fer}},
  \bibinfo{author}{\bibfnamefont{W.}~\bibnamefont{Wang}}, \bibnamefont{and}
  \bibinfo{author}{\bibfnamefont{S.}~\bibnamefont{Zhao}}
  (\bibinfo{year}{2018}), \eprint{1808.10824}.

\bibitem[{\citenamefont{Boer et~al.}(2016)\citenamefont{Boer, Cotogno, van
  Daal, Mulders, Signori, and Zhou}}]{Boer:2016dlh}
\bibinfo{author}{\bibfnamefont{D.}~\bibnamefont{Boer}},
  \bibinfo{author}{\bibfnamefont{S.}~\bibnamefont{Cotogno}},
  \bibinfo{author}{\bibfnamefont{T.}~\bibnamefont{van Daal}},
  \bibinfo{author}{\bibfnamefont{P.~J.} \bibnamefont{Mulders}},
  \bibinfo{author}{\bibfnamefont{A.}~\bibnamefont{Signori}}, \bibnamefont{and}
  \bibinfo{author}{\bibfnamefont{Y.}~\bibnamefont{Zhou}},
  \bibinfo{journal}{PoS} \textbf{\bibinfo{volume}{DIS2016}},
  \bibinfo{pages}{207} (\bibinfo{year}{2016}), \eprint{1609.02788}.

\bibitem[{\citenamefont{Jackiw}()}]{Jackiw}
\bibinfo{author}{\bibfnamefont{R.}~\bibnamefont{Jackiw}}.

\bibitem[{\citenamefont{Sheikholeslami and
  Wohlert}(1985)}]{Sheikholeslami:1985ij}
\bibinfo{author}{\bibfnamefont{B.}~\bibnamefont{Sheikholeslami}}
  \bibnamefont{and} \bibinfo{author}{\bibfnamefont{R.}~\bibnamefont{Wohlert}},
  \bibinfo{journal}{Nucl. Phys.} \textbf{\bibinfo{volume}{B259}},
  \bibinfo{pages}{572} (\bibinfo{year}{1985}).

\bibitem[{\citenamefont{L{\"u}scher and Weisz}(1985)}]{LuscherWeisz}
\bibinfo{author}{\bibfnamefont{M.}~\bibnamefont{L{\"u}scher}} \bibnamefont{and}
  \bibinfo{author}{\bibfnamefont{P.}~\bibnamefont{Weisz}},
  \bibinfo{journal}{Communications in Mathematical Physics}
  \textbf{\bibinfo{volume}{97}}, \bibinfo{pages}{59} (\bibinfo{year}{1985}),
  ISSN \bibinfo{issn}{1432-0916},
  \urlprefix\url{http://dx.doi.org/10.1007/BF01206178}.

\bibitem[{\citenamefont{Meinel}()}]{stefan}
\bibinfo{author}{\bibfnamefont{S.}~\bibnamefont{Meinel}},
  \emph{\bibinfo{title}{Private communication}}.

\bibitem[{\citenamefont{Orginos et~al.}(2015)\citenamefont{Orginos, Parre\~no,
  Savage, Beane, Chang, and Detmold}}]{PhysRevD.92.114512}
\bibinfo{author}{\bibfnamefont{K.}~\bibnamefont{Orginos}},
  \bibinfo{author}{\bibfnamefont{A.}~\bibnamefont{Parre\~no}},
  \bibinfo{author}{\bibfnamefont{M.~J.} \bibnamefont{Savage}},
  \bibinfo{author}{\bibfnamefont{S.~R.} \bibnamefont{Beane}},
  \bibinfo{author}{\bibfnamefont{E.}~\bibnamefont{Chang}}, \bibnamefont{and}
  \bibinfo{author}{\bibfnamefont{W.}~\bibnamefont{Detmold}}
  (\bibinfo{collaboration}{NPLQCD Collaboration}), \bibinfo{journal}{Phys. Rev.
  D} \textbf{\bibinfo{volume}{92}}, \bibinfo{pages}{114512}
  (\bibinfo{year}{2015}),
  \urlprefix\url{http://link.aps.org/doi/10.1103/PhysRevD.92.114512}.

\bibitem[{\citenamefont{L{\"u}scher}(2010)}]{Luscher:2010iy}
\bibinfo{author}{\bibfnamefont{M.}~\bibnamefont{L{\"u}scher}},
  \bibinfo{journal}{JHEP} \textbf{\bibinfo{volume}{08}}, \bibinfo{pages}{071}
  (\bibinfo{year}{2010}), \bibinfo{note}{[Erratum: JHEP03,092(2014)]},
  \eprint{1006.4518}.

\bibitem[{\citenamefont{Detmold and Shanahan}(2016)}]{Detmold:2016gpy}
\bibinfo{author}{\bibfnamefont{W.}~\bibnamefont{Detmold}} \bibnamefont{and}
  \bibinfo{author}{\bibfnamefont{P.~E.} \bibnamefont{Shanahan}},
  \bibinfo{journal}{Phys. Rev.} \textbf{\bibinfo{volume}{D94}},
  \bibinfo{pages}{014507} (\bibinfo{year}{2016}), \eprint{1606.04505}.

\bibitem[{\citenamefont{Detmold et~al.}(2017)\citenamefont{Detmold, Pefkou, and
  Shanahan}}]{Detmold:2017oqb}
\bibinfo{author}{\bibfnamefont{W.}~\bibnamefont{Detmold}},
  \bibinfo{author}{\bibfnamefont{D.}~\bibnamefont{Pefkou}}, \bibnamefont{and}
  \bibinfo{author}{\bibfnamefont{P.~E.} \bibnamefont{Shanahan}},
  \bibinfo{journal}{Phys. Rev.} \textbf{\bibinfo{volume}{D95}},
  \bibinfo{pages}{114515} (\bibinfo{year}{2017}), \eprint{1703.08220}.

\bibitem[{\citenamefont{Mandula et~al.}(1983)\citenamefont{Mandula, Zweig, and
  Govaerts}}]{Mandula:1983ut}
\bibinfo{author}{\bibfnamefont{J.~E.} \bibnamefont{Mandula}},
  \bibinfo{author}{\bibfnamefont{G.}~\bibnamefont{Zweig}}, \bibnamefont{and}
  \bibinfo{author}{\bibfnamefont{J.}~\bibnamefont{Govaerts}},
  \bibinfo{journal}{Nucl. Phys.} \textbf{\bibinfo{volume}{B228}},
  \bibinfo{pages}{91} (\bibinfo{year}{1983}).

\bibitem[{\citenamefont{G{\"o}ckeler et~al.}(1996)\citenamefont{G{\"o}ckeler,
  Horsley, Ilgenfritz, Perlt, Rakow, Schierholz, and
  Schiller}}]{Gockeler:1996mu}
\bibinfo{author}{\bibfnamefont{M.}~\bibnamefont{G{\"o}ckeler}},
  \bibinfo{author}{\bibfnamefont{R.}~\bibnamefont{Horsley}},
  \bibinfo{author}{\bibfnamefont{E.-M.} \bibnamefont{Ilgenfritz}},
  \bibinfo{author}{\bibfnamefont{H.}~\bibnamefont{Perlt}},
  \bibinfo{author}{\bibfnamefont{P.~E.~L.} \bibnamefont{Rakow}},
  \bibinfo{author}{\bibfnamefont{G.}~\bibnamefont{Schierholz}},
  \bibnamefont{and} \bibinfo{author}{\bibfnamefont{A.}~\bibnamefont{Schiller}},
  \bibinfo{journal}{Phys. Rev.} \textbf{\bibinfo{volume}{D54}},
  \bibinfo{pages}{5705} (\bibinfo{year}{1996}), \eprint{hep-lat/9602029}.

\bibitem[{\citenamefont{Martinelli et~al.}(1993)\citenamefont{Martinelli,
  Petrarca, Sachrajda, and Vladikas}}]{Martinelli:1993dq}
\bibinfo{author}{\bibfnamefont{G.}~\bibnamefont{Martinelli}},
  \bibinfo{author}{\bibfnamefont{S.}~\bibnamefont{Petrarca}},
  \bibinfo{author}{\bibfnamefont{C.~T.} \bibnamefont{Sachrajda}},
  \bibnamefont{and} \bibinfo{author}{\bibfnamefont{A.}~\bibnamefont{Vladikas}},
  \bibinfo{journal}{Phys. Lett.} \textbf{\bibinfo{volume}{B311}},
  \bibinfo{pages}{241} (\bibinfo{year}{1993}), \bibinfo{note}{[Erratum: Phys.
  Lett.B317,660(1993)]}.

\bibitem[{\citenamefont{Yang et~al.}(2018{\natexlab{a}})\citenamefont{Yang,
  Gong, Liang, Lin, Liu, Pefkou, and Shanahan}}]{Yang:2018bft}
\bibinfo{author}{\bibfnamefont{Y.-B.} \bibnamefont{Yang}},
  \bibinfo{author}{\bibfnamefont{M.}~\bibnamefont{Gong}},
  \bibinfo{author}{\bibfnamefont{J.}~\bibnamefont{Liang}},
  \bibinfo{author}{\bibfnamefont{H.-W.} \bibnamefont{Lin}},
  \bibinfo{author}{\bibfnamefont{K.-F.} \bibnamefont{Liu}},
  \bibinfo{author}{\bibfnamefont{D.}~\bibnamefont{Pefkou}}, \bibnamefont{and}
  \bibinfo{author}{\bibfnamefont{P.}~\bibnamefont{Shanahan}}
  (\bibinfo{year}{2018}{\natexlab{a}}), \eprint{1805.00531}.

\bibitem[{\citenamefont{Yang et~al.}(2018{\natexlab{b}})\citenamefont{Yang,
  Liang, Bi, Chen, Draper, Liu, and Liu}}]{Yang:2018nqn}
\bibinfo{author}{\bibfnamefont{Y.-B.} \bibnamefont{Yang}},
  \bibinfo{author}{\bibfnamefont{J.}~\bibnamefont{Liang}},
  \bibinfo{author}{\bibfnamefont{Y.-J.} \bibnamefont{Bi}},
  \bibinfo{author}{\bibfnamefont{Y.}~\bibnamefont{Chen}},
  \bibinfo{author}{\bibfnamefont{T.}~\bibnamefont{Draper}},
  \bibinfo{author}{\bibfnamefont{K.-F.} \bibnamefont{Liu}}, \bibnamefont{and}
  \bibinfo{author}{\bibfnamefont{Z.}~\bibnamefont{Liu}}
  (\bibinfo{year}{2018}{\natexlab{b}}), \eprint{1808.08677}.

\bibitem[{\citenamefont{Yang et~al.}(2016)\citenamefont{Yang, Glatzmaier, Liu,
  and Zhao}}]{Yang:2016xsb}
\bibinfo{author}{\bibfnamefont{Y.-B.} \bibnamefont{Yang}},
  \bibinfo{author}{\bibfnamefont{M.}~\bibnamefont{Glatzmaier}},
  \bibinfo{author}{\bibfnamefont{K.-F.} \bibnamefont{Liu}}, \bibnamefont{and}
  \bibinfo{author}{\bibfnamefont{Y.}~\bibnamefont{Zhao}}
  (\bibinfo{year}{2016}), \eprint{1612.02855}.

\bibitem[{\citenamefont{Collins and Scalise}(1994)}]{Collins:1994ee}
\bibinfo{author}{\bibfnamefont{J.~C.} \bibnamefont{Collins}} \bibnamefont{and}
  \bibinfo{author}{\bibfnamefont{R.~J.} \bibnamefont{Scalise}},
  \bibinfo{journal}{Phys. Rev.} \textbf{\bibinfo{volume}{D50}},
  \bibinfo{pages}{4117} (\bibinfo{year}{1994}), \eprint{hep-ph/9403231}.

\bibitem[{\citenamefont{G{\"o}ckeler et~al.}(2010)}]{Gockeler:2010yr}
\bibinfo{author}{\bibfnamefont{M.}~\bibnamefont{G{\"o}ckeler}}
  \bibnamefont{et~al.}, \bibinfo{journal}{Phys. Rev.}
  \textbf{\bibinfo{volume}{D82}}, \bibinfo{pages}{114511}
  (\bibinfo{year}{2010}), \bibinfo{note}{[Erratum: Phys.
  Rev.D86,099903(2012)]}, \eprint{1003.5756}.

\bibitem[{\citenamefont{Herren and Steinhauser}(2018)}]{Herren:2017osy}
\bibinfo{author}{\bibfnamefont{F.}~\bibnamefont{Herren}} \bibnamefont{and}
  \bibinfo{author}{\bibfnamefont{M.}~\bibnamefont{Steinhauser}},
  \bibinfo{journal}{Comput. Phys. Commun.} \textbf{\bibinfo{volume}{224}},
  \bibinfo{pages}{333} (\bibinfo{year}{2018}), \eprint{1703.03751}.

\bibitem[{\citenamefont{Schmidt and Steinhauser}(2012)}]{Schmidt:2012az}
\bibinfo{author}{\bibfnamefont{B.}~\bibnamefont{Schmidt}} \bibnamefont{and}
  \bibinfo{author}{\bibfnamefont{M.}~\bibnamefont{Steinhauser}},
  \bibinfo{journal}{Comput. Phys. Commun.} \textbf{\bibinfo{volume}{183}},
  \bibinfo{pages}{1845} (\bibinfo{year}{2012}), \eprint{1201.6149}.

\bibitem[{\citenamefont{Chetyrkin et~al.}(2000)\citenamefont{Chetyrkin, Kuhn,
  and Steinhauser}}]{Chetyrkin:2000yt}
\bibinfo{author}{\bibfnamefont{K.~G.} \bibnamefont{Chetyrkin}},
  \bibinfo{author}{\bibfnamefont{J.~H.} \bibnamefont{Kuhn}}, \bibnamefont{and}
  \bibinfo{author}{\bibfnamefont{M.}~\bibnamefont{Steinhauser}},
  \bibinfo{journal}{Comput. Phys. Commun.} \textbf{\bibinfo{volume}{133}},
  \bibinfo{pages}{43} (\bibinfo{year}{2000}), \eprint{hep-ph/0004189}.

\bibitem[{\citenamefont{Falcioni et~al.}(1985)\citenamefont{Falcioni, Paciello,
  Parisi, and Taglienti}}]{Falcioni:1984ei}
\bibinfo{author}{\bibfnamefont{M.}~\bibnamefont{Falcioni}},
  \bibinfo{author}{\bibfnamefont{M.~L.} \bibnamefont{Paciello}},
  \bibinfo{author}{\bibfnamefont{G.}~\bibnamefont{Parisi}}, \bibnamefont{and}
  \bibinfo{author}{\bibfnamefont{B.}~\bibnamefont{Taglienti}},
  \bibinfo{journal}{Nucl. Phys.} \textbf{\bibinfo{volume}{B251}},
  \bibinfo{pages}{624} (\bibinfo{year}{1985}).

\bibitem[{\citenamefont{Capitani et~al.}(2012)\citenamefont{Capitani,
  Della~Morte, von Hippel, J{\"a}ger, J{\"u}ttner, Knippschild, Meyer, and
  Wittig}}]{Capitani:2012gj}
\bibinfo{author}{\bibfnamefont{S.}~\bibnamefont{Capitani}},
  \bibinfo{author}{\bibfnamefont{M.}~\bibnamefont{Della~Morte}},
  \bibinfo{author}{\bibfnamefont{G.}~\bibnamefont{von Hippel}},
  \bibinfo{author}{\bibfnamefont{B.}~\bibnamefont{J{\"a}ger}},
  \bibinfo{author}{\bibfnamefont{A.}~\bibnamefont{J{\"u}ttner}},
  \bibinfo{author}{\bibfnamefont{B.}~\bibnamefont{Knippschild}},
  \bibinfo{author}{\bibfnamefont{H.~B.} \bibnamefont{Meyer}}, \bibnamefont{and}
  \bibinfo{author}{\bibfnamefont{H.}~\bibnamefont{Wittig}},
  \bibinfo{journal}{Phys. Rev.} \textbf{\bibinfo{volume}{D86}},
  \bibinfo{pages}{074502} (\bibinfo{year}{2012}), \eprint{1205.0180}.

\bibitem[{\citenamefont{Hill and Paz}(2010)}]{Hill:2010yb}
\bibinfo{author}{\bibfnamefont{R.~J.} \bibnamefont{Hill}} \bibnamefont{and}
  \bibinfo{author}{\bibfnamefont{G.}~\bibnamefont{Paz}},
  \bibinfo{journal}{Phys. Rev.} \textbf{\bibinfo{volume}{D82}},
  \bibinfo{pages}{113005} (\bibinfo{year}{2010}), \eprint{1008.4619}.

\bibitem[{\citenamefont{Tanabashi et~al.}(2018)}]{Tanabashi:2018oca}
\bibinfo{author}{\bibfnamefont{M.}~\bibnamefont{Tanabashi}}
  \bibnamefont{et~al.} (\bibinfo{collaboration}{Particle Data Group}),
  \bibinfo{journal}{Phys. Rev.} \textbf{\bibinfo{volume}{D98}},
  \bibinfo{pages}{030001} (\bibinfo{year}{2018}).

\bibitem[{\citenamefont{Deka et~al.}(2015)}]{Deka:2013zha}
\bibinfo{author}{\bibfnamefont{M.}~\bibnamefont{Deka}} \bibnamefont{et~al.},
  \bibinfo{journal}{Phys. Rev.} \textbf{\bibinfo{volume}{D91}},
  \bibinfo{pages}{014505} (\bibinfo{year}{2015}), \eprint{1312.4816}.

\bibitem[{\citenamefont{Horsley et~al.}(2012)\citenamefont{Horsley, Millo,
  Nakamura, Perlt, Pleiter, Rakow, Schierholz, Schiller, Winter, and
  Zanotti}}]{Horsley:2012pz}
\bibinfo{author}{\bibfnamefont{R.}~\bibnamefont{Horsley}},
  \bibinfo{author}{\bibfnamefont{R.}~\bibnamefont{Millo}},
  \bibinfo{author}{\bibfnamefont{Y.}~\bibnamefont{Nakamura}},
  \bibinfo{author}{\bibfnamefont{H.}~\bibnamefont{Perlt}},
  \bibinfo{author}{\bibfnamefont{D.}~\bibnamefont{Pleiter}},
  \bibinfo{author}{\bibfnamefont{P.~E.~L.} \bibnamefont{Rakow}},
  \bibinfo{author}{\bibfnamefont{G.}~\bibnamefont{Schierholz}},
  \bibinfo{author}{\bibfnamefont{A.}~\bibnamefont{Schiller}},
  \bibinfo{author}{\bibfnamefont{F.}~\bibnamefont{Winter}}, \bibnamefont{and}
  \bibinfo{author}{\bibfnamefont{J.~M.} \bibnamefont{Zanotti}}
  (\bibinfo{collaboration}{UKQCD, QCDSF}), \bibinfo{journal}{Phys. Lett.}
  \textbf{\bibinfo{volume}{B714}}, \bibinfo{pages}{312} (\bibinfo{year}{2012}),
  \eprint{1205.6410}.

\bibitem[{\citenamefont{G{\"o}ckeler et~al.}(1997)\citenamefont{G{\"o}ckeler,
  Horsley, Ilgenfritz, Oelrich, Perlt, Rakow, Schierholz, Schiller, and
  Stephenson}}]{Gockeler:1996zg}
\bibinfo{author}{\bibfnamefont{M.}~\bibnamefont{G{\"o}ckeler}},
  \bibinfo{author}{\bibfnamefont{R.}~\bibnamefont{Horsley}},
  \bibinfo{author}{\bibfnamefont{E.-M.} \bibnamefont{Ilgenfritz}},
  \bibinfo{author}{\bibfnamefont{H.}~\bibnamefont{Oelrich}},
  \bibinfo{author}{\bibfnamefont{H.}~\bibnamefont{Perlt}},
  \bibinfo{author}{\bibfnamefont{P.~E.~L.} \bibnamefont{Rakow}},
  \bibinfo{author}{\bibfnamefont{G.}~\bibnamefont{Schierholz}},
  \bibinfo{author}{\bibfnamefont{A.}~\bibnamefont{Schiller}}, \bibnamefont{and}
  \bibinfo{author}{\bibfnamefont{P.}~\bibnamefont{Stephenson}},
  \bibinfo{journal}{Nucl. Phys. Proc. Suppl.} \textbf{\bibinfo{volume}{53}},
  \bibinfo{pages}{324} (\bibinfo{year}{1997}), \eprint{hep-lat/9608017}.

\bibitem[{\citenamefont{Dulat et~al.}(2016)\citenamefont{Dulat, Hou, Gao,
  Guzzi, Huston, Nadolsky, Pumplin, Schmidt, Stump, and Yuan}}]{Dulat:2015mca}
\bibinfo{author}{\bibfnamefont{S.}~\bibnamefont{Dulat}},
  \bibinfo{author}{\bibfnamefont{T.-J.} \bibnamefont{Hou}},
  \bibinfo{author}{\bibfnamefont{J.}~\bibnamefont{Gao}},
  \bibinfo{author}{\bibfnamefont{M.}~\bibnamefont{Guzzi}},
  \bibinfo{author}{\bibfnamefont{J.}~\bibnamefont{Huston}},
  \bibinfo{author}{\bibfnamefont{P.}~\bibnamefont{Nadolsky}},
  \bibinfo{author}{\bibfnamefont{J.}~\bibnamefont{Pumplin}},
  \bibinfo{author}{\bibfnamefont{C.}~\bibnamefont{Schmidt}},
  \bibinfo{author}{\bibfnamefont{D.}~\bibnamefont{Stump}}, \bibnamefont{and}
  \bibinfo{author}{\bibfnamefont{C.~P.} \bibnamefont{Yuan}},
  \bibinfo{journal}{Phys. Rev.} \textbf{\bibinfo{volume}{D93}},
  \bibinfo{pages}{033006} (\bibinfo{year}{2016}), \eprint{1506.07443}.

\bibitem[{\citenamefont{Meyer and Negele}(2008)}]{Meyer:2007tm}
\bibinfo{author}{\bibfnamefont{H.~B.} \bibnamefont{Meyer}} \bibnamefont{and}
  \bibinfo{author}{\bibfnamefont{J.~W.} \bibnamefont{Negele}},
  \bibinfo{journal}{Phys. Rev.} \textbf{\bibinfo{volume}{D77}},
  \bibinfo{pages}{037501} (\bibinfo{year}{2008}), \eprint{0707.3225}.

\bibitem[{\citenamefont{Gl{\"u}ck et~al.}(1999)\citenamefont{Gl{\"u}ck, Reya,
  and Schienbein}}]{Gluck:1999xe}
\bibinfo{author}{\bibfnamefont{M.}~\bibnamefont{Gl{\"u}ck}},
  \bibinfo{author}{\bibfnamefont{E.}~\bibnamefont{Reya}}, \bibnamefont{and}
  \bibinfo{author}{\bibfnamefont{I.}~\bibnamefont{Schienbein}},
  \bibinfo{journal}{Eur. Phys. J.} \textbf{\bibinfo{volume}{C10}},
  \bibinfo{pages}{313} (\bibinfo{year}{1999}), \eprint{hep-ph/9903288}.

\bibitem[{\citenamefont{Brommel}(2007)}]{Brommel:2007zz}
\bibinfo{author}{\bibfnamefont{D.}~\bibnamefont{Brommel}}, Ph.D. thesis,
  \bibinfo{school}{Regensburg U.} (\bibinfo{year}{2007}).

\bibitem[{\citenamefont{Polyakov}(1999)}]{Polyakov:1998ze}
\bibinfo{author}{\bibfnamefont{M.~V.} \bibnamefont{Polyakov}},
  \bibinfo{journal}{Nucl. Phys.} \textbf{\bibinfo{volume}{B555}},
  \bibinfo{pages}{231} (\bibinfo{year}{1999}), \eprint{hep-ph/9809483}.

\bibitem[{\citenamefont{Edwards and Joo}(2005)}]{Edwards:2004sx}
\bibinfo{author}{\bibfnamefont{R.~G.} \bibnamefont{Edwards}} \bibnamefont{and}
  \bibinfo{author}{\bibfnamefont{B.}~\bibnamefont{Joo}}
  (\bibinfo{collaboration}{SciDAC, LHPC, UKQCD}), \bibinfo{journal}{Nucl. Phys.
  Proc. Suppl.} \textbf{\bibinfo{volume}{140}}, \bibinfo{pages}{832}
  (\bibinfo{year}{2005}), \eprint{hep-lat/0409003}.

\end{thebibliography}

\end{document}